\documentclass[12pt]{article}
\usepackage{amsmath}
\usepackage{graphicx}
\usepackage{enumerate}
\usepackage{natbib}
\usepackage{url} 
\usepackage{color}
\usepackage{natbib,amsmath,amssymb,amsthm,graphicx,setspace,paralist,booktabs,rotating,times}
\usepackage{subfigure}
\usepackage{mathrsfs}
\usepackage{verbatim}
\usepackage{bbm}
\usepackage{url}
\usepackage[ruled,vlined]{algorithm2e}
\usepackage{algpseudocode}
\usepackage{mathrsfs}
\usepackage{float}
\usepackage{multirow}
\usepackage{multicol}
\usepackage{makecell}
\usepackage{array}
\usepackage{lipsum}

\newcommand{\blind}{1}

\newtheorem{proposition}{Proposition}
\newtheorem{theorem}{Theorem}
\newtheorem{lemma}{Lemma}
\newtheorem{remark}{Remark}

\theoremstyle{definition}

\newcommand \onecolumngrid{
	\do@columngrid{one}{\@ne}%
	\def\set@footnotewidth{\onecolumngrid}
	\def\footnoterule{\kern-6pt\hrule width 1.5in\kern6pt}%
}
\algnewcommand{\algorithmicand}{\textbf{ and }}
\algnewcommand{\algorithmicor}{\textbf{ or }}
\algnewcommand{\OR}{\algorithmicor}
\algnewcommand{\AND}{\algorithmicand}

\DeclareMathOperator\tr{tr}

\DeclareMathOperator*\argmin{\arg\min}
\def\pr{\mathbb{P}}
\def\bA{\mathbf{A}}
\def\bB{\mathbf{B}}
\def\bF{\mathbf{F}}
\def\bI{\mathbf{I}}
\def\by{\mathbf{y}}
\def\bQ{\mathbf{Q}}
\def\bS{\mathbf{S}}
\def\bH{\mathbf{H}}
\def\bT{\mathbf\top}
\def\bU{\mathbf{U}}
\def\bW{\mathbf{W}}
\def\bX{\mathbf{X}}
\def\bY{\mathbf{Y}}
\def\bZ{\mathbf{Z}}
\def\bP{\mathbf{P}}
\def\bx{\mathbf{x}}
\def\bM{\mathbf{M}}
\def\bS{\mathbf{S}}
\def\bG{\mathbf{G}}
\def\bL{\mathbf{L}}
\def\bN{\mathbf{N}}
\def\bR{\mathbf{R}}
\def\bD{\mathbf{D}}
\def\bc{\mathbf{c}}
\def\bzero{\mathbf{0}}
\def\bnu{\boldsymbol{\nu}}
\def\bGamma{\boldsymbol{\Gamma}}
\def\bOmega{\boldsymbol{\Omega}}
\def\bSigma{\boldsymbol{\Sigma}}
\def\balpha{\boldsymbol{\alpha}}
\def\bgamma{\boldsymbol{\gamma}}
\def\bomega{\boldsymbol{\omega}}
\def\bTheta{\boldsymbol{\Theta}}

\def\btheta{\boldsymbol{\theta}}
\def\bLambda{\boldsymbol{\Lambda}}
\def\blambda{\boldsymbol{\lambda}}
\def\bbeta{\boldsymbol{\eta}}
\def\bpi{\boldsymbol{\pi}}
\def\bl{\boldsymbol{l}}
\def\bmu{\boldsymbol{\mu}}
\def\bzero{\mathbf{0}}
\def\bone{\mathbf{1}}
\def\cA{\mathcal{A}}
\def\cM{\mathcal{M}}
\def\cU{\mathcal{U}}
\def\what{\widehat}
\def\wtilde{\widetilde}
\def\ve{\varepsilon}
\def\thefootnote{}
\def\qedsymbol{}
\def\vecone{{\rm vech_1}}
\def\vectwo{{\rm vech_2}}
\def\Y{{\bf Y}}
\def\K{{\bf K}}
\def\R{{\bf R}}
\def\trans{^{\rm T}}
\def\mR{\mathbf{R}}
\def\vecone{{\rm vech_1}}
\def\vectwo{{\rm vech_2}}
\def\Y{{\bf Y}}
\def\K{{\bf K}}
\def\R{{\bf R}}
\def\trans{^{\rm T}}
\def\mR{\mathbf{R}}

\begin{document}

\def\spacingset#1{\renewcommand{\baselinestretch}%
{#1}\small\normalsize} \spacingset{1}


\if1\blind
{
  \title{\bf Single-cell gene regulatory network analysis for mixed cell populations}
  \author{Junjie Tang \hspace{.2cm}\\
    School of Mathematical Sciences, Peking University\\
    Center for Statistial Sciences, Peking University\\
    and \\
    Changhu Wang\hspace{.2cm}\\
    School of Mathematical Sciences, Peking University\\
    and \\
    Feiyi Xiao\hspace{.2cm}\\
    School of Mathematical Sciences, Peking University\\
    and \\
    Ruibin Xi \thanks{
    	The authors gratefully acknowledge \textit{the National Key Basic Research Project of China (2020YFE0204000), the National Natural Science Foundation of China (11971039), and Sino-Russia  Mathematics Center.}}\\
    School of Mathematical Sciences, Peking University\\
    Center for Statistial Sciences, Peking University}
  \maketitle
} \fi

\if0\blind
{
  \bigskip
  \bigskip
  \bigskip
  \begin{center}
    {\LARGE\bf  Single-cell gene regulatory network analysis for mixed cell populations}
\end{center}
  \medskip
} \fi

\bigskip
\begin{abstract}
Gene regulatory network (GRN) refers to the complex network formed by regulatory interactions between genes in living cells. In this paper, we consider inferring GRNs in single cells based on single cell RNA sequencing (scRNA-seq) data. In scRNA-seq, single cells are often profiled from mixed populations and their cell identities are unknown. A common practice for single cell GRN analysis is to first cluster the cells and infer GRNs for every cluster separately. However, this two-step procedure ignores uncertainty in the clustering step and thus could lead to inaccurate estimation of the networks. To address this problem, we propose to model scRNA-seq by the mixture multivariate Poisson log-normal (MPLN) distribution. The precision matrices of the MPLN are the GRNs of different cell types and can be jointly estimated by maximizing MPLN’s lasso-penalized  log-likelihood. We show that the MPLN model is identifiable and the resulting penalized log-likelihood estimator is consistent. To avoid the intractable optimization of the MPLN’s log-likelihood, we develop an algorithm called VMPLN based on the variational inference method. Comprehensive simulation and real scRNA-seq data analyses reveal that VMPLN performs better than the state-of-the-art single cell GRN methods.  
\end{abstract}

\noindent%

\vfill

\newpage
\spacingset{1.9} 
\section{Introduction}
\label{sec:intro}
Gene regulatory network (GRN), representing the regulatory relationships between genes, is important for understanding the complex biological system \citep{arendt2016origin}. GRNs can be inferred based on gene expression data such as RNA sequencing (RNA-seq) data. Bulk expression data are most commonly used for GRN inference and numerous methods have been developed (see \citet{marbach2012wisdom} and references therein). However, bulk data are profiled from pooled cell populations and thus can only provide average expressions of many cells. The recent development of single-cell RNA-seq (scRNA-seq) technologies can measure gene expression at the single-cell level \citep{gohil2021applying,nam2021integrating}, thus offering unprecedented opportunity for single-cell GRN inference. 

To account for the unique features of scRNA-seq data, a number of GRN inference methods based on scRNA-seq data have been developed \citep{aibar2017scenic,specht2017leap,chan2017gene}. These methods often make the implicit assumption that all cells share the same GRN. However, single cells in scRNA-seq data usually belong to multiple cell types and each cell type has its own specific GRN and expression pattern. The cell identities are unknown and have to be determined using scRNA-seq data. To infer GRNs of different cell types, one has to first assign single cells to different cell types (e.g. by clustering) and then estimate the GRNs using available methods. This two-step procedure can provide accurate GRN estimation if different cell types are well separated. If, instead, different cell types have a higher mixing degree, a large proportion of cells cannot be confidently assigned to a cell type and the ambiguity of the cell type assignment could seriously influence the performance of GRN inference. 

Here, we consider developing a GRN inference method based on scRNA-seq data with mixed cell populations using mixture models. One major advantage of mixture models is that cell types need not be predetermined before GRN inference. Instead, mixture models can allow joint analyses of clustering and GRN inference, and thus could give better GRN estimation when different cell types are poorly separated. Since scRNA-seq data are count data, often rather small count data with many zeros (the high dropout problem), graphical models for count data would be more suitable than the widely used Gaussian graphical model (GGM) \citep{meinshausen2006high,friedman2008sparse} or its mixture version. Available graphical models for count data include Poisson graphical models (PGMs) \citep{yang2012graphical,allen2013local} and Poisson log-normal (PLN) models \citep{Wu2018,chiquet2019variational,silva2019multivariate}. Compared with PGMs, the PLN models can model the over-dispersion commonly observed in scRNA-seq data \citep{ziegenhain2017comparative}. Therefore, we propose to use the mixture PLN (MPLN) model for GRN inference in single cells.

A non-negative integer random vector $\bY = (Y_1,\cdots,Y_p)^T \in \mathbf{R}^p$ follows a PLN distribution, if conditional on a latent random vector $\bX = (X_1,\cdots,X_p)^T \in \mathbf{R}^p$ with $\bX \sim \mbox{N}(\bmu,\bSigma)$, each element of $\bY$ independently follows the univariate Poisson distribution, i.e. $Y_j \sim \mbox{Poisson}(\exp(X_j))$ ($j = 1,2,\cdots,p$). Similar to the GGM, the network of the PLN model is the precision matrix $\bTheta = {\bSigma}^{-1}$ of the latent variable $\bX$. The MPLN model is a mixture of $G$ different PLN models. The precision matrix of each component of the MPLN model represents the network of a cell type. Assuming that the networks are sparse, we can maximize the lasso-penalized log-likelihood of the MPLN model to estimate the networks. 

We first establish the basic properties of the MPLN model: the MPLN model is identifiable, and its Fisher information matrix is positive definite. We further show that the network estimator by maximizing the lasso-penalized log-likelihood of the MPLN model is consistent. As a special case, this result also establishes the consistency result for the PLN model, which has been lacking in the literature. Directly maximizing the lasso-penalized log-likelihood of the MPLN is computationally intractable. We adopt the variational inference approach \citep{jordan1999introduction,wainwright2008graphical} and develop an algorithm called variational mixture Poisson log-normal (VMPLN) for simultaneous analyses of clustering and network inference. 

We compare VMPLN with popular graphical methods and state-of-the-art single cell regulatory network inference methods. Comprehensive simulation shows that VMPLN achieves better performance especially in the scenarios that different cell types have a high mixing degree. Benchmarking on real scRNA-seq data also demonstrates that VMPLN can provide more accurate network estimation in most cases. Finally, we apply VMPLN to a large scRNA-seq dataset from patients infected with severe acute respiratory syndrome coronavirus 2 (SARS-CoV-2) and find that VMPLN identifies critical differences of regulatory networks in immune cells between patients with moderate and severe symptoms.

The paper is organized as follows. Section \ref{sec:Model} presents the MPLN model and its theoretical properties. Section \ref{sec:Algorithm} derives the VMPLN algorithm. Simulation and real data analyses are in Section \ref{sec:simu} and \ref{sec:appl}, respectively. All proofs of the theoretical properties are in supplementary material. All data used and corresponding source codes are available and can be accessed at https://github.com/XiDsLab/scGeneNet.

\section{Model and Theoretical properties}
\label{sec:Model}
\subsection{The Mixture Poisson log-normal model for scRNA-seq data}
Suppose that a scRNA-seq dataset consists of $n$ cells and $p$ genes. Let $\mathbf{Y}_{i}=\left(Y_{i 1}, \cdots, Y_{i p}\right)^T$ be the observed expression vector of the $i$th cell, where $Y_{ij}$'s are all non-negative integers. Let $\bY = (\bY_1,\cdots,\bY_n)^T$ be the observed count matrix. Single cells in scRNA-seq data belong to $G$ different cell types and each cell type has its own unique mean gene expression and regulatory network. The cell identities are unknown and have to be determined based on the observed data $\bY$. The observed expression $Y_{ij}$ is a noisy measurement of the true expression $\exp\left(X_{ij}\right)$ of the $i$th cell at the $j$th gene. Conditional on $X_{ij}$, we assume that $Y_{ij}$ follows a Poisson distribution with a mean $\lambda_{i j} = l_i \exp\left(X_{ij}\right)$, where $l_i$ is the library size of the $i$th cell and can be readily estimated using available methods \citep{hafemeister2019normalization,lun2016pooling}. Denote $\bX_i = \left(X_{i 1}, \cdots, X_{i p}\right)^T$ be the logarithm of the true expression vector of the $i$th cell. We assume that the logarithm of expressions of single cells in the $g$th cell type are normally distributed with a mean $\bmu_g$ and a covariance $\bTheta_g^{-1}$. Thus, given the underlying cell type $Z_i = g$ ($g=1,\cdots,G$) of the $i$th cell, the conditional distribution of $\bX_i$ is $\mbox{N}\left( \bmu_g ,{\bTheta_g}^{-1} \right)$. We further assume that $Z_i$ follows a multinomial distribution $\mbox{Multinomial} (1,\bpi)$, where $\bpi = (\pi_1,\cdots,\pi_{G})^T$ is the proportion parameter representing the composition of cell types. In summary, we have the following MPLN model 
\begin{equation}\label{equ0}
\begin{aligned}
\bY_i |\bX_i &\sim \prod_{j=1}^p\mbox{Poisson}\left[l_i\ \exp \left(X_{i j}\right)\right], \\
\bX_i|Z_i = g &\sim \mbox{N}\left( \bmu_g ,{\bTheta_g}^{-1} \right), \bTheta_g \succ 0,\\
Z_i &\sim \mbox{Multinomial} (1,\bpi),
\end{aligned}
\end{equation}
where $\bTheta_g \succ 0$ means that $\bTheta_g$ is positive definite.

Denote $\btheta = \big(\bpi, \bmu=\{\bmu_g\}_{g = 1}^G,\bTheta = \{\bTheta_g \}_{g = 1}^G\big)$ as the set of unknown model parameters. Let $p(\bY_i|\bX_i) = \prod_{j = 1}^{p} \left\{\left[l_i \exp \left(X_{i j}\right)\right]^{Y_{i j}} \exp\left[-l_i \exp \left(X_{i j}\right)\right](Y_{i j }!)^{-1}\right\}$ be the conditional probability mass function of $\bY_i$ given $\bX_i$. Suppose that $p(\bX;\bmu_g,\bTheta_g)$ is the density function of the normal distribution with mean $\bmu_g$ and covariance $\bTheta_g^{-1}$,
$$p(\bX;\bmu_g,\bTheta_g) = \left(2\pi\right)^{-p/2} |\bTheta_g|^{1/2} \exp \left[-\frac{1}{2} \left( \bX - \bmu_g\right)^T\bTheta_g\left(\bX - \bmu_g\right)\right].$$
The conditional density function $p(\bX_i|Z_i;\bmu,\bTheta)$ of $\bX_i$ given $Z_i$ can be written as $p(\bX_i|Z_i;\bmu,\bTheta) = \prod_{g = 1}^{G} \left[p(\bX_i;\bmu_g,\bTheta_g)\right]^{I\left(Z_i = g\right)}$. Denote $p(Z_i ; \bpi) = \prod_{g = 1}^{G} \pi_{g}^{I(Z_i = g)}$ as the probability mass function of the multinomial distribution. The log-likelihood of the MPLN model (\ref{equ0}) is
\begin{equation}\label{equ_likelihood}
	\ell_n(\btheta) = \sum_{i = 1}^{n} \log\left(p(\bY_i;\btheta)\right) = \sum_{i = 1}^{n}\log \iint p(\bY_i|\bX_i) p(\bX_i|Z_i;\bmu,\bTheta) p(Z_i;\bpi) \mathrm{d} \mathbf{\bX_i} \mathrm{d} Z_i\ ,
\end{equation}
where $p(\bY_i;\btheta)$ is the marginal probability mass function of $\bY_i$. Similar to the GGM, the precision matrix $\bTheta_g$ of the $g$th cell type represents the cell-type-specific regulatory network. Let $\Theta_{g,l m}$ be the element of the $l$th row and the $m$th column of $\bTheta_g$. The regulatory networks are sparse and can be estimated by minimizing the lasso-penalized negative log-likelihood 
\begin{equation}\label{penalizedloglike}
- n^{-1}\ell_n(\btheta)+\lambda_{n} \sum_{g=1}^{G}\| {\bTheta_{g}}\|_{1, \text { off }},
\end{equation}
where $\lambda_{n}>0$ is a tuning parameter and $\| {\bTheta_{g}}\|_{1, \text { off }} = \sum_{l \neq m} |{\Theta_{g,l m}}|$  is the off-diagonal $l_1$-norm of the matrix $\bTheta_{g}$. In the following, we first establish the consistency of the estimator obtained by minimizing the objective function (\ref{penalizedloglike}) and then derive an algorithm for estimating the precision matrices based on the variational inference method.

\subsection{Theoretical properties}

In this section, we always assume that the true means $\bmu_g^*$ and proportions $\pi_g^*$ ($g=1,\cdots,G$) are known. Let $\bnu_g = {\rm vech}(\bTheta_g) $ be the vectorization of the precision matrix $\bTheta_g$ and ${\bnu} =  (\bnu_1^{\rm T}, \ldots, \bnu_G^{\rm T})^{\rm T}$  (see supplementary material for the exact definition of $\rm vech(\cdot)$). In this case, the log-likelihood $\ell_n(\btheta)$ can be viewed as a function of ${\bnu}$, also denoting as $\ell_n(\bnu)$, and we consider the estimator $\hat{\bnu}_n$ that minimizes $-n^{-1}\ell_n(\bnu) + \lambda_{n} \sum_{g=1}^{G}\| {\bTheta_{g}}\|_{1, \text { off }}$ subject to $\bTheta_{g} \succ 0\  (g=1,\cdots,G)$.  Denote $\bnu^*$ as the true value of the unknown parameter $\bnu$, $S(\bnu) = \{i | \ \bnu_{i} \neq 0\}$ as the support of $\bnu$, and $S^{*} = S({\bnu}^{*})$. Suppose that $\by$ follows the MPLN model with its log-likelihood function $\ell(\bnu,\by)$. Denote $\bGamma^* = \bGamma(\bnu^*) = -{\rm E}\left[\frac{\partial^2\ell(\bnu, \by)}{\partial \bnu \partial \bnu^T} \right]\bigg|_{\bnu = \bnu^*}$ as the Fisher information matrix of the MPLN at $\bnu^*$, and ${\bGamma^*}_{T_1 T_2}$ as the submatrix of ${\bGamma^*}$ with rows and columns index by sets $T_1$ and $T_2$, respectively. Before presenting the theoretical properties, we give the following conditions:

\begin{itemize}
	\item[(C1)] 
	The eigenvalues of the precision matrices are bounded away from zero and infinity, i.e. 
	there are two constants $0<m<M$ such that 
	$ m \leq \lambda_{min}(\bTheta_g)\leq\lambda_{max}(\bTheta_g)\leq M$ for $g=1,\cdots,G$, where $\lambda_{min}(\bTheta_g)$ and $\lambda_{max}(\bTheta_g)$ are the minimum and maximum eigenvalue of the precision matrix $\bTheta_g$. 
	
	\item[(C2)]  	
	The library sizes $l_i>0$ ($i = 1,\cdots,n$) are independent and identically distributed random variables with a bounded support.
	
	\item[(C3)] 
	The true mean vectors $\bmu_g^*$ ($g = 1,\cdots,G$) are bounded and different from each other.  
	
	\item[(C4)] 
	The irrepresentability condition:  $||\bGamma^*_{{S^{*}}^c S^{*}}{(\bGamma^{*}_{S^{*} S^{*}})}^{-1}||_{\infty} < 1$. 
\end{itemize}
Define $\mathcal{D} = \{ {\bnu} =  (\bnu_1^{\rm T}, \bnu_2^{\rm T}, \ldots, \bnu_G^{\rm T})^{\rm T}|~\bnu_g = {\rm vech}(\bTheta_g), m \leq \lambda_{min}(\bTheta_g)\leq\lambda_{max}(\bTheta_g)\leq M \}$ and $ \kappa =  \lambda_{min}(\bGamma^*)$ be the minimum eigenvalue of the Fisher information matrix at $\bnu^{*}$. Condition (C1-C2) are commonly used in the literature \citep{cai2011constrained, li2020transfer}. Condition (C3) is to ensure that different components of the MPLN model can be distinguished from each other. Under Condition (C3), the MPLN model is identifiable and its Fisher information $\bGamma^*$ is positive definite, and thus $\kappa>0$. The irrepresentability condition (C4) is also commonly used \citep{zhao2006model, ravikumar2011high}. For a GGM with a covariance $\bSigma$ and a known mean, its Fisher information matrix is $\bSigma\otimes \bSigma$, where $\otimes$ represents the Kronecker product. There is an intuitive explanation for the irrepresentability condition of the GGM \citep{ravikumar2011high}. However, the Fisher information matrix $\bGamma^*$ of the MPLN model has no closed form and we do not have an intuitive explanation for the irrepresentability condition (C4). Based on the above conditions, we present the theoretical properties of the MPLN and the estimator $\hat{\bnu}_n$ in the following theorems.

\begin{theorem} \label{thm:basic}
	Under Condition (C1-C3), the MPLN model is identifiable, and its Fisher information matrix  $\bGamma^*$ at $\bnu^*$ is positive definite.
\end{theorem}

Theorem \ref{thm:basic} establishes basic properties of the MPLN model and ensures that the MPLN model is well-behaved under rather mild conditions. The proof of this theorem is nontrivial because the PLN distribution has no finite moment generating function and its density function is rather complex. However, its moments are finite and have closed forms. We use its moments to prove Theorem \ref{thm:basic}. For the identifiability, the basic idea of the proof is that  identifiability of a mixture model is equivalent to linear independence of its components. Using moments of the PLN distribution, we can show that only a zero vector can make the linear combination of the components of the MPLN model as zero. To prove the positive definiteness of the Fisher information, we also use the moments of the PLN distribution and convert the problem to showing that a set of equations only have zero solutions. Based on this result, we can further prove the consistency and the sign consistency of the estimator $\hat{\bnu}_n$.
\begin{theorem} \label{thmConvergenceRate}
	Under Conditions (C1-C3), we have
	$$  \pr\left[||\hat{\bnu}_n-\bnu^*||_2 \leq  \frac{3}{\kappa} \sqrt\frac{G{p(p+1)}}{{2}}\big(n^{-1} ||\nabla{\ell}_n(\bnu^*)||_{\infty} + 2{\lambda_n}\big)\right] \rightarrow 1, \mbox{ as } n \rightarrow \infty,$$
	where $\nabla{\ell}_n(\bnu^*)$ is the gradient of the log-likelihood $\ell_n(\btheta)$ at $\bnu^*$ and $\lambda_n > 0$ is the regularization parameter.
\end{theorem}

\begin{theorem} \label{signconsistency}
	Under the assumptions of Theorem \ref{thmConvergenceRate} and Condition (C4), choosing $\lambda_n > 0$ such that $\lambda_n \rightarrow 0$ and ${\sqrt{n}\lambda_n} \rightarrow \infty$, we have 
	$$\pr\left[S(\hat{\bnu}_n) = S^{*} \right] \rightarrow 1 \mbox{ as } n \rightarrow \infty.$$
\end{theorem}
Theorem \ref{signconsistency} says that if we choose $\lambda_n > 0$ such that it does not converge to 0 too fast, $\hat{\bnu}_n$ can consistently recover the nonzero elements of $\bnu^*$. For example, we can choose $\lambda_n = \sqrt{\log n / n}$. The consistency theory established in this paper is only for fixed dimension $p$. Currently, we cannot prove consistency results for the high dimensional setting. One major difficulty is that, we can prove that the expectation of the negative log-likelihood function $-n^{-1}{\ell}_n(\bnu)$ at $\bnu^*$ is strong-$\kappa$ convex, but we can not characterize how $\kappa$ changes as $p \rightarrow \infty$. The PLN graphical model corresponds to the case of $G=1$. Theorem \ref{thmConvergenceRate} and \ref{signconsistency} imply that, for the PLN graphical model, the network estimated by minimizing the lasso-penalized negative log-likelihood is a consistent estimator. 

\section{Algorithm}
\label{sec:Algorithm}
\subsection{Variational inference for the MPLN model}
The log-likelihood (\ref{equ_likelihood}) of the MPLN model involves an intractable integration and thus directly minimizing (\ref{penalizedloglike}) is computationally very difficult. We therefore adopt the variational inference approach to estimate the networks \citep{jordan1999introduction, wainwright2008graphical}. We approximate the log-likelihood by the evidence low bound (ELBO) $\ell_{\text{E}}\left(\bbeta,\btheta \right)$ and estimate $\btheta$ by minimizing $-\ell_{\text{E}}\left(\btheta,\bbeta \right)+\lambda_{n} \sum_{g=1}^{G}\left\| {\bTheta_{g}}\right\|_{{1, \text { off }}}$, where $\bbeta\in \mathscr{H}$ is the parameter of the variational distribution family $\mathscr{L}=\{q(\bX,\bZ;\bbeta): \bbeta \in \mathscr{H}\}$. For $\bbeta\in \mathscr{H}$,  the ELBO $\ell_{\text{E}}\left(\bbeta,\btheta \right)$ is defined as $\ell_{\text{E}}\left(\bbeta,\btheta \right) = {\mbox{E}}_{q(\mathbf{\bX,\bZ};\bbeta)}\left[\log p(\mathbf{\bX,\bZ}, \mathbf{\bY};\btheta) - \log q(\mathbf{\bX,\bZ};\bbeta)\right]$. 

For computational considerations, we consider the following variational distribution family. Conditional on $Z_i=g$, this distribution family assumes that $X_{ij}$ ($j=1,\cdots,p$) are independent normal variables with a mean $M_{g,ij}$ and a variance $S_{g,ij}$. The distribution of $Z_i$ is a multinomial distribution with probabilities $\bP_i = (P_{i1},\cdots,P_{iG})^T$ as proportion parameters. Denote $\bM_{g}=(M_{g,ij})_{n\times p}$, $\bS_{g}=(S_{g,ij})_{n\times p}$ and $\bP = (\bP_1,\cdots,\bP_n)^T$. The variational parameters are $\bbeta = \left\{\bbeta_g\right\}_{g=1}^{G} = \left\{\bM_{g}, \bS_{g}, \{\bP_{ig}\}_{i = 1}^{n}\right\}_{g = 1}^{G}$ with $\bbeta\in\mathscr{H} = \left\{\bbeta \big| \ S_{g,ij}>0,~P_{ig}\geq0,~\sum_{g=1}^G P_{ig} =1\right\}$. Thus, the uncentered variational distribution family is
$$
	\mathscr{Q} = \left\{q(\mathbf{\bX,\bZ};\bbeta) = \prod_{i=1}^{n} \left[q(Z_i;\bP_i) \prod_{j=1}^{p} \prod_{g=1}^{G} q\big(X_{i j}|Z_{i} = g; M_{g,i j},S_{g,i j}\big)\right],\ \bbeta \in \mathscr{H}\right\},
$$
where $q\big(X_{i j}|Z_{i} = g; M_{g,i j},S_{g,i j}\big)$ is the density function of $N(M_{g,i j},S_{g,i j})$ and $q(Z_i;\bP_i)$ is the density function of $\mbox{Multinomial} (1,\bP_i)$. 

Given two matrices $\bA$ and $\bB$, let  $\bA\odot\bB$ be the Hadamard product of $\bA$ and $\bB$, $\bA_{i \cdot}$ be the vector of the $i$th row of $\bA$ and $\bA_{\cdot j}$ be the vector of the $j$th column of $\bA$. Given a vector $\bc$, define $\bD(\bc)$ as the diagonal matrix whose diagonal elements are $\bc$. Denote $\bl = \left(l_1,l_2,\cdots,l_n\right)^T$, $\bSigma_{g i} = (\bM_{g,i\cdot} - \bmu_g){(\bM_{g, i\cdot} - \bmu_g)}^T + \bD \left(\bS_{g ,i\cdot}\right)$, $F_1\left(l_i,M_{g,i j},S_{g,i j}\right) = \exp\left(M_{g,i j} + 2^{-1}S_{g,i j} + \log l_i\right)$, $F_2\left(\bTheta_{g},\bM_{g,i \cdot},\bS_{g,i \cdot},\bmu_{g}\right) = 2^{-1} \left[\log \det \bTheta_g - \tr (\bTheta_g \bSigma_{g i})\right]$ and $\btheta_g = \left(\bpi_{g}, \bmu_g, \bTheta_g\right)$ as the set of the unknown model parameters of the $g$th cell type. With the above variational distribution family $\mathscr{L}$, the ELBO can be written as $\ell_{\text{E}}\left(\bbeta,\btheta \right) = \sum_{g = 1}^{G} \ell_{\text{E}}^{(g)} \left( \bbeta_g,\btheta_g \right)$ with 
$$
\ell_{\text{E}}^{(g)} \left(\bbeta_g,\btheta_g \right) = \bP_{\cdot g}^{T} \left(\bLambda_{g}^{(1)} - \bLambda_{g}^{(2)} + \bLambda_{g}^{(3)}\right) \mathbf{1}_{p} + \bP_{\cdot g}^{T} \left(\bLambda_{g}^{(4)} + \bLambda_{g}^{(5)}\right) + K_g\left(\bY\right),
$$
where $\bLambda_{g}^{(1)} = \mathbf{Y} \odot \bM_{g}$, $\bLambda_{g}^{(2)} = \left(\bLambda_{g,i j}^{(2)}\right)_{n \times p} = \big(F_1\left(l_i,M_{g,i j},S_{g,i j}\right)\big)_{n \times p}$, $\bLambda_{g}^{(3)} = 2^{-1} \log \bS_{g}$, $\bLambda_{g}^{(4)} = \log (\pi_{g}) \mathbf{1}_{n} - \log \left(\bP_{\cdot g}\right)$, $\bLambda_{g}^{(5)} = \left(\bLambda_{g,i}^{(5)}\right)_{n} = \big(F_2\left(\bTheta_{g},\bM_{g,i \cdot},\bS_{g,i \cdot},\bmu_{g}\right)\big)_{n}$ and $K_g(\mathbf{Y}) = \sum_{i, j} P_{i g} \left[ - \log \left(Y_{i j} !\right) + Y_{i j}\log l_i\right]$.

In real applications, we may have prior knowledge that some gene pairs cannot have direct interactions. In this case, we can directly set the corresponding edges as zero. Denote $E_p$ as the set of the edges that are priorly known to be zero. Generally, we consider the following optimization problem
\begin{equation}\label{con:VMPLNloss}
	\begin{aligned}
		\underset{\bbeta,\btheta}{\min}\left\{-\ell_{\text{E}}\left(\btheta,\bbeta \right)+\lambda_{n} \sum_{g=1}^{G}\left\| {\bTheta_{g}}\right\|_{{1, \text { off }}}: ~ {\bTheta}_{g} \succ 0,\ {\Theta_{g,l m}} = 0 ~ \mbox{for}~(l,m) \in E_{p}, \bbeta \in \mathscr{H}\right\}.
	\end{aligned}
\end{equation}

\subsection{The optimization process}
We develop a block-wise descent algorithm called VMPLN to optimize (\ref{con:VMPLNloss}). Let
$$
\begin{aligned}
	L_{1}\left(\bM_{g, i \cdot}, \bTheta_{g}, \bmu_g\right) &= 2^{-1} {\left(\bM_{g, i \cdot} - \bmu_g\right)}^T \bTheta_{g} \left(\bM_{g, i \cdot} - \bmu_g\right),\\
	L_{2}\left(M_{g,i j}, l_i, S_{g,i j}\right) &= - \Lambda_{g,i j}^{(1)} + F_{1}\left(l_i, M_{g,i j}, S_{g,i j}^{(k+1)}\right),\\
	\bSigma \left(\bP_{\cdot g}, \bmu_{g},\bM_{g},\bS_{g}\right) &= \left\{\sum_{i = 1}^{n} P_{i g} \left[(\bM_{g,i\cdot} - \bmu_g){(\bM_{g, i\cdot} - \bmu_g)}^T + \bD \left(\bS_{g ,i\cdot}\right)\right]\right\} \bigg/ \sum_{i = 1}^{n} P_{i g}.
\end{aligned}
$$
Let $\mathscr{N} = \left\{1,2,\cdots,n\right\}, \mathscr{G} = \left\{1,2,\cdots,G\right\}, \mathscr{P} = \left\{1,2,\cdots,p\right\}$. Our proposed VMPLN algorithm is summarized in Algorithm \ref{alg:algorithm1}. Given initial values, we iteratively update $\bP$, $\bpi$, $\left\{\bM_{g}\right\}_{g = 1}^{G}$, $\left\{\bS_{g}\right\}_{g = 1}^{G}$, $\bmu$ and $\left\{\bTheta_{g}\right\}_{g = 1}^{G}$ and terminate the iteration if the ELBO and the network estimations only have very small changes between two successive update steps. The step for updating $\bM_g$ ($g=1,\cdots,G$) is presented in the next subsection \ref{subsec:Mstep}. The other steps are straightforward and presented below.

\begin{algorithm}

	\footnotesize
	\caption{Framework of VMPLN.}
	\label{alg:algorithm1}
	\KwIn{Single cell RNA-seq count data $\bY$, the pre-estimated library size $\bl$, the number of cell types $G$, the tunning parameter $\lambda_{n}$, the optional prior set of zero edges $E_{p}$, the maximum iteration number $K>0$ and the convergence thresholds $\epsilon_{L},\epsilon_{s}$.}
	\KwOut{$\hat{\bpi}, \hat{\bmu}, \{\hat{\bTheta}_g \}_{g = 1}^G, \{\hat{\bM}_{g}\}_{g = 1}^{G}, \{\hat{\bS}_{g}\}_{g = 1}^{G}, \hat{\bP}$.}
	\BlankLine
	\underline{Initialization step}. Let $\delta_{L} = 10^{6}, \delta_{s} = p(p+1)/2, k = 0$ and initialize $\btheta, \bbeta$ as $\btheta^{(0)}, \bbeta^{(0)}$ by equation (\ref{equ_init}).
	
	\While{\textnormal{ $\left(\delta_{L} > \epsilon_{L} \ \OR \ \delta_{s} > \epsilon_{s}\right) \ \AND \ k \leq K $ }}{
		\underline{$\bP$-step}. For each $\left(i,g\right) \in \mathscr{N} \times \mathscr{G}$, compute $U_{i g}^{(k)} = F_2\left(\hat{\bTheta}_{g}^{(k)}, \hat{\bM}_{g,i \cdot}^{(k)}, \hat{\bS}_{g,i \cdot}^{(k)},\hat{\bmu}_{g}^{(k)}\right)$ and update
		$${\hat{P}_{i g}}^{(k+1)}  = \hat{\pi}_{g}^{(k)} \exp \left({U}_{i g}^{(k)}\right) \bigg/ \sum_{l = 1}^{G} \left[\hat{\pi}_{l}^{(k)} \exp \left({U}_{i l}^{(k)}\right)\right],$$
		
		\underline{$\bpi$-step}. For each $g \in \mathscr{G}$, update
		$${\hat{\pi}_{g}}^{(k+1)} = n^{-1}\sum_{i = 1}^{n} {\hat{P}_{i g}}^{(k+1)}$$
		
		\underline{$\bM$-step}. For each $\left(i,g\right) \in \mathscr{N} \times \mathscr{G}$, update
		\begin{equation}\label{opti_M}
			{\hat{\bM}_{g,i\cdot}}^{(k+1)} = \underset{\bM_{g,i\cdot}}{\argmin} \left\{ L_{1} \left(\bM_{g, i \cdot}, \hat{\bTheta}_{g}^{(k)}, \hat{\bmu}_g^{(k)}\right) + \sum_{j = 1}^{p} L_{2}\left(M_{g,i j}, l_i, \hat{S}_{g,i j}^{(k)}\right) \right\}.
		\end{equation}
		
		\underline{$\bS$-step}. For each $\left(i,g,j\right) \in \mathscr{N} \times \mathscr{G} \times \mathscr{P}$, update
		$${\hat{\bS}_{g,i j}}^{(k+1)} = \underset{S_{g,i j} > 0}{\argmin} \left\{F_1\left(l_i,\hat{M}_{g,i j}^{(k+1)},S_{g,i j}\right) + 2^{-1} \hat{\bTheta}_{g,j j}^{(k)} S_{g,i j} - 2^{-1} \log S_{g,i j} \right\}.$$
		
		\underline{$\bmu$-step}. For each $\left(g,j\right) \in \mathscr{G} \times \mathscr{P}$, update
		$$\hat{\mu}_{g j}^{(k+1)} = \sum_{i = 1}^{n} \left(\hat{P}_{i g}^{(k+1)} \hat{M}_{g,i j}^{(k+1)}\right)\bigg/ \sum_{i = 1}^{n} \hat{P}_{i g}^{(k+1)}.$$
		
		\underline{$\bTheta_{g}$-step}. \ For each $g \in \mathscr{G}$, compute $\hat{\bSigma}_g^{(k+1)} = \bSigma\left(\hat{\bP}_{.g}^{(k+1)}, \hat{\bmu}_{g}^{(k+1)},\hat{\bM}_{g}^{(k+1)},\hat{\bS}_{g}^{(k+1)} \right)$ and update
		\begin{equation*}
			\begin{aligned}
				\hat{\bTheta}_{g}^{(k+1)} = \underset{\bTheta_{g}}{\argmin} \bigg\{&- 2^{-1} \log \det \bTheta_{g } + 2^{-1} \tr\big(\bTheta_{g} \hat{\bSigma}_{g}^{(k+1)}\big) + \left(\lambda_{n} / \mathbf{1}_{n}^{T} \hat{\bP}_{\cdot g}^{(k+1)} \right) \left\| {\bTheta_{g}}\right\|_{{1, \text { off }}}:\\
				&{\bTheta}_{g} \succ 0,\ {\Theta_{g,l m}} = 0 ~ \mbox{for}~(l,m) \in E_{p}\bigg\}.
			\end{aligned}
		\end{equation*}
		
		\underline{Evaluation step}. \ Update \begin{equation*}
				\begin{aligned}
				\delta_{s} &= \underset{g \in \mathscr{G}}{\max} \left\{\sum_{l<m} \left|\text{sign}\left(\hat{\bTheta}_{g,lm}^{(k+1)}\right) - \text{sign}\left(\hat{\bTheta}_{g,lm}^{(k)}\right) \right|  {\tbinom{p}{2}}^{-1} \right\},\\
				\delta_{L} &= \delta\left(\ell_{\text{E}}\left(\hat{\bbeta}^{(k+1)},\hat{\btheta}^{(k+1)} \right),  \ell_{\text{E}}\left(\hat{\bbeta}^{(k)},\hat{\btheta}^{(k)} \right)\right),\ \text{with} \ \delta \left(a,b\right) = \left| a - b\right|/b,\\
				k &= k+1.
				\end{aligned}
			\end{equation*}
		\BlankLine
	}
\end{algorithm}

The parameters $\bP$, $\bpi$ and $\bmu$ all have explicit updating formulas and can be efficiently calculated. The updating form of $\hat{\bmu}_g$ is different from that of the centered variational approach used by VPLN \citep{chiquet2019variational}, which involves the logarithm of the sum of $Y_{i j}$ and may be numerically unstable when the data is of high dropout.

For the parameter $\bS$, given all other parameters, the ELBO loss function (\ref{con:VMPLNloss}) can be decomposed into the sum of $npG$ functions, each of which only involves one $S_{g,ij}$. Thus, in the $\bS$-step of Algorithm \ref{alg:algorithm1}, the optimization problem can be decomposed into $npG$ one-dimensional convex optimization problems. Each of these can be efficiently solved by the Newton-Raphson algorithm.

For the network parameters $\bTheta_g$ ($g=1,\cdots,G$), given all other parameters, the corresponding sub-optimization problem is equivalent to $G$ independent Glasso problems \citep{meinshausen2006high,friedman2008sparse}, and $\bTheta_g$ can be updated by solving the corresponding Glasso problem with its covariance matrix as $\hat{\bSigma}_g^{(k+1)}$ and its penality parameter as $\lambda_{n} \big/ \mathbf{1}_{n}^{T} \hat{\bP}_{.g}^{(k+1)}$.

\subsection{The optimization of the $\bM$-step in Algorithm \ref{alg:algorithm1}}
\label{subsec:Mstep}
The sub-optimization problem corresponding to $\bM$ is $$\underset{\bM}{\argmin} \left\{\sum_{g \in \mathscr{G}, i \in \mathscr{N}} \bP_{i g}\left[L_{1} \left(\bM_{g, i \cdot}, \hat{\bTheta}_{g}^{(k)}, \hat{\bmu}_g^{(k)}\right) + \sum_{j = 1}^{p} L_{2}\left(M_{g,i j}, l_i, \hat{S}_{g,i j}^{(k)}\right)\right]\right\},$$
which is equivalent to $nG$ independent optimization problems (\ref{opti_M}). 
The p-dimensional optimization problem (\ref{opti_M}) can be solved by the $p$-dimensional Newton-Raphson algorithm or coordinate descent algorithm. However, the $p$-dimensional Newton-Raphson algorithm involves inverting $p\times p$ Heissen matrices. In each of the $nG$ optimization problems (\ref{opti_M}) and at each step of the $p$-dimensional Newton-Raphson algorithm, the Heissen matrices are unique and different from each other. In total, the $p$-dimensional Newton-Raphson algorithm would involve $O(nGt)$ $p \times p$ matrix inversions, where $t$ is the number of iterations in the Newton-Raphson algorithm, and is computationally very expensive. The coordinate descent algorithm is also slow because of its inferior convergence rate and the large number of optimization problems. 

Here, we instead develop a more efficient algorithm based on the alternating direction method of multipliers (ADMM) algorithm \citep{boyd2011distributed} for the optimization problem (\ref{opti_M}). Specifically, we introduce an auxiliary matrix $\bN_{g}$ for $\bM_{g}$, and denote $\ell_{\bM}\left(\bM_{g, i \cdot},\bN_{g, i \cdot}\right) = L_{1} \left(\bN_{g, i \cdot}, \hat{\bTheta}_{g}^{(k)}, \hat{\bmu}_g^{(k)}\right)$ $ + \sum_{j = 1}^{p} L_{2}\left(M_{g,i j}, l_i, \hat{S}_{g,i j}^{(k)}\right)$. Solving  (\ref{opti_M}) is equivalent to solving the following problem
\begin{equation}\label{opti_MN}
	\underset{\bM_{g, i \cdot}=\bN_{g,i\cdot}}{\argmin} \left\{\ell_{\bM}\left(\bM_{g, i \cdot},\bN_{g, i \cdot}\right) \right\}.
\end{equation}
The augmented Lagrangian of \eqref{opti_MN} is 
$$
\ell_{\bM}\left(\bM_{g, i \cdot},\bN_{g, i \cdot}\right) +\sum_{j = 1}^{p} {\alpha}_j \left(M_{g,i j} - N_{g,i j}\right) + \rho/2 \sum_{j = 1}^{p} {\left(M_{g,i j} - N_{g,i j}\right)}^2,
$$
where $\balpha = \left(\alpha_{1},\cdots,\alpha_{p}\right)$ is the Lagrangian multiplier, and $\rho$ is the step size. The corresponding ADMM algorithm is detailed in Algorithm \ref{alg:algorithm2}. Given initial values, we iteratively update $\bM_{g, i \cdot}$, $\bN_{g, i \cdot}$ and $\balpha$, and terminate the iteration if the ELBO only has a very small change between two successive update steps. 

\begin{algorithm}
	\footnotesize
	\caption{ADMM algorithm for updating $\bM_{g,i \cdot}$.}
	\label{alg:algorithm2}
	\KwIn{Single cell RNA-seq count data $\bY$, the pre-estimated library size $\bl$, the current estimation of parameters $\hat{\btheta}, \hat{\bbeta}$, the maximum iteration number $T>0$ and the convergence threshold $\epsilon_{M}$.}
	\KwOut{$\hat{\bM}_{g,i \cdot}$.}
	\underline{Initialization step}. Let $t = 0, \delta_{M} = 10^{6}$, and initialize $\hat{\bM}_{g,i\cdot}^{\left(0\right)} = \hat{\bN}_{g,i\cdot}^{\left(0\right)} = \hat{\bM}_{g,i\cdot}$, $\hat{\balpha}^{(0)} = \mathbf{0}$.
	
	\While{$t \leq T$ \AND $\delta_{M} > \epsilon_{M}$}{
		
		\underline{Step 1}. For each $j \in \mathscr{P}$, update $$\hat{M}_{g,i j}^{(t+1)} = \underset{M_{g,i j}}{\argmin} \bigg\{L_{2}\left(M_{g,i j}, l_i, \hat{S}_{g,i j}\right) + {\hat{\alpha}}_{j}^{(t)} (M_{g,i j} - \hat{N}_{g,i j}^{(t)}) +\frac{\rho}{2} {(M_{g,i j} - \hat{N}_{g,i j}^{(t)})}^2\bigg\}.$$
		
		\underline{Step 2}. Update
		$$\hat{\bN}_{g,i \cdot}^{(t+1)} = {\left(\rho \bI + \hat{\bTheta}_{g}\right)}^{-1} \left(\rho \hat{\bM}_{g,i \cdot}^{(t+1)} - \hat{\balpha}^{(t)} + \hat{\bTheta}_{g} \hat{\bmu}_g  \right).$$
		
		\underline{Step 3}. Update 
		$${\hat{\balpha}}^{(t+1)} = {\hat{\balpha}}^{(t)} + \rho \left(\hat{\bM}_{g,i \cdot}^{(t+1)} - \hat{\bN}_{g, i \cdot}^{(t+1)}\right).$$
		
		\underline{Evaluation step}. \ Update \begin{equation*}
			\begin{aligned}
				\delta_{M} &= \delta\left(\ell_{\bM}\left(\hat{\bM}_{g, i \cdot}^{(t+1)},\hat{\bN}_{g, i \cdot}^{(t+1)}\right),  \ell_{\bM}\left(\hat{\bM}_{g, i \cdot}^{(t)},\hat{\bN}_{g, i \cdot}^{(t)}\right)\right),\ \text{with} \ \delta \left(a,b\right) = \left| a - b\right|/b,\\
				t & = t + 1.
			\end{aligned}
		\end{equation*} 
	}
	\BlankLine
\end{algorithm}

Given all other variables, the optimization problem for $\bM_{g,i\cdot}$ can be decomposed into $p$ independent one-dimensional optimization problems, which can be easily optimized by the one-dimensional Newton-Raphson algorithm (Step 1 in Algorithm 2). The optimization problem for $\bN_{g,i\cdot}$  has an explicit solution and involves inverting the matrix $\rho \bI + \hat{\bTheta}_{g}^{(k)}$. Note that in this ADMM algorithm, we only need to calculate the matrix inversion $(\rho \bI + \hat{\bTheta}_{g}^{(k)})^{-1}$ once. Thus, this ADMM algorithm is more efficient than the $p$-dimensional Newton-Raphson algorithm. Moreover, the $\bM$-step of Algorithm 1 involves $nG$ optimization problems (7), but we only need to calculate $G$ matrix inversions $(\rho \bI+\hat{\bTheta}_{g}^{(k)})^{-1}$ in the  $nG$ runs of the ADMM Algorithm 2. In total, using this ADMM algorithm to solve (7) makes Algorithm 1 much more efficient than using the Newton-Raphson algorithm.

\subsection{Initialization, tunning parameters selection} 
To initialize the parameters in the initialization step of Algorithm \ref{alg:algorithm1}, we perform dimension reduction using principal component analysis (PCA) on the normalized data $\tilde{\bY} \triangleq \log (\bY+1) - (\log \bl) \mathbf{1}_{p}^{T} $. Then, we use K-means \citep{hartigan1979algorithm} to cluster single cells in the low dimensional space. Let $\tilde{Z}_i \in \{1,2,\cdots,G\}$ be the clustering label of the $i$th cell. The parameters are initialized as

\begin{equation}\label{equ_init}
	\begin{aligned}
		\hat{P}_{i g}^{(0)} &= I(\tilde{Z}_i = g),~ \hat{M}_{g,i j}^{(0)} = \tilde{Y}_{i j},~ \hat{S}_{g,i j}^{(0)} = 10^{-5},\\
		\hat{\pi}_g^{(0)} &= n^{-1}\sum_{i = 1}^{n} \hat{P}_{i g}^{(0)}, ~ \hat{\mu}_{g j}^{(0)} = \bigg(\sum_{i = 1}^{n} \hat{P}_{i g}^{(0)}\bigg)^{-1}\sum_{i = 1}^{n} \hat{P}_{i g}^{(0)} \hat{M}_{g,i j}^{(0)} ,\\
		& \text{with}\ g = 1,\cdots,G;\ i = 1,\cdots,n;\ j = 1,\cdots,p.
	\end{aligned}
\end{equation}
The precision matrix ${ \hat{\bTheta}_g }^{(0)}$ of the $g$th cell type is initialized by the Glasso algorithm with its covariance matrix as ${\hat{\bSigma}}_{g}^{(0)} = \bSigma\left(\hat{\bP}_{.g}^{(0)}, \hat{\bmu}_{g}^{(0)},\hat{\bM}_{g}^{(0)},\hat{\bS}_{g}^{(0)} \right)$ and its penalty parameter as $10^{-6}$.

The tunning parameter $\lambda_n > 0$ is selected for each cell type independently by the integrated complete likelihood (ICL) criterion \citep{biernacki2000assessing}, which selects the tunning parameter $\lambda_n > 0$ for $\bTheta_g$ by minimizing
\begin{equation}
	-2\ {l^{(g)}_{\text{E}}} \left(\hat{\bbeta},\hat{\btheta} \right) + \log \left(\mathbf{1}_{n}^{T} \hat{\mathbf{P}}_{\cdot g}\right) s(\hat{\bTheta}_{g}),
\end{equation}
where $s(\hat{\bTheta}_{g})$ denote the number of non-zero elements in $\hat{\bTheta}_{g}$. As another choice, we can select the tuning parameter $\lambda_n$ such that the estimated network has a desired density.

\section{Simulation}
\label{sec:simu}

In this section, we perform simulation to evaluate the performance of VMPLN and compare with state-of-the-art algorithms, including VPLN \citep{chiquet2019variational}, Glasso \citep{meinshausen2006high,friedman2008sparse}, LPGM \citep{allen2013local}, PPCOR \citep{kim2015ppcor}, GENIE3 \citep{huynh2010inferring} and PIDC \citep{chan2017gene}. PPCOR infers the regulatory network based on the partial correlation coefficients. LPGM is an extension of the PGM. GENIE3, a random-forest-based regulatory network method, is the best performing algorithm in the DREAM4 GRN inference challenge \citep{greenfield2010dream4}. PIDC is a recently developed algorithm for single cell data based on the mutual information.

\subsection{Simulation setups}
We first generate simulation data based on the MPLN model. The number of components is set as $G=3$ and the proportion parameter $\bpi$ is set as $(1/3,1/3,1/3)$. The number of observations is $n=3000$. We consider 48 different simulation scenarios, which are 3 cell-population-mixing levels (low, middle and high) $\times$ 2 dropout levels (low and high) $\times$  2 dimension setups ($p=100$ and $300$) $\times$ 4 graph structures. In each scenario, we generate 50 datasets. The four graph structures are as follows.

1. Random Graph: Pairs of nodes are connected with probability 0.1. The nonzero edges are randomly set as 0.3 or -0.3.

2. Hub Graph: 20 $\%$ nodes are set as hub nodes. A hub node is connected with another node with probability 0.1. Non-hub nodes are not connected with each other. The nonzero edges are randomly set as 0.3 or -0.3.

3. Blocked random Graph: The nodes are divided into 5 blocks of equal sizes. Pairs of nodes within the same block are connected with probability 0.1.  Nodes in different blocks are not connected. The nonzero edges are randomly set as 0.3 or -0.3.

4. Scale-free Graph: The Barabasi-Albert model \citep{barabasi1999emergence} is used to generate a scale-free graph with power 1. The nonzero edges are randomly set as 0.3 or -0.3.

Details of the data generation process are shown in supplementary material (Section S1.2.1). To test the performance of the proposed method under the  misspecified model setting, we also generate simulation data using a mixture multinomial log-normal distribution. In this model, the conditional Poisson layer of the MPLN is replaced with a conditional multinomial distribution, $\bY_i |\bX_i \sim \mbox{Multinomial}(\lfloor\sum_{j = 1}^{p} \lambda_{i j}\rfloor, \blambda_{i} \big/ \sum_{j = 1}^{p} \lambda_{i j})$, where $\lambda_{i j} = l_i\ \exp(X_{i j})$, $\blambda_{i} = \left(\lambda_{i 1},\cdots,\lambda_{i p} \right)$ and $\lfloor\sum_{j = 1}^{p} \lambda_{i j}\rfloor$ represents the maximum integer that does not exceed $\sum_{j = 1}^{p} \lambda_{i j}$. The simulation data are similarly generated.

\subsection{Performance Comparison}
For the proposed VMPLN algorithm, we use clustering results given by the K-means algorithm as the initial value and infer the regulatory networks jointly for all cell types. For the other algorithms, we use K-means to assign cell types of the single cells and infer regulatory network for each cell type separately. We compare VMPLN with other algorithms in terms of their accuracies of network inference.

Following \citet{pratapa2020benchmarking},  we define confidence scores of the predicted edges for each method based on its reported statistics (supplementary material, Section S1.2.2), and evaluate the algorithms by the area under the precision-recall curve (AUPRC) ratios and the early precision (EP) ratios based on the confidence scores. For VMPLN, VPLN and Glasso, suppose that $\hat{\bTheta}$ is its estimation of a network, we define a confidence score for the edge $(i,j)$ ($i\neq j$) as its absolute partial correlation coefficient, i.e. $\big|-(\hat{\Theta}_{ii} \hat{\Theta}_{jj})^{-\frac{1}{2}} \hat{\Theta}_{ij}\big|$. For LPGM, we define a confidence score for each edge as its stability score. For PPCOR, GENIE3 and PIDC, we define a confidence score for each edge as its estimated connected weight. By varying the confidence score, we can obtain the AUPRC for each algorithm on each simulation data. The AUPRC ratio is defined as the ratio between the AUPRC of an algorithm and the AUPRC of the random predictor. The EP is defined as the precision of the top $\min\{K,s\}$ edges in the inferred network, where $K$ is the number of edges in the true network, and $s$ is the number of edges in the inferred network. The EP ratio is defined as the ratio between the EP of the algorithm and the EP of the random predictor. 

\begin{figure}[htb]
	\begin{center}
		\subfigure[AUPRC ratios]{
			\includegraphics[height = 6.6 cm,width = 14.5 cm]{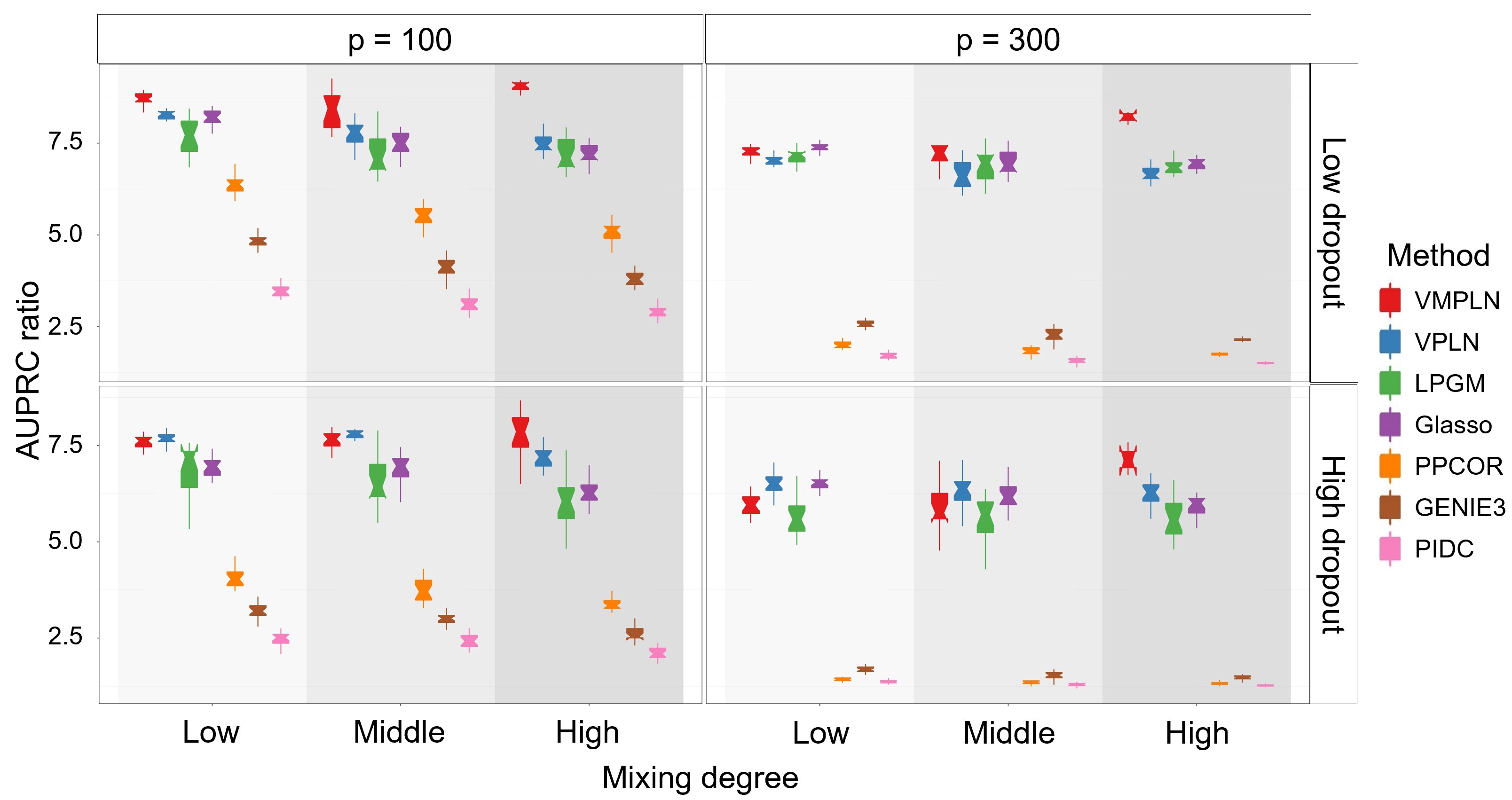}\label{subfigure1}}
		\quad
		\subfigure[EP ratios]{
			\includegraphics[height = 6.6 cm,width = 14.5 cm]{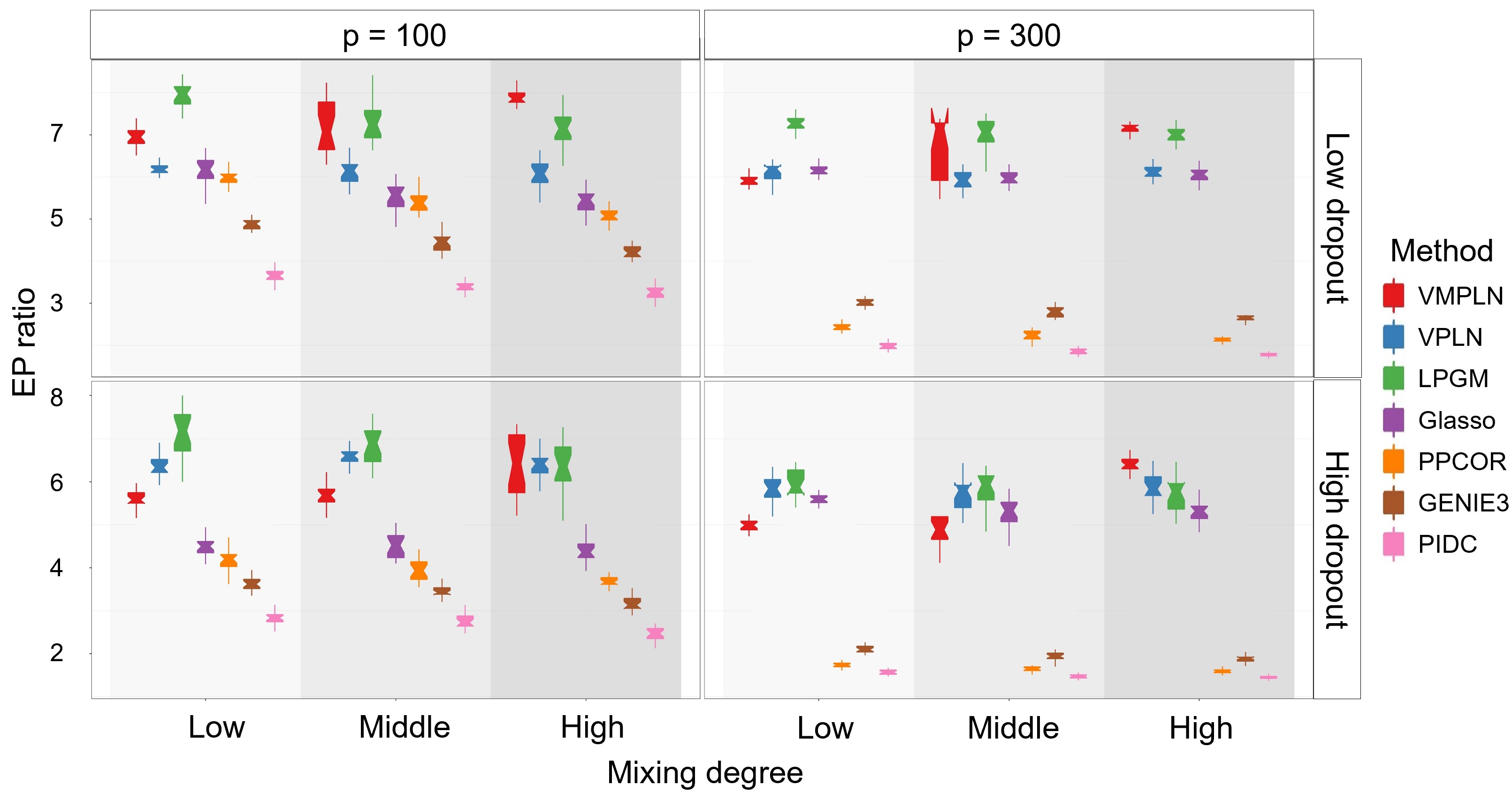}\label{subfigure2}}
	\end{center}
	\caption{The AUPRC ratios and EP ratios for the hub graph. The parameters are set as their default values or are tuned by their default methods.}\label{AUPRC-EPR-1}
\end{figure}

We first compare the estimated networks given by different algorithms with their default parameters or default ways of selecting tuning parameters (supplementary material, Section S1.2.2). Figure \ref{AUPRC-EPR-1} shows the boxplots of AUPRC ratios and EP ratios of different algorithms for the hub graph when the parameters are selected using their default methods. The results for the random graph, blocked random graph and scale-free graph are shown in supplementary material (Figure S1-S3). Overall, VMPLN is the best performing algorithm in terms of the AUPCR ratio. The advantage of VMPLN is more pronounced when the cell types have a higher mixing level, suggesting that compared with the two-step procedure, the joint analysis of network inference and clustering can help to improve the network inference. In terms of the EP ratio, VMPLN is also among the best performing algorithms. LPGM often has larger EP ratios than VMPLN in simulation settings with low cell-type mixing degrees. This is because LPGM uses stability to select the tuning parameter and thus is very conservative. LPGM often only reports a few edges, which are mostly true discoveries, leading to its high precision (as measured by the EP ratio) but a low sensitivity (supplementary material, Figure S4). When the cell-type mixing level is high, VMPLN also has higher EP ratio values than other methods including LPGM. 

\begin{figure}[htb]
	\begin{center}
		\subfigure[AUPRC ratios]{
			\includegraphics[height = 6.6 cm,width = 14.5 cm]{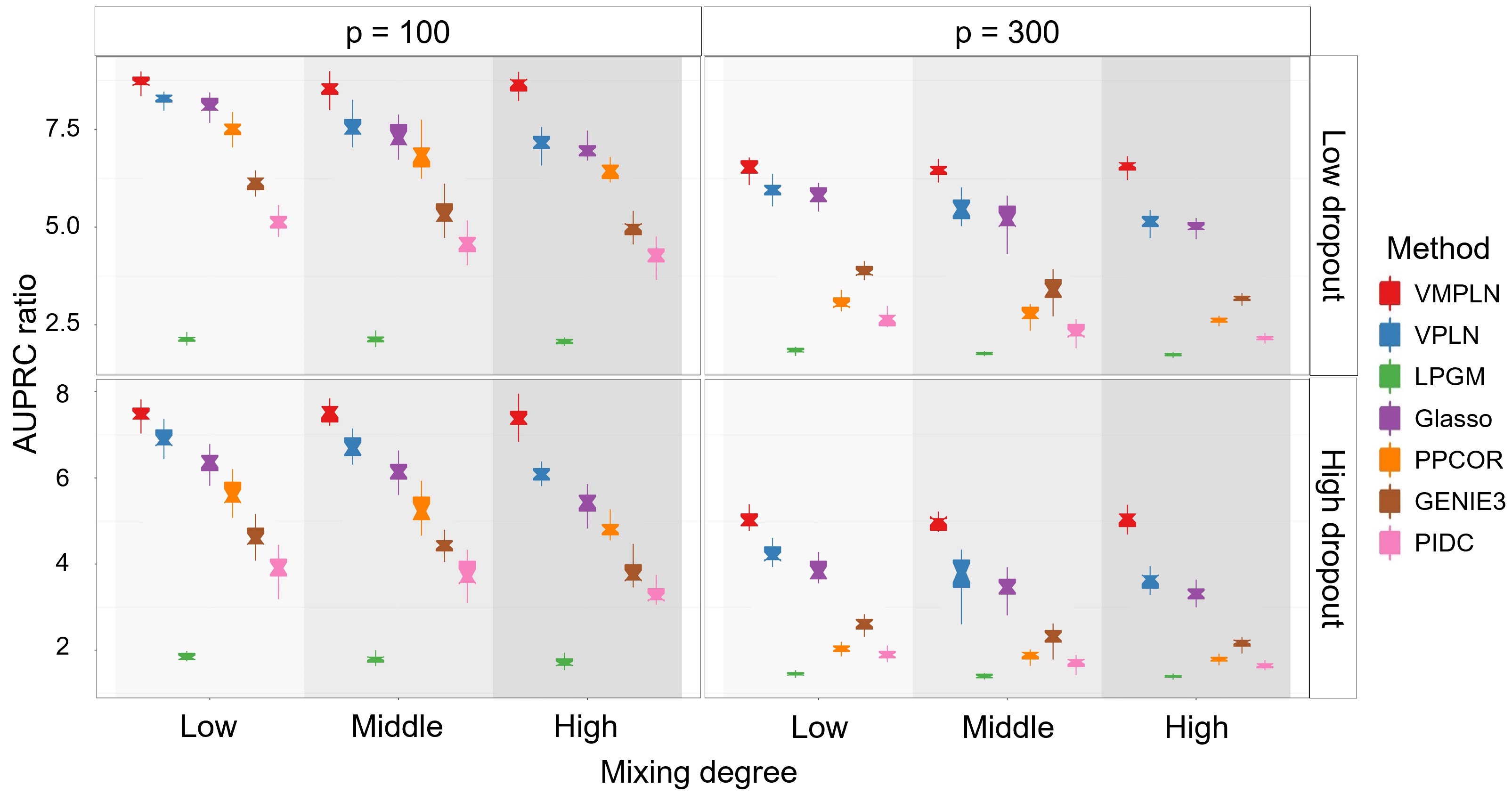}\label{subfigure1}}
		\quad
		\subfigure[EP ratios]{
			\includegraphics[height = 6.6 cm,width = 14.5 cm]{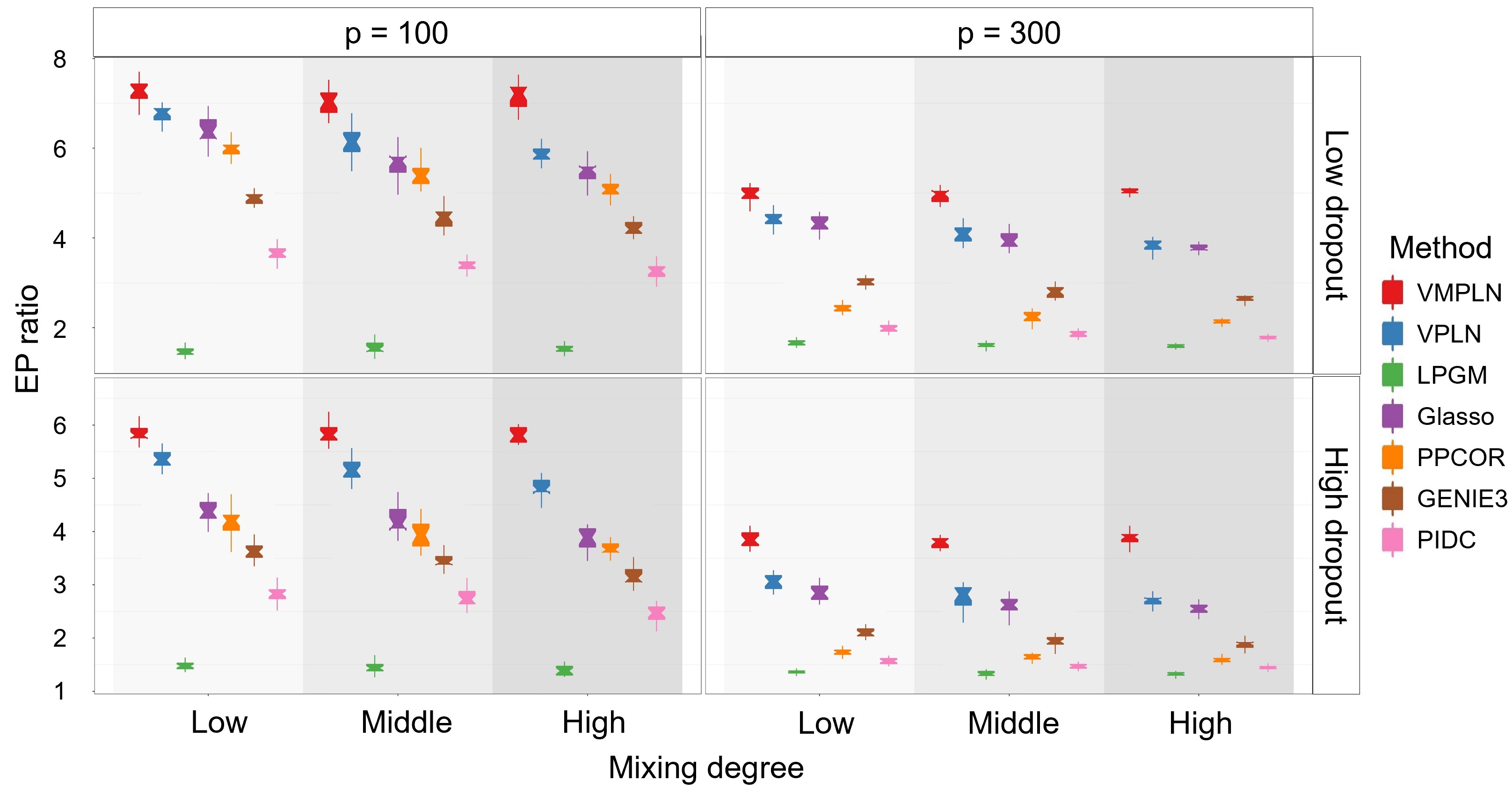}\label{subfigure2}}
	\end{center}
	\caption{The AUPRC ratios and EP ratios for the hub graph. The confidence score cutoffs or the tuning parameters are selected such that the network density is 20\%. }\label{AUPRC-EPR-2}
\end{figure}

Different ways of selecting the tuning parameters can lead to network estimations having very different densities. Dense network predictions usually have a  high sensitivity and a low specificity, but sparse network predictions have a low sensitivity and a high specificity. Thus, dense and sparse network predictions may not be directly comparable. To eliminate the influence of the tuning parameter selection methods, we further compare the algorithms at the same network density 20\%  (2 $\times$ the density of the true network) (supplementary material, Section S1.2.2). Figure \ref{AUPRC-EPR-2} shows the boxplots of AUPRC ratios and EP ratios for the hub graph at the same network density. The results for the random graph, blocked random graph and scale-free graph are shown in supplementary material (Figure S5-S7). Similarly, VMPLN has the best performance in most scenarios, especially in the cases with high cell-type mixing levels. For example, in the simulation of the hub graph with $p=100$ and low-dropout, VMPLN has mean AUPRC ratios $8.63$ and $8.72$ in the high and low cell-type mixing scenarios, respectively, about $21\%$ and $5\%$ larger than the AUPRC ratios ($7.12$ and $8.26$) of VPLN under the same scenarios. All algorithms tend to have decreased performances in higher dimensions or with high dropout rates and VMPLN consistently has better performances in these more difficult settings. Even in the low-mixing level case, where the two-step methods should work well, VMPLN still performs better than VPLN (supplementary material, Figure S8).  

The simulation results for the data generated from the misspecified model (the compositional model) are shown in supplementary material (Figure S9-S16). Similarly, VMPLN also performs better in most simulation settings.

\section{Real data analysis} 
\label{sec:appl}
\subsection{Benchmarking on scRNA-seq data}
In this section, we evaluate VMPLN and compare with other algorithms using two real scRNA-seq datasets. LPGM is not included in this comparison because it was unable to finish computation in a reasonable amount of time (7 days). One dataset is the scRNA-seq of human peripheral blood mononuclear cells (PBMC) profiled by \citet{kang2018multiplexed} (Kang data) and another dataset is the scRNA-seq data of human PBMC cells profile by \citet{zheng2017massively} (Zheng data). The Kang data consists of two batches, the interferon Beta 1 (IFNB1)-stimulated and control groups. The Zheng data also consists of two batches, which are respectively sequenced by $3^\prime$ and $5^\prime$ scRNA-seq technologies. We consider 10 cell types (7217 cells) from the Kang data and 6 cell types (5962 cells) from the Zheng data for network analysis. The other cell types of the Kang and Zheng datasets have less than 150 cells and are not considered for the network analysis. The number of cells in each cell type is listed in supplementary material (Table S1).

In both datasets, we first use data from one of the two batches (construction data) and public GRN databases (supplementary material, Table S2) to construct a silver standard (supplementary material, Section S1.3.1). Then, we test different algorithms using the another batches (testing data). For the testing data, we select the top 1000 highly variable genes (HVGs) using Seurat \citep{stuart2019comprehensive} and cluster the cells using these HVGs. We select the top 500 HVGs as the gene sets of interest for GRN inference. The top 500 HVGs contain 39 and 19 TFs for the Kang and Zheng data, respectively. For algorithms other than VMPLN, the cell types are first identified using clustering results of the top 1000 HVGs and gene regulatory relationships between genes in the gene sets of interest are inferred for each cell type. To make the comparison fair, for VMPLN, clustering and network inference are simultaneously performed using top 1000 HVGs and network inference focuses on genes in the gene sets of interest.

\begin{figure}
	\begin{center}
		\includegraphics[scale = 0.25]{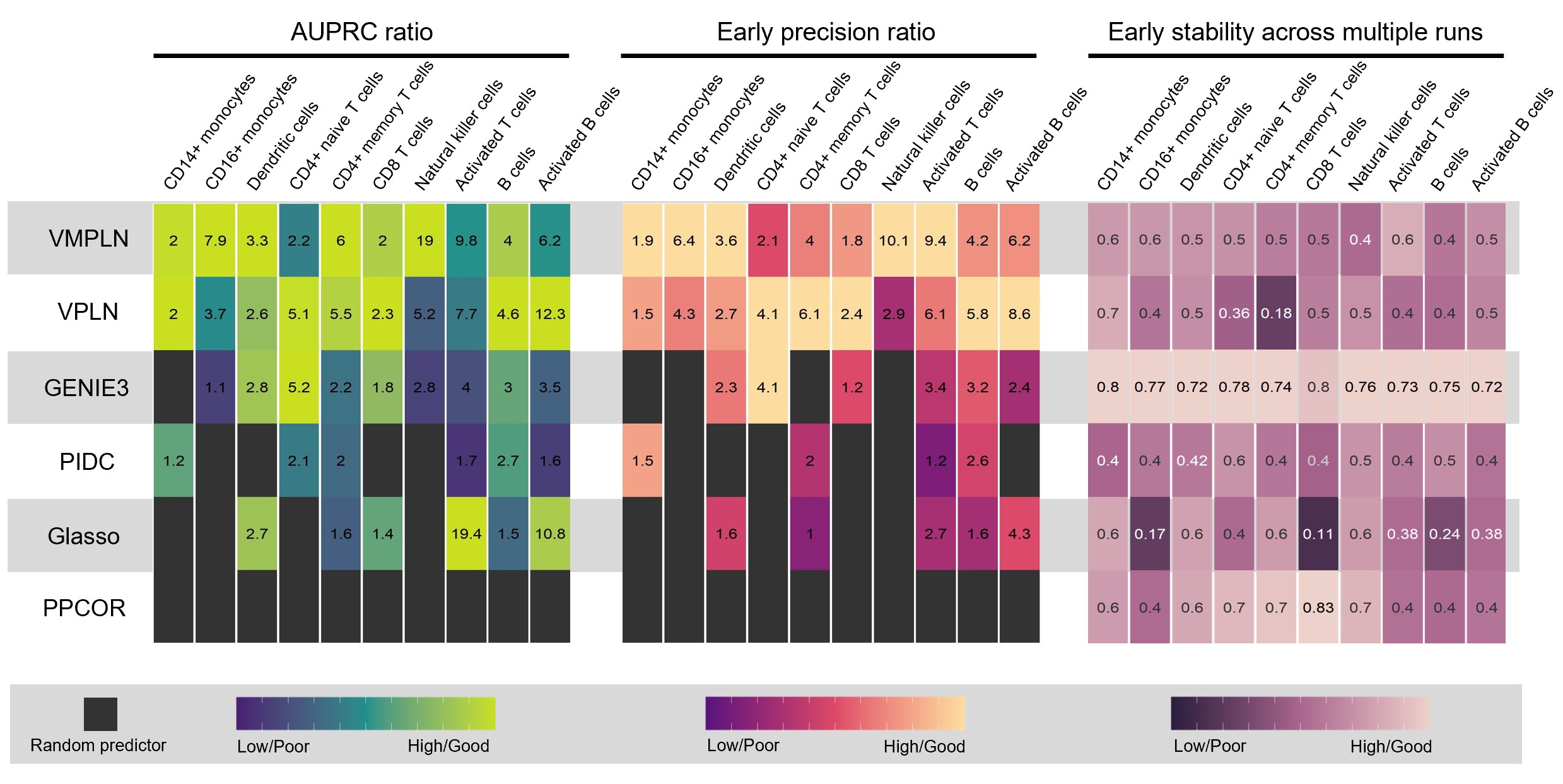}
	\end{center}
	\caption{The performance of the netwok inference algorithms in the Kang data. The gene set of interest for GRN inference consists of 39 TFs and 461 non-TFs. The colors represent the scaled values of these metrics (scaled to between 0 and 1 within each cell type) and the actual values are marked in the boxes. Black color in the boxes: random predictor performs better.}\label{realdata1}
\end{figure}

We compare the algorithms at the same network density ($5\%$) in terms of the AUPRC ratio, the EP ratio and the early stability. The AUPRC ratios and EP ratios are defined in the simulation section and are computed by comparing the estimated network with the silver standard. Since the regulatory relationships in the silver standard sets all involve TFs, we only consider edges involving TFs for performance evaluation. The early stability is to measure the stability of each algorithm by perturbing the input data. To calculate the stability, we down-sample $90\%$ of the original data, estimate the network using the same estimating procedure based on the down-sampled data, and calculate the pairwise Jaccard index between the networks estimated based on the down-sampled data. This process is repeated 100 times and the early stability is defined as the median of the Jaccard indexes.  

Figure \ref{realdata1} and supplementary material (Figure S17) show the heatmap of the AUPRC ratio, the EP ratio and the early stability of the 6 methods for these two benchmarking datasets.  The algorithms are ordered by the overall AUPRC ratio across all cell types. VMPLN has the highest AUPRC ratios and EP ratios in most cases and is the only method that consistently performs better than the random predictor. The stability of VMPLN is also reasonable and roughly similar to GENIE3. Note that GENIE3 is a tree-ensemble based method that uses data perturbation for network estimation, and thus should have good stability. 

\subsection{Application to scRNA-seq data from COVID-19 patients}
In this section, we consider the scRNA-seq data of bronchoalveolar lavage fluid macrophage cells from coronavirus disease 2019 (COVID-19) patients \citep{liao2020single}. The data consists of 29,980 single cells from 8 patients, including 2 patients with moderate COVID-19 infection and 6 patients with severe infection (supplementary material, Table S3). \citet{liao2020single} clustered the macrophages to four clusters including two classic M1-like macrophage groups (Group1 and Group2), the alternative M2-like macrophages (Group3) and the alveolar macrophages (Group4).

\begin{figure}[H]
	\begin{center}
		\includegraphics[scale = 0.46]{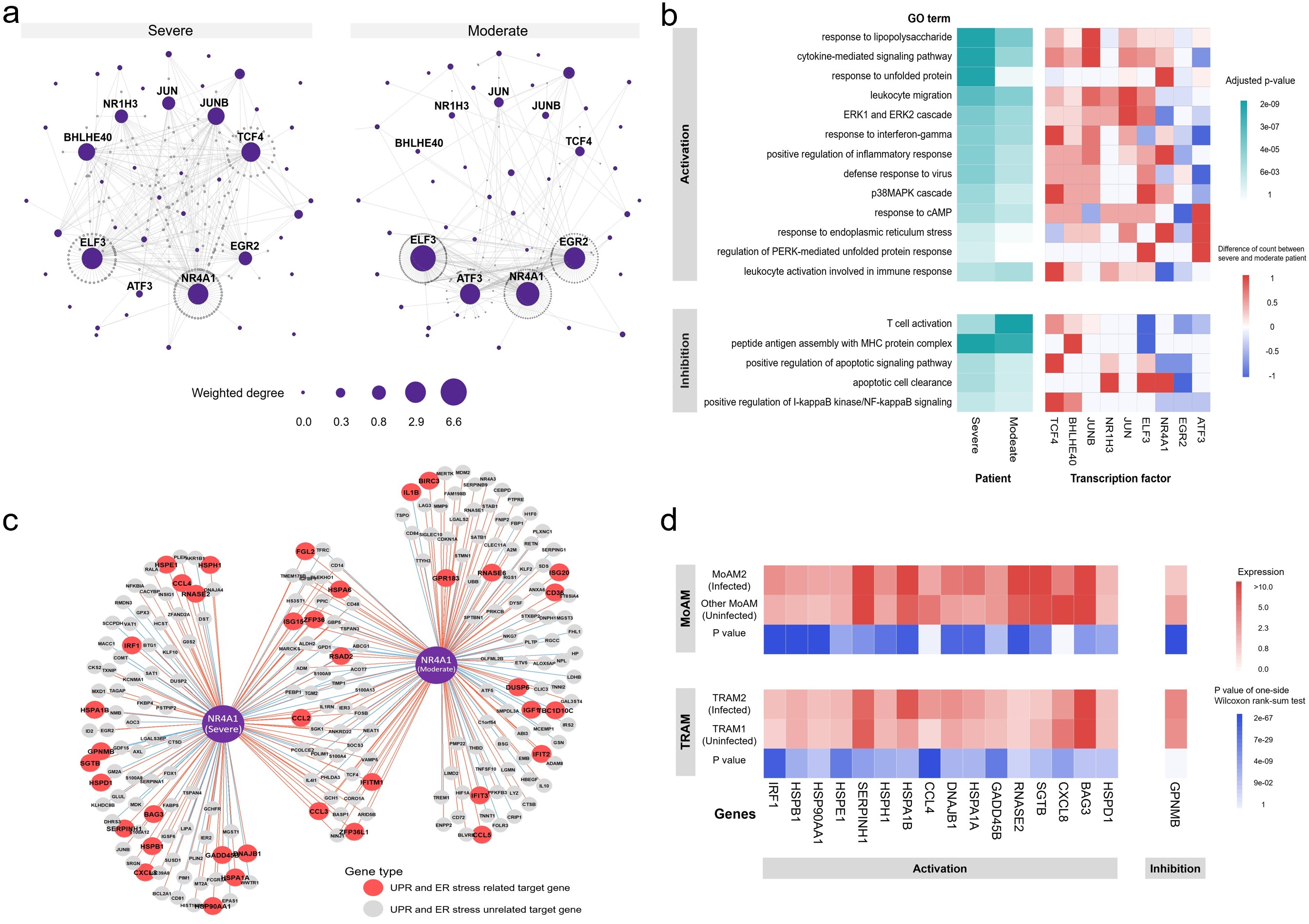}
	\end{center}
	\caption{\textbf{The GRN analysis of the COVID-19 data}. (a) The inferred GRNs of Group4 macrophages in severe and moderate patients. The size of the node represents the weighted node degree. (b) GO enrichment analysis of  TFs' target genes. TFs are selected as those with a large weighted degree differences ($> 0.2$) between severe and moderate patients. Activation and Inhibition mean that the regulatory relationships are positive and negative, respectively.  Left panel: p-values of GO terms. Right panel: differences of number of genes in the GO terms between severe and moderate patients. (c) The genes regulated by NR4A1. (d) Differential gene expression analysis between cells with and without SARS-CoV-2 infection. MoAM2: monocyte derived aleovar macrophage 2.  TRAM: tissue resident aleovar macrophage 2. One-sided Wilcoxon rank-sum test is used.}\label{COVID_19_whole}
\end{figure}

We perform the GRN analysis of the top 1000 HVG genes for the four macrophage groups using VMPLN  (Figure \ref{COVID_19_whole} a; supplementary material, Section S1.3.2 and Figure S18-S20) and compare networks of the patients with severe and moderate infection. We focus on the alveolar macrophages, since unlike other macrophage groups, the proportion of the alveolar macrophages tends to be smaller in patients with severe infection \citep{liao2020single}. A number of TFs exhibit a large weighted degree difference between the GRNs in the moderate and severe patients (weighted degree difference greater than 0.2; supplementary material, Figure S21). The weighted degree of a node is defined as the summation of the absolute partial correlation between the node and all other nodes that are connected with it. The node degrees of TCF4, BHLHE40, JUNB, NR1H3, and JUN in severe patients are substantially larger than in moderate patients, while the node degrees of ELF3, NR4A1, EGR2 and ATF3 are substantially larger in moderate patients. Gene oncology (GO) enrichment analysis of their target genes (Figure \ref{COVID_19_whole} b) shows that, as expected, many of target genes are involved in immune responses such as leukocyte migration and regulation of T cell activation. Interestingly, we observe that a number of GO terms are only enriched in severe patients such as response to unfolded protein (UPR) and response to endoplasmic reticulum (ER) stress. Similar analyses of the other macrophage groups also show that UPR is more enriched in severe patients than in moderate patients (supplementary material, Figure S22-S24). The enrichment of the UPR process is mainly due to the activation of target genes of NR4A1 (Figure \ref{COVID_19_whole} b; supplementary material, Figure S22-S24). The UPR and ER stress processes are frequently activated in cells infected by viruses \citep{janssens2014emerging} including coronavirus \citep{chan2006modulation}, indicating that alveolar macrophages might be infected by SARS-CoV-2. In fact, a recent study showed that macrophages can be infected by SARS-CoV-2 and the infected macrophages activate T-cells to promote alveolitis in patients with severe COVID-19 \citep{grant2021circuits}. These data imply that the UPR and ER stress related genes might be activated in SARS-CoV-2-infected macrophages through modulation of TFs such as NR4A1. Macrophage single cell data in \citet{grant2021circuits} has SARS-CoV-2 infection information for each single cell and thus allows expression comparison between cells with or without SARS-CoV-2 infection.  We take the UPR and ER stress related genes that are regulated by NR4A1 only  in severe patients (Figure \ref{COVID_19_whole} c) and compare their expressions in cells with and without SARS-CoV-2 infection. We find that most of these genes are indeed significantly differentially expressed (Figure \ref{COVID_19_whole} d).  In light of these findings, we reason that NR4A1 might play an important role in regulating cellular responses to SARS-CoV-2 infection. A number of NR4A1's target genes, including IRF1 and HSP90, have recently been discovered to be potential therapeutic targets for COVID-19 \citep{echavarria2021manipulation,shaban2021multi}. NR4A1 and its target genes that we identified here may also serve as potential therapeutic targets.

\section{Discussion}
\label{sec:disc}

In this paper, we develop a regulatory network inference method called VMPLN for scRNA-seq data. Instead of using the two-step procedure for network inference, VMPLN performs clustering and network inference simultaneously, and thus are especially suitable for scRNA-seq with mixed cell types. Most of the scRNA-seq data contain multiple cell types. We expect that VMPLN will have many applications in single cell studies.

A potential limitation of VMPLN is that it assumes that the regulatory relationships are linear. If the regulatory relationships are far away from being linear, VMPLN will not perform well. Methods like GENIE3 can allow the regulatory relationships to be nonlinear, but they require pre-clustering before network inference. One important research direction is to develop network methods that can account for nonlinear regulatory relationship as well as mixed cell populations. In addition, VMPLN is developed for scRNA-seq data. Other data and information can only be incorporated by setting the prior edges. Currently, single cell multi-omics technologies have been developed \citep{chappell2018single}. Developing network inference methods that can integrate multi-omics data can help to improve sensitivity and reduce false discoveries.

\section{Acknowledgments}
\label{sec:ackn}
This work was supported by the National Key Basic Research Project of China (2020YFE0204000), the
National Natural Science Foundation of China (11971039), and Sino-Russian Mathematics Center.


\bigskip


\pagebreak
\onecolumngrid
\begin{center}
	{\large\bf SUPPLEMENTARY MATERIAL}
\end{center}
\theoremstyle{definition}
\newtheorem{definition}{Definition}
\renewcommand{\thefigure}{S\arabic{figure}}
\renewcommand{\thetable}{S\arabic{table}}
\renewcommand{\thelemma}{S\arabic{lemma}}
\renewcommand{\theequation}{S\arabic{equation}}
\renewcommand{\thecondition}{S\arabic{condition}}
\renewcommand{\thedefinition}{S\arabic{definition}}
\renewcommand{\theproposition}{S\arabic{proposition}}
\renewcommand{\thesection}{S\arabic{section}}
\def\bA{\mathbf{A}}
\def\X{\mathbf{X}}
\def\Ex{{\rm E}}
\def\bt{\mathbf{t}}
\def\bT{\mathbf{T}}
\def\pr{\mathbb{P}}
\def\bB{\mathbf{B}}
\def\bF{\mathbf{F}}
\def\bI{\mathbf{I}}
\def\bQ{\mathbf{Q}}
\def\bK{\mathbf{K}}
\def\bS{\mathbf{S}}
\def\bT{\mathbf{T}}
\def\bU{\mathbf{U}}
\def\bW{\mathbf{W}}
\def\bX{\mathbf{X}}
\def\bY{\mathbf{Y}}
\def\bZ{\mathbf{Z}}
\def\bP{\mathbf{P}}
\def\bx{\mathbf{x}}
\def\by{\mathbf{y}}
\def\bz{\mathbf{z}}
\def\bM{\mathbf{M}}
\def\bS{\mathbf{S}}
\def\bG{\mathbf{G}}
\def\bL{\mathbf{L}}
\def\bN{\mathbf{N}}
\def\bzero{\mathbf{0}}
\def\bGamma{\boldsymbol{\Gamma}}
\def\bnu{\boldsymbol{\nu}}
\def\bOmega{\boldsymbol{\Omega}}
\def\bSigma{\boldsymbol{\Sigma}}
\def\bsigma{\boldsymbol{\sigma}}
\def\balpha{\boldsymbol{\alpha}}
\def\bgamma{\boldsymbol{\gamma}}
\def\bomega{\boldsymbol{\omega}}
\def\bTheta{\boldsymbol{\Theta}}
\def\btheta{\boldsymbol{\theta}}
\def\bbeta{\boldsymbol{\eta}}
\def\bxi{\boldsymbol{\xi}}
\def\bpi{\boldsymbol{\pi}}
\def\bl{\boldsymbol{l}}
\def\bmu{\boldsymbol{\mu}}
\def\bzero{\mathbf{0}}
\def\bone{\mathbf{1}}
\def\cA{\mathcal{A}}
\def\cM{\mathcal{M}}
\def\cU{\mathcal{U}}
\def\what{\widehat}
\def\wtilde{\widetilde}
\def\ve{\varepsilon}
\def\thefootnote{}
\def\qedsymbol{}
\def\vecone{{\rm vech}}
\def\vectwo{{\rm vech_2}}
\def\Y{{\bf Y}}
\def\K{{\bf K}}
\def\R{{\bf R}}
\def\D{{\bf D}}
\def\F{{\bf F}}
\def\bH{{\bf H}}
\def\trans{^{\rm T}}
\def\mR{\mathbb{R}}
\setcounter{equation}{0}
\setcounter{figure}{0}
\setcounter{table}{-1}
\setcounter{page}{1}
\setcounter{section}{0}
\makeatletter

\section{Supplementary Text}

\subsection{Technical proofs}
\setcounter{table}{-1}
\subsubsection{Notation}
For notational simplicity, we do not distinguish different lower (upper) bounds in Condition (C1-C3), and always use $m$ to denote the lower bound and $M$ to denote the upper bound. We always assume $M>0$. When $m$ is the lower bound for eigenvalues of precision matrices, we always assume $m>0$.  	We define two vectorization operators,  $\vecone$ and $\vectwo$. For a symmetric matrix $\bA \in \mathbb{R}^{p \times p} = [a_{ij}]$, $\vecone(\bA)$ is defined as
$$\vecone(\bA) = (a_{11}, a_{12}, a_{13},\ldots,a_{1p},  a_{22}, a_{23}, \ldots, a_{2p},\ldots,a_{(p-1)(p-1)}, a_{(p-1)p}, a_{pp})\trans,$$
and $\vectwo(\bA)$ is 
$$\vectwo(\bA) = (a_{11}, 2a_{12},2 a_{13},\ldots,2a_{1p},  a_{22}, 2a_{23}, \ldots, 2a_{2p},\ldots,a_{(p-1)(p-1)}, 2a_{(p-1)p}, a_{pp})\trans.$$
Note that  $\vecone$ and $\vectwo$  only differ at off-diagonal elements. Define ${\bnu} =  (\bnu_1^{\rm T},\ldots,  \bnu_g^{\rm T}, \ldots, \bnu_G^{\rm T})^{\rm T},$ where  $\bnu_g = {\rm vech}(\bTheta_g)$. We assume the true parameter $\bnu^*$  is an interior point of $\mathcal{D}$. 
For the PLN, we can write the log-likelihood as follows,
$$\log\, p(\by,  \bl;\bTheta,  \bmu) = \sum_{i=1}^{n} \log \int f(\bx;\bTheta,  \bmu)p(\by_i|\bx, l_i) d\bx +\sum_{i=1}^{n}\log \left(p(l_i)\right).$$
where $f(\bx; \bTheta,  \bmu)$ is the probability density function of ${\rm N}(\bTheta, \bmu)$. 
Note that $\sum_{i=1}^{n}\log \left(p(l_i)\right)$ does not depend on $\bTheta$.  We only need to consider the conditional distribution $p(\by_i | l_i ;\bTheta, \bmu_g)$.
Since $p(l)$ is independent of the  unknown parameters and has a bounded support,  
it can be seen from the following section that the library size $l$ does not have essential influence on the proof. For brevity, we assume that the library size is the constant $1$ in the proof. 

In the following sections,  we always use  $p(\by; \bTheta, \bmu)$ and  $p\left(\by; \bnu, \{\bmu_g\}_{g=1}^G\right)$ to represent the  density of  the PLN distribution ${\rm PLN }\left(\bTheta, \bmu\right)$ and  the  density of  the MPLN distribution  MPLN $\left(\bnu, \{\bmu_g\}_{g=1}^G\right)$, respectively.
Given a single sample  $i$, we write the log-likelihood function  of the PLN at $\by_i$ as 
\begin{align*}
	\ell(\bTheta, \by_i)& = \log\left(p(\by_i ;\bTheta, \bmu)\right) =\frac{1}{2} \log \det(\bTheta) \\ &  + \log 
	\int  \det(\bTheta) ^{\frac{1}{2}} \exp\left(- \frac{1}{2} (\bx- \bmu)\trans \bTheta (\bx- \bmu)\right) {h(\by_i,\bx)}d\bx + C(\by_i),
\end{align*}
where 
\begin{equation} \label{equ:h}
	h(\by_i,\bx) = \prod_{j=1}^{p} \exp(x_{j} y_{ij}) \exp\left(-\exp(x_j)\right)
\end{equation}
and $C(\by_i) = \sum_{j=1}^p \log y_{ij}! - 2^{-1}p \log(2\pi)$. In the following sections, we always write $h(\by,\bx) = \prod_{j=1}^{p} \exp(x_{j} y_{j}) \exp(-\exp(x_j))$. Also, we define $\ell_n(\bTheta) = \sum_{i=1}^{n} \ell(\bTheta, \by_i)$ as the log-likelihood in the PLN. 
For the MPLN,  its log-likelihood function at $\by_i$ is
\begin{align*}
	\ell(\bnu, \by_i)& = \log \left(\sum_{g=1}^{G}\pi_g  p(\by_i ;\bTheta_g, \bmu_g)\right)    \\ &  = \log \left(\sum_{g=1}^{G}\pi_g  \int \det\left(\bTheta_g\right) ^{\frac{1}{2}}\exp \left(- \frac{1}{2} (\bx- \bmu_g)\trans \bTheta_g (\bx- \bmu_g)\right)   {h(\by_i,\bx)}d\bx\right) + C(\by_i),
\end{align*}
where  $C(\by_i) = \sum_{j=1}^p \log y_j! - 2^{-1}p \log(2\pi)$. 
The log-likelihood of the MPLN model is
${\ell}_n(\bnu) =  \sum_{i=1}^{n}	\ell(\bnu, \by_i).$
If we define 
$$L_g(\bnu_g, \by) = \int \det(\bTheta_g) ^{\frac{1}{2}}\exp \left(- \frac{1}{2} (\bx- \bmu_g)\trans \bTheta_g (\bx- \bmu_g)\right) 
{h(\by,\bx)}d\bx ,$$
and $L_M(\bnu, \by) = \sum_{g=1}^{G}\pi_g L_g(\bnu_g, \by)$, then ${\ell}(\bnu, \by_i)  = \log \left(L_M(\bnu, \by_i)\right) + C(\by_i)$.
Note that  the function $L_g(\bnu_g, \by)$  is proportional to the density 
$p(\by ;\bTheta_g, \bmu_g)$.
Let 
$\mathcal{L}_n(\bnu) = -n^{-1}\ell_n (\bnu)$.
The optimization problem (3) in the main manuscript can be written as 
\begin{equation} \label{prob:opt}
	\hat{\bnu}_{n} \in \argmin_{\bnu \in {\mathcal{D}}} \left\{\mathcal{L}_n(\bnu) + {\lambda_n} \mathcal{R}(\bnu)\right\},
\end{equation}
where $\mathcal{R}(\bnu) = \sum_{g=1}^G ||\bTheta_g||_{1,\rm off}.$

For the PLN model, denote the derivative (the score function) and the Hessian matrix of its log-likelihood as 
$$\mathcal{S}(\bTheta, \by) = \frac{\partial\ell(\bTheta , \by)}{\partial \vecone(\bTheta)} , \bH(\bTheta, \by) = \frac{\partial^2 \ell(\bTheta, \by)}{\partial \vecone(\bTheta)\partial\vecone(\bTheta)\trans}.$$ 
For the MPLN model, we can similarly define its  score function $\mathcal{S}^M(\bnu, \by)$, its Hessian matrix $\F (\bnu, \by )$, and  its Fisher information matrix $\bGamma^*= \bGamma(\bnu^*)$.    
\begin{equation} \label{equ:D}
	\D(\bnu)=\Ex_{\bnu^*} (\F (\bnu, \by )). 
\end{equation}
Note that $\D(\bnu^*) = -\bGamma(\bnu^*)$. These notations are summarized in Table \ref{tab:notation}.
\begin{table} [htb]
	\centering
	\caption{Summary of  notations.}
	\renewcommand\arraystretch {1.5}
	\begin{tabular}{l l}
		\hline
		Name & Definition \\
		\hline
		The  density of  the PLN distribution ${\rm PLN }\left(\bTheta, \bmu\right)$ & $p(\by; \bTheta, \bmu)$ \\
		
		The log-likelihood function  of the PLN at $\by$ & $\ell(\bTheta, \by)  = \log\left(p(\by ;\bTheta, \bmu)\right)$   \\ 
		
		The log-likelihood function  of the PLN & $\ell_n(\bTheta) =  \sum_{i=1}^{n} \ell(\bTheta, \by_i)$  \\  
		
		The derivative of  $\ell(\bTheta, \by)$ &  $\mathcal{S}(\bTheta, \by) = \frac{\partial\ell(\bTheta , \by)}{\partial \vecone(\bTheta)} $\\
		
		The Hessian matrix of $\ell(\bTheta, \by)$ & $\bH(\bTheta, \by) = \frac{\partial^2 \ell(\bTheta, \by)}{\partial \vecone(\bTheta)\partial\vecone(\bTheta)\trans}$ \\
		
		The  density of  the MPLN distribution MPLN $\left(\bnu, \{\bmu_g\}_{g=1}^G\right)$ & $p(\by; \bnu, \{\bmu_g\}_{g=1}^G)$ \\
		
		The log-likelihood function  of the MPLN at $\by$ & $\ell(\bnu, \by)  = \log\left(p\left(\by; \bnu, \{\bmu_g\}_{g=1}^G\right)\right)$   \\ 
		
		The log-likelihood function  of the MPLN & $\ell_n(\bnu) =  \sum_{i=1}^{n} \ell(\bTheta, \by_i)$  \\  
		
		The first part of the objective function  & $\mathcal{L}_n(\bnu) = -n^{-1}\ell_n(\bnu) $  \\  
		The second part of the objective function  & 						$\mathcal{R}(\bnu) = \sum_{g=1}^G ||\bTheta_g||_{1,\rm off}$   \\  
		
		The derivative of  $\ell(\bnu, \by)$ &  $\mathcal{S}^M(\bnu, \by) = \frac{\partial\ell(\bnu , \by)}{\partial \bnu} $\\
		
		The Hessian matrix of $\ell(\bnu, \by)$ & $\F(\bnu, \by) = \frac{\partial^2 \ell(\bnu, \by)}{\partial \bnu\partial\bnu\trans}$ \\
		
		The expectation of $\F(\bnu, \by) $ & 		$\D(\bnu)=\Ex_{\bnu^*} (\F (\bnu, \by ))$  \\
		
		The Fisher information matrix of MPLN  &$\bGamma^* =    \bGamma(\bnu^*) = -\D(\bnu^*)$ \\
		\hline
	\end{tabular}
	\label{tab:notation}
\end{table}
Finally, we denote  $\mathbb{N}^p$ as the set of all $p$-dimensional non-negative integer vectors.  For a vector $\mathbf{a} = (a_1,\ldots,a_p)$, we denote $||\mathbf{a}||_2 = \sqrt{\sum_{j=1}^p a_j^2}$ as its $L_2$-norm and $||\mathbf{a}||_{\infty} = \max_{j} |a_j|$  as its $L_{\infty}$-norm. For a matrix $\bA$, we denote $||\mathbf{A}||_2$ as its largest singular value of $\mathbf{A}$ and $||\mathbf{A}||_{1,\infty}$ as its maximum absolute row sum of $\mathbf{A}$.  Given $\bTheta$ and $\bmu$, we define an operator $\mathcal{T}$ that maps functions in $x$ to functions in $y$,
$$\mathcal{T}(f) =  \int \exp\left(-\frac{1}{2} (\bx-\bmu)\trans \bTheta (\bx-\bmu) \right)f(\bx) h(\by,\bx)d\bx .$$ We use $\mathbb{I}(\bx) \equiv 1$ as the constant function taking value 1.

\subsubsection{Some Lemmas }
\begin{lemma}\label{lem:poly}
	Let $\by \sim {\rm PLN} (\bTheta, \bmu)$. 
	For any $n,y \in \mathbb{N}$,
	we define
	$$
	\phi(y,n) = 
	\left\{  
	\begin{aligned}
		& 1 & & n=  0, \\
		& y(y-1)\cdots(y-n+1) & & n > 0.
	\end{aligned}
	\right.  
	$$
	Then, for ${\mathbf n} = (n_1,\cdots, n_p)\trans$,
	we have 
	$$
	\Ex\left(\prod_{j=1}^p \phi({n_j},y_j)\right) = \exp \left({{\mathbf n}}\trans \bmu + {{\mathbf n}}\trans \bTheta^{-1} {{\mathbf n}}/2 \right).
	$$  
\end{lemma}

\begin{lemma}[1-dimensional dominating function] \label{1dim}
	Suppose $\theta \in [m,M]$ ($m, M> 0$), and $|\mu| \leq M$.   Let 
	$$f^1(y,\theta,\mu) = {\int \exp\left(-\frac{1}{2}\theta(x -  \mu)^2\right) \exp\left(-\exp(x)\right) \exp(xy)dx}.$$ 
	Then, for any large enough positive integer $y$, we have 
	$$
	f^1(y,\theta,\mu)
	\geq  C \exp(y\log(y+1)/2),
	$$
	where $C>0$ is  constant depending on $m,M$.
	
\end{lemma}

\begin{lemma}[$p$-dimensional dominating function] \label{pdim}
	Let 
	\begin{equation} \label{pdimequa}
		f^p(\by,\bTheta ,\bmu) = \frac{\mathcal{T}\left((x_{1} - \mu_1)^4\right)}{\mathcal{T}(\mathbb{I})}, 
	\end{equation}
	where $\by=(y_1, \dots, y_p) \in \mathbb{N}^p$. Assuming $\by \sim {\rm PLN}(\bTheta^*, \bmu)$, under Condition (C1-C3), there exists a polynomial function $g(\by)$ with a constant $C_p$ only depending on $p,m$ and $M$,
	$$g(\by) = \left(4\frac{||\by||_2}{m}\right)^4 + C_p$$
	such that  $\Ex_{\bTheta^*}(g(\by))<\infty$ and $|f^p(\by, \bTheta,\bmu)| \leq g(\by)$ for any $\bTheta, \bmu$ satisfying Condition (C1-C3).  
\end{lemma}
\begin{remark}\label{pdimremark}
	Applying the same proof as in Lemma \ref{pdim}, the polynomial $(x_{1} - \mu_1)^4 $ can be replaced by any polynomial with respect to $\bx$, e.g. $(x_{i} -\mu_i)^2(x_{j} - \mu_j)^2$ and $(x_{i} - \mu_i)(x_{j} - \mu_j)(x_{k} - \mu_k)^2$. Further, for any two polynomial functions ${\psi}_1 (\bx) , {\psi}_2 (\bx)$ and 
	$$f^p(\by, \bTheta ,\bmu) = \frac{\mathcal{T}\left({\psi}_1 (\bx)\right) \mathcal{T}\left({\psi}_2 (\bx)\right)}{\mathcal{T}^2(\mathbb{I})}, 
	$$ 
	there exists a polynomial function $g(\by)$ with $\Ex_{\Theta^*}(g(\by))<\infty$ such that  
	$|f^p(\by,\bTheta,\bmu)| \leq g(\by)$ for any $\bTheta,\bmu$ satisfying Condition (C1-C3).  
\end{remark}

~\\
\noindent
{\bf  Proof of Lemma \ref{lem:poly}. }By the property of conditional expectation, we have 
$$
\Ex\left(\prod_{j=1}^p \phi({n_j},y_j)\right) = \Ex_{\bx}\Ex_{\by}\left(\prod_{j=1}^p \phi({n_j},y_j)|\bx\right). 
$$
From the moments of the Poisson distribution, we have 
$$\Ex_{\by}\left(\prod_{j=1}^p \phi({n_j},y_j)|\bx\right) = \prod_{j=1}^p \exp(n_j x_j) .$$
Further, since $\bx \sim {\rm N}(\bmu, \bTheta^{-1})$, we have
$$\Ex_{\bx}\left( \prod_{j=1}^p \exp(n_j x_j)\right) = \Ex_{\bx}(\exp({{\mathbf n}}\trans\bx)) = \exp\left({{\mathbf n}}\trans \bmu +  {{\mathbf n}}\trans \bTheta^{-1} {{\mathbf n}}/2\right),$$
and the conclusion follows.

~\\
\noindent
{\bf{Proof of Lemma \ref{1dim}. }}Since for any $y \in \mathbb{N}$, $f^{1}(y, \theta, \mu) > 0$,  we only need to consider $y$ large enough.
Let 
$$g(x,y) = \exp\left(-\frac{1}{2}\theta(x - \mu)^2\right) \exp(-\exp(x)) \exp(xy).$$
Clearly, we have, for $y$ large enough,
$$\int g(x,y)dx 
\geq \int_{\log(y)-1}^{\log(y)} g(x,y)dx.$$
Let $t = x-\log(y)$, then 
\begin{align}\label{1dimlower}
	&\quad \int_{\log(y)-1}^{\log(y)} g(x,y)dx  \notag \\
	&= \int_{-1}^{0} \exp\left(-\frac{1}{2}\theta(\log(y)+t- \mu)^2\right) \exp(-\exp(\log(y)+t)) \exp((\log(y)+t )y)dt \notag\\
	&= \exp(y\log(y))\int_{-1}^{0} \exp\left(-\frac{1}{2}\theta(\log(y)+t- \mu)^2\right) \exp(-\exp(\log(y)+t)) \exp(yt)dt \notag\\
	&\geq \exp(y\log(y)) \min_{t\in[-1,0]} \left\lbrace \exp\left(-\frac{1}{2}\theta(\log(y)-t- \mu)^2\right)\right\rbrace 
	\int_{-1}^0 \exp(y(t-\exp(t)))dt.\notag\\
	&\geq \exp(y\log(y)) \min_{t\in[-1,0]} \left\lbrace \exp\left(-\frac{1}{2}\theta(\log(y)-t- \mu)^2\right)\right\rbrace
	\int_{-1}^0 \exp(y(-1-t^2/2))dt. \notag \\
	&\geq \exp(y\log(y)-y) \min_{t\in[-1,0]} \left\lbrace \exp\left(-\frac{1}{2}\theta(\log(y)-t- \mu)^2\right)\right\rbrace
	\int_{-1}^0 \exp(-t^2/2)dt /\sqrt{y}. \notag\\
	&\geq \exp(y\log(y)-y) C_2 \exp(-C_3\log^2(y)+C_4\log(y))/\sqrt{y},
\end{align}
where $C_2$, $C_3$ and $C_4$ are three constants only depending on $m,M$.
Since the leading order of (\ref{1dimlower}) is $\exp(y\log(y)-y) $ , when $y\geq 1$, there exists a constant $C$ depending on $m,M$ such that for any large enough positive integer $y$, we have 
$$	f^1(y,\theta,\mu) \geq C\exp(y\log(y+1)/2).$$

~\\
\noindent
{\bf{Proof of Lemma \ref{pdim}. }}Similarly, we only need to consider $||\by||_2$ large enough.  The proof consists of the following three steps.

\noindent
{\bf Step 1.}  We first give a lower bound for the denominator of (\ref{pdimequa}). 
Note that 
$$\exp\left(-\frac{1}{2}  (\bx - \bmu)\trans \bTheta  (\bx - \bmu) \right) \geq \exp\left(-\frac{1}{2} M (\bx - \bmu)\trans   (\bx - \bmu) \right).$$ We have
\begin{align*}
	&\quad \int \exp\left(-\frac{1}{2}  (\bx - \bmu)\trans \bTheta  (\bx - \bmu) \right)h(\by,\bx)d\bx \\
	&\geq \int\exp\left(-\frac{1}{2} M (\bx - \bmu)\trans   (\bx - \bmu) \right)h(\by,\bx)d\bx \\
	&= \prod_{j=1}^p \int \exp\left(-\frac{1}{2} M(x_j - \mu_j)^2\right)\exp(-\exp(x_j)) \exp(x_jy_j)dx_j
\end{align*}
By Lemma \ref{1dim}, we have, for any fixed $j$, $ \int \exp\left(-\frac{1}{2}  M(x_j -\mu_j)^2 \right)\exp(-\exp(x_j)) \exp(x_jy_j)dx_j$ is greater than $C\exp(y_j\log(y_j + 1)/2)$. Hence,  we have
$$\int \exp\left(-\frac{1}{2} (\bx - \bmu)\trans \bTheta  (\bx - \bmu)\right)h(\by,\bx)d\bx \geq C^p  \exp\left(\sum_j y_j\log(y_j + 1)/2\right).$$
\noindent
{\bf Step 2.}  When $||\bx- \bmu||_2 \leq A(\by) = 4\frac{||\by||_2}{m}$, we have $(x_1 - \mu_1)^4 \leq ||\bx- \bmu||_2^4\leq A^4(\by)$, where $x_1$ is the first element of $\bx$. Then, we have 
$$
\frac{\int_{||\bx-\bmu||_2 \leq A} \exp\left(-\frac{1}{2}  (\bx - \bmu)\trans \bTheta  (\bx - \bmu)\right)  (x_{1} - \mu_1)^4h(\by,\bx)d\bx}{\int \exp\left(-\frac{1}{2}  (\bx - \bmu)\trans \bTheta  (\bx - \bmu) \right)h(\by,\bx)d\bx} \leq A^4(\by). 
$$
Since $A^4(\by)$ is a polynomial function of $\by$, we have $\Ex_{\bTheta^{*}}(A^4(\by)) < \infty.$

\noindent
{\bf Step 3.}  When $||\bx-\bmu||_2 > A(\by)$,  we have
\begin{align*}
	&\quad \int_{||\bx-\bmu||_2 > A(\by)} \exp\left(-\frac{1}{2}  (\bx - \bmu)\trans \bTheta  (\bx - \bmu)\right) (x_{1} -\mu_1)^4 h(\by,\bx)d\bx \\
	&\leq  \int_{||\bx - \bmu||_2 > A(\by)} \exp\left(-\frac{1}{2}  (\bx - \bmu)\trans \bTheta  (\bx - \bmu)\right) (x_{1} - \mu_1)^4 \exp\left(\sum_{j=1}^{p}x_{j}{y_{j}}\right) d\bx\\
	&= \prod_{j=1}^{p} \exp(\mu_j{y_{j}}) \int_{||{\mathbf{u} }||_2 > A(\by)} \exp\left(-\frac{1}{2}  {\mathbf{u} }\trans \bTheta {\mathbf{u} }\right) u_1^4\prod_{j=1}^{p} \exp(u_{j}{y_{j}}) d{\mathbf{u} } \quad({\mathbf{u} }= \bx - \bmu)\\
	&\leq \prod_{j=1}^{p} \exp(\mu_j{y_{j}})\int_{||{\mathbf{u} }||_2 > A(\by)} \exp\left(-\frac{1}{2} m ||{\mathbf{u} }||_2^2\right) ||{\mathbf{u} }||_2^4 \exp(||{\mathbf{u} }||_2||\by||_2)d{\mathbf{u} },
\end{align*}
where the last inequality is by Cauchy's inequality. Note that the area of the $p$-dimensional sphere of radius $r$ is $M_p r^{p-1}$ with $M_p$ being a constant only depending on $p$. Let $r = ||{\mathbf{u} }||_2$. By the polar decomposition, the $p$-dimensional integral can be rewritten as
\begin{align*}
	&\quad \int_{||{\mathbf{u} }||_2 > A(\by)} \exp(-\frac{1}{2} m ||{\mathbf{u} }||_2^2) ||{\mathbf{u} }||_2^4 \exp(||{\mathbf{u} }||_2||\by||_2)d{\mathbf{u} } \\
	&= \int_{r>A(\by)} \exp(-\frac{1}{2} m r^2) r^4 \exp(r||\by||_2) M_p r^{p-1}dr \\
	&= \int_{r>A(\by)} \exp(-\frac{1}{2} r (m r - 2||\by||_2)) M_p r^{p+3}dr \\
	&\leq \int_{r>A(\by)} \exp(-\frac{1}{2} r (2||\by||_2)) M_p r^{p+3}dr  \quad (r>4\frac{||\by||_2}{m})\\
	&\leq \int_{r>A(\by)} \exp(-r) M_p r^{p+3}dr \\
	&\leq \int_{r>0} \exp(-r) M_p r^{p+3}dr \\
	&= C'_p,
\end{align*}
where $C'_p$ is a constant  only depending on $p$. Then, under Condition (C1-C3), we have 
\begin{align*}
	&\quad 	\frac{\int_{||\bx-\bmu||_2 \leq A} \exp\left(-\frac{1}{2}  (\bx - \bmu)\trans \bTheta  (\bx - \bmu)\right)  (x_{1} - \mu_1)^4h(\by,\bx)d\bx}{\int \exp\left(-\frac{1}{2}  (\bx - \bmu)\trans \bTheta  (\bx - \bmu) \right)h(\by,\bx)d\bx} \\ \\
	& \leq  \frac{ \prod_{j=1}^{p} \exp(\mu_j{y_{j}}) C'_p}{\exp\left(\sum_j y_j\log(y_j + 1)/2 \right) } \leq C_p, 
\end{align*}
where $C_p$ is a constant depending on $p,m$ and $M$. 

Finally, combining the results in Step 2 and 3,  we get the dominating function 
$$g(\by) = \left( 4\frac{||\by||_2}{m}\right) ^4 +C_p$$
which is a polynomial. By Lemma \ref{lem:poly}, we have $\Ex_{\bTheta^{*}}(g(\by)) < \infty$.

\subsubsection{Proof of Theorem  1,  part I}

To prove the the first conclusion of Theorem 1, we introduce the following definition and give two lemmas.

\begin{definition}[Good vector]
	We call a vector $\bxi = (\xi_1, \cdots, \xi_G)\trans \in \mathbb{R}^G$ as a \textbf{good vector} if one $\xi_g$ only appears once  in $\bxi$, i.e. $\xi_{g'} \neq \xi_{g}$ for all $g' \neq g$. We call the index $g$ as a \textbf{good index} with respect to $\bxi$.
\end{definition}
\begin{lemma}\label{goodvec}
	Let $\bxi= (\xi_1, \cdots, \xi_G)\trans $ be a good vector with a good index $s$,  $\bsigma = (\sigma_1,\cdots, \sigma_G)\trans$ satisfy $\sigma_g >0$ for $g=1,\cdots,G$ and $\balpha = (\alpha_1,\cdots, \alpha_G)\trans$. If for any $z \in \mathbb{N}$
	$$\sum_{g=1}^{G}\alpha_g \exp(\xi_g z +  \sigma_g z^2/2) = 0,$$
	then $\alpha_s = 0$.
\end{lemma} 
\begin{lemma} \label{lem:subspace}
	For any $n>0$, let $\mathcal{M}_i \subset \mathbb{R}^p, i=1,\cdots, n$ be $n$ linear proper subspaces. Then, there  exists a non-negative integer vector $\bgamma$ such that $\bgamma \not\in \bigcup_{i=1}^n \mathcal{M}_i$.
\end{lemma}
\begin{proposition}\label{prop:density}
	$p\left(\by; \bTheta_1, \bmu_1\right), \cdots, p\left(\by; \bTheta_G, \bmu_G\right)$ are linearly independent for $\bmu_g$ ($g=1,\cdots, G$) that are bounded and different from each other. 
\end{proposition} 

\noindent
{\bf  Proof of Theorem  1  part I. }By \cite{yakowitz1968identifiability}, under Condition (C1-C3), the identifiability of the MPLN model is equivalent to the linear independence of the PLN components. Thus, we aim to prove   Proposition \ref{prop:density}.
We prove this by mathematical induction.

The independence for $G=1$ is trivial.  
Now we assume that Proposition \ref{prop:density} holds for $G-1$. For any $\bmu_1,\cdots, \bmu_G$ that are bounded and different from each other, if  we can prove that there exists $\balpha = (\alpha_1, \cdots, \alpha_{G})\trans$ and an index $s \in \{1,\cdots,G\}$ such that 
$$\sum_{g=1}^G \alpha_g p\left(\by;\bTheta_g, \bmu_g\right) = 0 \mbox{ and } \alpha_s=0,$$
then by induction, we have  $\balpha = 0$ and hence $p(\by;  \bTheta_g, \bmu_g)$ ($g=1,\cdots,G$) are linearly independent. So our goal is to prove that if $\sum_{g=1}^G \alpha_g p(\by;\bTheta_g, \bmu_g) = 0$, then we can always find an index $s$ such that $\alpha_s = 0$.

Let ${{\mathbf n}} = (n_1,\cdots, n_p)\trans$ be any non-negative integer vector.  Then, for any positive integer $z$,  by Lemma \ref{lem:poly},  there exists a polynomial function $p_z(\by) = \prod_{j=1}^p  \phi(z n_j,y_j), z\in \mathbb{N} $ such that $$\sum_{g=1}^G \alpha_g \Ex_g\left(p_z(\by)\right) = 0 \mbox{ and } \Ex_g\left(p_z(\by)\right) = \exp\left(z {\mathbf n}\trans \bmu_g + z^2{{\mathbf n}}\trans \bTheta^{-1}_g {{\mathbf n}} /2\right),$$
where $\Ex_g$ represents taking expectation with respect to  ${\rm PLN}\left(\bTheta_g, \bmu_g\right)$.  Let 
$$\bxi = \left({\mathbf n}\trans \bmu_1, \cdots,  {\mathbf n}\trans \bmu_G\right) \mbox{ and } \bsigma = \left({{\mathbf n}}\trans \bTheta^{-1}_1 {{\mathbf n}}/2 , \cdots, {{\mathbf n}}\trans \bTheta^{-1}_G {{\mathbf n}}/2\right).$$ By  Lemma \ref{goodvec}, if  there exists an ${\mathbf n}$ such that $\bxi$ is a good vector with good index $s$, then $\alpha_s = 0$ and we complete the proof. If, on the other hand,  $\bxi = \left({\mathbf n}\trans \bmu_1, \cdots,  {\mathbf n}\trans \bmu_G\right)$ is not a good vector for any non-negative integer vector ${\mathbf n}$.  Therefore, for any ${\mathbf n}$, there exists $s
\neq 1$ such that $ {\mathbf n}\trans \bmu_1 =  {\mathbf n}\trans \bmu_s$. Thus, ${\mathbf n}$ is the solution to the linear equation 
${\mathbf x}\trans( \bmu_1 - \bmu_s) = 0$. We define $\mathcal{M}_g$ as the linear space consisting of solutions to the linear equation ${\mathbf x}\trans\left(\bmu_1 - \bmu_g\right) = 0$ ($g \neq 1$) and $\mathcal{M} = \cup_{g=2}^G \mathcal{M}_g$.  Thus, for any non-negative integer vector ${\mathbf n}$, we have ${\mathbf n} \in \mathcal{M}$. Since $\bmu_g$ are different form each other, then $\dim{\mathcal{M}_g} = p-1$ and $\mathcal{M}_g$ is a proper subspace of $\mathbb{R}^p$. This is contradictory to Lemma \ref{lem:subspace}.  So there exists an ${\mathbf n}$ such that $\bxi = \left({\mathbf n}\trans \bmu_1, \cdots,  {\mathbf n}\trans \bmu_G\right)$ is  a good vector and hence for any $\bmu_g$ ($g=1,\cdots,G$) that are different from each other, $p(\by;\bTheta_1, \bmu_1), \cdots, p(\by; \bTheta_{G}, \bmu_{G})$ are linearly independent. 

~\\
\noindent
{\bf  Proof  of Lemma \ref{goodvec}. }Without loss of generality,  we assume that $(\xi_g,\sigma_g )$ ($g=1,\cdots,G$) are increasingly ordered  (first by $\xi$ then by $\sigma$). We say that $(\xi_g,\sigma_g)$ and $(\xi_s,\sigma_s)$ are equivalent if $(\xi_g,\sigma_g) = (\xi_s,\sigma_s)$. By this equivalence relationship, 
$\{(\xi_g,\sigma_g ) \}_{g=1}^G$ can be partitioned into $Q$ groups ($Q\geq 1$). Let $S_q$ be the index set of the $q$-th group. We have 
$$\sum_{q=1}^{Q} \sum_{j \in S_q} \alpha_j \exp(\xi_j z +  \sigma_j z^2/2) = 0$$
for all $z \in \mathbb{N}$.  
Dividing $\exp(\xi_{G}z +  \sigma_{G} z^2/2) $  on both sides of the above equation, we get
\begin{equation}\label{equ3}
	\sum_{q=1}^{Q-1} \sum_{j \in S_q} \alpha_j\exp(\xi_j z + \sigma_j z^2/2 - \xi_{G}z-  \sigma_{G} z^2/2) + \sum_{j \in S_{Q}} \alpha_j = 0
\end{equation}  
for all $z \in \mathbb{N}$.  By the choice of $\sigma_{G}, \xi_{G} $, the first summation of  (\ref{equ3}) converges to zero when $z$ goes to infinity. So, we have $\sum_{j \in S_{Q}} \alpha_j = 0$. By  mathematical induction, we have 
$\sum_{j \in S_q} \alpha_j  = 0$ for $q = 1,\cdots, Q$.  Since $\bxi$ is a good vector with a good index $s$, ($\xi_s,\sigma_s$) itself forms a group, and hence $\alpha_s = 0$.  

~\\
\noindent
{\bf  Proof of Lemma \ref{lem:subspace}. }We prove by mathematical induction. The conclusion clearly holds for $n=1$.  Now we assume that Lemma \ref{lem:subspace}  holds for $n$ and we aim to prove that it also holds for $n+1$. 

By induction hypothesis, we can take $\balpha \in \mathbb{N}^p	\setminus \bigcup_{i=1}^n \mathcal{M}_i$. If $\balpha \not\in \mathcal{M}_{n+1}$, we have $\balpha  \not\in \bigcup_{i=1}^{n+1} \mathcal{M}_i$, and the proof is finished.  Thus, we only need to consider  $\balpha \in \mathcal{M}_{n+1}$. Similarly, we can take ${\boldsymbol{\beta}} \in \mathbb{N}^p	\setminus \bigcup_{i=2}^{n+1} \mathcal{M}_i$ and ${\boldsymbol{\beta}} \in \mathcal{M}_1$.  
For any $i\neq 1$, we can prove that  there is at most one $k_1$ such that $\balpha+k_1\boldsymbol{\beta}\in  \mathcal{M}_i$. 
In fact, if there are $k_1, k_2$ such that $k_1 \neq k_2$ and $\balpha + k _1{\boldsymbol{\beta}}  \in  \mathcal{M}_i, \balpha + k _2{\boldsymbol{\beta}}  \in  \mathcal{M}_i$, then 
${\boldsymbol{\beta}} \in   \mathcal{M}_i$, which is contradictory to the fact that ${\boldsymbol{\beta}} \in \mathbb{N}^p	\setminus \bigcup_{i=2}^{n+1} \mathcal{M}_i$.   Furthermore, there is no $k\in \mathbb{N}$ such that $\balpha+k{\boldsymbol{\beta}}\in \mathcal{M}_1$. If otherwise, there exists a  $k \in \mathbb{N}$ such that $\balpha + k {\boldsymbol{\beta}}  \in  \mathcal{M}_1$. Then, we have
$\balpha \in  \mathcal{M}_1$, which is also a contradiction. So we could find at most $n$ positive integers $k$  such that $\balpha + k {\boldsymbol{\beta}} \in  \bigcup_{i=1}^{n+1} \mathcal{M}_i$. Since 
there are infinitely many non-negative numbers, 
we prove that
there exists $k \in \mathbb{N}$ such that $\balpha + k {\boldsymbol{\beta}} \not\in  \bigcup_{i=1}^{n+1} \mathcal{M}_i$, and Lemma \ref{lem:subspace}  is proved. 
\subsubsection{Proof of Theorem  1,  part II}
To prove this conclusion, we need to give the explicit formula for the score functions and the Fisher information matrices of the PLN  and MPLN.
The Hessian matrix $\bH\left(\bTheta, \by\right)$  of the PLN is a $p(p+1)/2 \times p(p+1)/2$ matrix. For  notational convenience, we let $\bH_{(i,j)(i',j')} = \frac{\partial^2 \ell(\bTheta, \by)}{\partial \Theta_{i'j'}\partial \Theta_{ij}}$, $(i \leq j, i' \leq j')$ as the element at the $(2p-i+1)i/2 - p + j$ row and $(2p-i'+1)i'/2 - p + j'$  column of the Hessian matrix. 
It is clear that the densities of the PLN and MPLN satisfy the regularity conditions in \cite{shao2003mathematical}. Then, we can calculate the score function and the Fisher information as follows.
The score function  of the PLN can be   written as  
$$\mathcal{S}(\bTheta, \by) = \frac{1}{2} \vectwo\left(\bTheta^{-1}\right) - \frac{1}{2} \frac{\int \exp\left(-\frac{1}{2} (\bx-\bmu)\trans \bTheta (\bx-\bmu)\right) \vectwo\left((\bx-\bmu)(\bx-\bmu)\trans\right)h(\by,\bx)d\bx}{\int \exp\left(-\frac{1}{2} (\bx-\bmu)\trans \bTheta (\bx-\bmu)\right)h(\by,\bx)d\bx}.$$
Especially,  at the true parameter  $\bTheta^*$ , we have
$\mathcal{S}(\bTheta^*,  \by) = \frac{1}{2} \Ex_{\bx}\left(\vectwo\left(\bTheta^{*-1} - (\bx-\bmu)(\bx-\bmu)\trans \right)| \by\right)$. 
We use $(i,j), i \leq j$ to index $\mathcal{S}$. Using the operator $\mathcal{T}$,  the element of the score function at $(2p-i+1)i/2 - p + j$ can be rewritten as
$$\mathcal{S}_{(i,j)}\left(\bTheta,  \by\right) =\frac{1}{2} \vectwo\left(\bTheta^{-1}\right)_{(i,j)} - \frac{1}{2} \frac{ \mathcal{T}\left(\vectwo\left((\bx-\bmu)(\bx-\bmu)\trans\right)_{(i,j)}\right)}{\mathcal{T}\left(\mathbb{I}\right)}.$$
Using the operator $\mathcal{T}$, the Fisher information matrix can be written as follows.  Let $\Sigma = \bTheta^{-1}.$

\noindent
When $i=j, i'=j'$, 
\begin{align*}
	\bH_{(i,i)(i',i')}\left(\bTheta, \by\right) &=  -\frac{1}{2}\Sigma_{ii'}\Sigma_{i'i} + \frac{1}{4} \frac{\mathcal{T}\left({({\bx - \bmu})}_{i}^2({\bx - \bmu})_{i'}^2\right) }{ \mathcal{T}\left(\mathbb{I}\right)} \\
	& - \frac{1}{4} \frac{ \mathcal{T}\left(({\bx - \bmu})_{i'}^2\right) \mathcal{T}\left({({\bx - \bmu})}_{i}^2\right) }{ \mathcal{T}^2\left(\mathbb{I}\right)}.
\end{align*}
When $i\neq j, i'=j'$, 
\begin{align*}
	\bH_{(i,j)(i',i')}\left(\bTheta, \by\right) &=  -\Sigma_{ii'}\Sigma_{i'j} + \frac{1}{2}  \frac{ \mathcal{T}\left({({\bx - \bmu})}_{i}{({\bx - \bmu})}_{j}({\bx - \bmu})_{i'}^2\right) }{ \mathcal{T}\left(\mathbb{I}\right)} \\
	&- \frac{1}{2} \frac{ \mathcal{T}\left(({\bx - \bmu})_{i'}^2\right) \mathcal{T}\left({({\bx - \bmu})}_{i}{({\bx - \bmu})}_{j}\right) }{ \mathcal{T}^2\left(\mathbb{I}\right)}.
\end{align*}
When $i=j, i'\neq j'$, 
\begin{align*}
	\bH_{(i,i)(i',j')}\left(\bTheta, \by\right) &=  -\Sigma_{ii'}\Sigma_{j'i} + 
	\frac{1}{2}  \frac{\mathcal{T}\left({({\bx - \bmu})}_{i'}{({\bx - \bmu})}_{j'}({\bx - \bmu})_{i}^2\right) }{ \mathcal{T}\left(\mathbb{I}\right))} \\
	&- \frac{1}{2} \frac{ \mathcal{T}\left(({\bx - \bmu})_{i}^2\right) \mathcal{T}\left({({\bx - \bmu})}_{i'}{({\bx - \bmu})}_{j'}\right) }{ \mathcal{T}^2(\mathbb{I})}.
\end{align*}
When $i\neq j, i'\neq j'$, 
\begin{align*}
	\bH_{(i,j)(i',j')}\left(\bTheta, \by\right) &= -(\Sigma_{ii'}\Sigma_{j'j} + \Sigma_{ij'}\Sigma_{i'j}) + 
	\frac{ \mathcal{T}\left({({\bx - \bmu})}_{i'}{({\bx - \bmu})}_{j'}({\bx - \bmu})_{i}({\bx - \bmu})_{j}\right) }{ \mathcal{T}\left(\mathbb{I}\right)} \\
	&-  \frac{ \mathcal{T}\left(({\bx - \bmu})_{i}({\bx - \bmu})_{j}\right) \mathcal{T}\left({({\bx - \bmu})}_{i'}{({\bx - \bmu})}_{j'}\right) }{ \mathcal{T}^2\left(\mathbb{I}\right)}.
\end{align*}
\begin{lemma}\label{lem:PLNFisher}
	Assume $\by \sim {\rm PLN}(\bTheta, \bmu)$. Under Condition (C1-C3),  there exist two polynomial functions $K_1(\by),K_2(\by)$  with $\Ex[K_1(y)]<\infty$ and $\Ex[K_2(y)]<\infty$ such that   for any $i, j, i^\prime, j^\prime$, $|\mathcal{S}_{(i,j)} (\bTheta, \by )|\leq K_1(\by)$, $|\bH_{(i,i)(i',i')}(\bTheta,\by)| \leq K_2(\by)$.
	
\end{lemma}
Now we consider the score function and the Fisher information matrix of the MPLN.  The score function  of the MPLN can be written as
$$\mathcal{S}^M(\bnu, \by) = L_M(\bnu, \by)^{-1}\left({\pi_1 \frac{\partial L_1(\bnu_1, \by)}{\partial\bnu_1}}, \ldots,{\pi_G \frac{\partial L_G(\bnu_G, \by)}{\partial\bnu_G}}\right) := \left(\mathcal{S}^M_1(\bnu,\by), \cdots, \mathcal{S}^M_G(\bnu, \by)\right).$$
Similarly, we use $(g,i,j), i\leq j$ to index $\mathcal{S}^M$.
Recall that $\bnu_g= \vecone(\bTheta_{g})$.  We use $(g,i,j),(g',i',j'), i\leq j, i'\leq j'$ to index the  element at the $(g-1)p(p+1)/2 + (2p-i+1)i/2 - p + j$ row and $(g'-1)p(p+1)/2 +(2p-i'+1)i'/2 - p + j'$  column of $F(\bnu, \by)$, respectively.  

\noindent
When $g = g'$
\begin{align}
	\F_{(g,i,j,g,i',j')}(\bnu,\by) &= \frac{\partial}{{\partial\bTheta_{g,i'j'}}}\mathcal{S}^M_{(g,i,j)}(\bnu, \by) \notag \\
	&= \frac{\pi_g L_g(\bTheta_g, \by)}{L_M(\bnu, \by)} \left(  H_{(i,j)(i',j')}(\bTheta_g,\by) +  \mathcal{S}_{(i',j')}(\bTheta_g, \by	) \mathcal{S}_{(i,j)}(\bTheta_g, \by) \right)\notag\\
	&- \frac{(\pi_gL_g(\bTheta_g, \by))^2}{L_M(\bnu, \by)^2} 
	\mathcal{S}_{(i',j')}(\bTheta_g, \by	) \mathcal{S}_{(i,j)}(\bTheta_g, \by).
\end{align}
When $g \neq g'$, we have
\begin{align}
	\F_{(g,i,j,g',i',j')}(\bnu,\by) &= \frac{\partial}{{\partial\bTheta_{g',i'j'}}}\mathcal{S}^M_{(g,i,j)}(\bnu, \by)\notag \\
	&=- \frac{\pi_g\pi_g'L_g(\bTheta_g, \by)L_{g'}(\bTheta_{g'}, \by)}{L_M(\bnu, \by)^2} 
	\mathcal{S}_{(i',j')}(\bTheta_{g'}, \by	) \mathcal{S}_{(i,j)}(\bTheta_g, \by).
\end{align}
Recall the definition (\ref{equ:D}) of $\D(\bnu)$. Then, 
$$\bGamma(\bnu^*) = -\D(\bnu^*) = \Ex \left(\mathcal{S}^M(\bnu^*, \by) \mathcal{S}^M(\bnu^*, \by)\trans\right)$$
is the Fisher information matrix  of the MPLN at $\bnu^*$.
\begin{lemma}\label{lem:MPLNFisher}
	Assume $\by \sim {\rm MPLN}(\bnu, \bmu)$. Under Condition (C1-C3),  there exists a polynomial function $K(\by)$  with $\Ex[K(y)]<\infty$ such that   for any $g,i, j, g^\prime, i^\prime, j^\prime$, 	$|\F_{(g,i,j,g',i',j')}(\bnu,\by)| \leq K(\by)$.
	
\end{lemma}
In addition, we require the  following two lemmas.
\begin{lemma}\label{lem:tr}
	Let $\by$ and $\bx$ be random variables as in the PLN model (1) in the main manuscript of the paper and $\phi(y,n)$ is the same as Lemma \ref{lem:poly}.
	Let ${{\mathbf{n}}} = (n_1,\cdots, n_p)\trans$ and $\bT$ be a $p\times p$ matrix.  $\bT \in \mathbb{R}^{p\times p}$.	We have 
	\begin{align*}
		&\quad \Ex\left(\prod_{j=1}^p \phi({n_j},y_j) \tr\left(\bT\left(\bTheta^{-1} -  ({\bx} -\bmu) ({\bx} -\bmu)\trans\right)\right) \right) \\ 
		&=\left({{\mathbf{n}}}\trans \bTheta^{-1}\bT \bTheta^{-1} {{\mathbf{n}}}\right) \exp\left({{\mathbf{n}}}\trans \bmu + {{\mathbf{n}}}\trans \bTheta^{-1} {{\mathbf{n}}}/2\right).
	\end{align*}
	
\end{lemma}

\begin{lemma} \label{lem:mixsubspace}
	For any $n>0$, let $\mathcal{M}_i \subset \mathbb{R}^p, i=1,\cdots, n$ be $n$ linear proper subspaces. Let $A \neq {\bf 0}$ be a $p\times p$ symmetric matrix and $\mathcal{V} = \{\by: \by\trans A\by = 0 \}$.  
	Then, there  exists a non-negative integer vector $\bgamma$ such that $\bgamma \not\in \bigcup_{i=1}^n \mathcal{M}_i$ and $\bgamma \not\in \mathcal{V}$. 
\end{lemma}

~\\
\noindent
{\bf Proof of  Theorem  1,  part II. }Note that 
$$\bGamma(\bnu^*) = \Ex_{\by}(\mathcal{S}_M(\bnu^*, \by) \mathcal{S}_M(\bnu^*, \by)\trans).$$
If there exists a non-zero vector $\bt = (\bt_1,\cdots, \bt_G)\trans$ such that 
$\Ex(\bt \trans \mathcal{S}^M(\bnu^*, \by) \mathcal{S}^M(\bnu^*, \by)\trans \bt) = 0$, then we aim to prove that $\bt = {\bf 0}$.  
Since y is a discrete random variable,  it follows that for any $\by$, $\bt \trans \mathcal{S}^M(\bnu^*, \by) = 0$.   Then, we have 
$${L_M(\bnu, \by)} ^{-1}\sum_{g=1}^{G} \bt_g  {\pi_g \frac{\partial L_g(\bnu_g, \by)}{\partial\bnu_g}}= 0.$$ 
Since $L_M(\bnu, \by) \neq 0$, we have 
$$\sum_{g=1}^{G} \bt_g \trans{\pi_g \frac{\partial L_g(\bnu_g, \by)}{\partial\bnu_g}} = 0.$$ 
Let $\psi(\by) = \prod_{j=1}^p  \phi(z n_j,y_j), z\in \mathbb{N}^p$.  Then 
$$\sum_{g=1}^{G} \bt_g \trans{\pi_g \frac{\partial \log L_g(\bnu_g, \by)}{\partial\bnu_g}} L_g(\bnu_g, \by)\psi(\by)= 0.$$ 
Since $L_r(\bnu_r, \by)$ is proportional to the density of the PLN  with parameters $\bTheta_g, \bmu_g$, 
the above equation can be rewritten as 
$$\sum_{g=1}^{G} \bt_g \trans{\pi_g \frac{\partial \log L_g(\bnu_g, \by)}{\partial\bnu_g}} p({\by} ; \bTheta_g, \bmu_g) \psi(\by)= 0.$$ 
Summing over $\by$, we get 
$$\sum_\by \sum_{g=1}^{G} \bt_g \trans{\pi_g \frac{\partial \log L_g(\bnu_g, \by)}{\partial\bnu_g}} p({\by} ; \bTheta_g, \bmu_g) \psi(\by)= 0.$$ 
By Fubini 's Theorem, we get

\begin{equation}\label{prop2.1}
	\sum_{g=1}^{G}\sum_{\by} \bt_g \trans{\pi_g \frac{\partial \log L_g(\bnu_g, \by)}{\partial\bnu_g}} p({\by} ; \bTheta_g, \bmu_g) \psi(\by)= 0. 
\end{equation}
Then, let ${\mathbf{n}} =  (n_1,\cdots, n_p)\trans$ and  $\bT_g$ be the symmetric matrix  such that $\vecone(\bT_g) = \bt_g$. For a fixed $g$, we have 
\begin{align*}
	&\quad \sum_{{\by}} \bt_g \trans{ \frac{\partial \log L_g(\bnu_g, {\by} )}{\partial\bnu_g}} p({\by} ; \bTheta_g, \bmu_g)\psi({\by} ) = \Ex_{{\by}_g} \left(\psi(\by_g) \bt_g \trans{ \frac{\partial \log L_g(\bnu_g, \by_g )}{\partial\bnu_g}}\right) \\
	&=2^{-1} \Ex_{\bx_g, \by_g} \left(\psi(\by_g) \tr\left\lbrace \bT_g\bTheta^{*-1}_g - \bT_g (\bx_g -\bmu) (\bx_g -\bmu)\trans\right\rbrace \right),
\end{align*}
where $\by_g$ follows the PLN distribution with parameters $\bTheta_g$ and $\bmu_g$,  $\bx_g \sim {\rm N}\left(\bmu_g, \bTheta_g^{-1}\right)$ is the corresponding latent variable.
By Lemma \ref{lem:tr}, we get 
$$\Ex_{\bx_g, \by_g}\left(\psi(\by_g) \bt_g \trans{ \frac{\partial \log L_g(\bnu_g, \by_g)}{\partial\bnu_g}}\right) = z^2\left({{\mathbf{n}}}\trans \bTheta^{-1}_g\bT_g  \bTheta^{-1}_g {{\mathbf{n}}}\right) \exp\left(z{{\mathbf{n}}}\trans \bmu_g +  z^2{{\mathbf{n}}}\trans \bTheta_g^{-1} {{\mathbf{n}}}/2\right).$$
Then, (\ref{prop2.1}) can be rewritten as, for all $z \in\mathbb{N}$, 
$$ \sum_{g=1}^{G}\pi_gz^2 \left({{\mathbf{n}}}\trans \bTheta^{-1}_g\bT_g  \bTheta^{-1}_g {{\mathbf{n}}}\right) \exp\left(z{{\mathbf{n}}}\trans \bmu_g+  z^2{{\mathbf{n}}}\trans \bTheta_g^{-1} {{\mathbf{n}}}/2\right)= 0.$$
In order to show $\bT_1 = 0$,  similar to the proof of the first conclusion,  we define $\mathcal{M}_g$ as   linear space consisting of  solutions to the linear equation ${\mathbf x}\trans( \bmu_1 - \bmu_g) = 0$ ($g= 2, \cdots, G$) and $\mathcal{M} = \cup_{g=2}^G \mathcal{M}_g$. For any ${\mathbf n} \not\in \mathcal{M} $, then $({\mathbf n}\trans \bmu_1, \ldots, {\mathbf n}\trans \bmu_G)$ is a good vector with a good index $1$.  Since $\pi_1 > 0$,  we must have ${{\mathbf{n}}}\trans \bTheta^{-1}_1\bT_1  \bTheta^{-1}_1 {{\mathbf{n}}} = 0$.  By Lemma \ref{lem:mixsubspace},  if $\bTheta^{-1}_1 \bT_1  \bTheta^{-1}_1$ is not a zero matrix, then there exists an ${{\mathbf{n}}}$ such that ${\mathbf n} \not\in \mathcal{M} $ and ${{\mathbf{n}}}\trans \bTheta^{-1}_1\bT_1  \bTheta^{-1}_1 {{\mathbf{n}}} \neq 0$, which is contradictory to the fact that ${\mathbf n} \not\in \mathcal{M} $ implies  ${{\mathbf{n}}}\trans \bTheta^{-1}_1\bT_1  \bTheta^{-1}_1 {{\mathbf{n}}} = 0$. Hence, we must have $\bTheta^{-1}_1 \bT_1  \bTheta^{-1}_1 = {\bf{0}}$ and thus $\bT_1 ={\bf{0}}$.  Similarly, we get $\bT_g ={\bf{0}}$ for all $g = 1,\ldots, G$.  It follows that $\bt  = {\bf 0}$, and we compete the proof. 

~\\
\noindent
{\bf{Proof of Lemma \ref{lem:PLNFisher}.  }}   We only prove that there is a dominating function for $\mathcal{S}_{(i,j)}(\bTheta,\by)$. Others can be proved similarly.  Since  $m \leq \lambda_{min}(\bTheta)\leq\lambda_{max}(\bTheta)\leq M$,  we have   $\lambda_{max}(\bSigma)\leq 1/m$ and thus 
$\vectwo(\bTheta^{-1})_{(i,j)}$ is bounded by a constant. 
Further, since
$\vectwo((\bx-\bmu)(\bx-\bmu)\trans)_{(i,j)}$ is a polynomial function of $\bx$,   by  Remark \ref{pdimremark}, we have 
$$\frac{1}{2} \frac{ \mathcal{T}\left(\vectwo\left((\bx-\bmu)(\bx-\bmu)\trans\right)_{(i,j)}\right)}{\mathcal{T}\left(\mathbb{I}\right)}$$
can be bounded by an integrable  polynomial function and we prove the  existence of $K_1(\by)$. 

~\\
\noindent
{\bf{Proof of Lemma \ref{lem:MPLNFisher}. }}If $g = g'$, since $ \pi_gL_g(\bnu, \by) / L_M(\bnu, \by)  \leq 1 $,  then we have, 	
$$
|\F_{(g,i,j,g,i',j')}(\bnu, \by)|  \leq    |\bH_{(i,j,i'j')}(\bTheta_g, \by )|  + 2|\mathcal{S}_{(i,j)}(\bTheta_g, \by	)\mathcal{S}_{(i',j')}(\bTheta_g, \by	)|.
$$
The, by Lemma \ref{lem:PLNFisher}, we know that, there exists a function $K(\by)$ such that 
$|\F_{(g,i,j,g,i',j')}(\bnu, \by)| \leq K( \by)$ and $\Ex(K(\by)) < \infty$. 
The same proof  can be applied to the $g \neq g'$ case.

~\\
\noindent
{\bf  Proof of Lemma \ref{lem:tr}. }By Lemma \ref{lem:poly}
$$
\Ex\left(\prod_{j=1}^p \phi({n_j},y_j) \tr\left(\bT\bTheta^{-1}\right)\right) = \tr\left(\bT\bTheta^{-1}\right)\exp\left({{\mathbf{n}}}\trans \bmu + {{\mathbf{n}}}\trans \bTheta^{-1} {{\mathbf{n}}}/2\right). 
$$
Similar to the proof of Lemma S1, by  the moment generating function of the normal distribution, we have
\begin{align*}
	&\quad \Ex\left(\prod_{j=1}^p \phi({n_j},y_j) \tr\left(\bT({\bx} -\bmu) ({\bx} -\bmu)\trans\right)\right) \\
	& = \Ex_\bx\left( \exp({\mathbf{n}}\trans \bx)\tr\left(\bT({\bx} -\bmu) ({\bx} -\bmu)\trans\right)\right) \\ 
	&= \left\lbrace \tr\left(\bT\bTheta^{-1}\right) + {{\mathbf{n}}}\trans \bTheta^{-1}\bT \bTheta^{-1} {{\mathbf{n}}}\right\rbrace \exp\left({{\mathbf{n}}}\trans \bmu + {{\mathbf{n}}}\trans \bTheta^{-1} {{\mathbf{n}}}/2\right).
\end{align*} 
Lemma \ref{lem:tr} follows from the above two equations. 

~\\
\noindent
{\bf  Proof of Lemma \ref{lem:mixsubspace}. }By the proof of Lemma \ref{lem:subspace}, there exists $\balpha\not\in  \bigcup_{i=1}^n \mathcal{M}_i$.  We can assume $\balpha \in \mathcal{V}$. Otherwise, we complete the proof.
Since $A$ is not a zero matrix, we can take $ {\boldsymbol{\beta}} \not\in \mathcal{V} $.
On the one hand, since ${\boldsymbol{\beta}} \not\in \mathcal{V} $,  there are at most two  integers $k \in \mathbb{N}$ satisfying  the quadratic equation $(\balpha + k{\boldsymbol{\beta}} )\trans A (\balpha + k{\boldsymbol{\beta}} ) = 0$.  
On the other hand,   for any $\mathcal{M}_i, i = 1 ,\ldots, n$, there is at most one  integer $k$ such that $\balpha + k{\boldsymbol{\beta}} \in \mathcal{M}_i$.  Otherwise, 
if there exist $k_1\neq k_2$ and $i$ satisfying
$$\balpha + k_1{\boldsymbol{\beta}} \in \mathcal{M}_i , \balpha + k_2{\boldsymbol{\beta}}\in \mathcal{M}_i ,$$ 
then we have $(k_2 - k_1){\boldsymbol{\beta}} \in \mathcal{M}_i $.  It follows that $\balpha \in \mathcal{M}_i$, which is contradictory to the  fact that $\balpha \not\in \bigcup_{i=1}^n \mathcal{M}_i$.  Hence, there are at most $(n+2)$ $k$ such that $\balpha + k{\boldsymbol{\beta}} \in  \bigcup_{i=1}^n \mathcal{M}_i  \bigcup  \mathcal{V} $. Since there are infinitely many non-negative  integers,  there exists an integer $k$ such that  $\balpha + k{\boldsymbol{\beta}} \not \in  \bigcup_{i=1}^n \mathcal{M}_i  \bigcup  \mathcal{V} $. 

\subsubsection{Proof of Theorem 2}
Since we only prove the positive definiteness of the Fisher information, we can only get the local strong convexity.   In order  to prove the convergence rate,   we first give the consistency of MLE.  The proof is based on the M-estimator theory.
For convenience, we introduce some notations in the M-estimator theory. We let $m_{\bnu}(\by_i)=\ell(\bnu, \by_i)$, $M_n(\bnu)=n^{-1}\sum_{i=1}^{n} m_{\bnu}(\by_i) = -\mathcal{L}_n(\bnu)$ and $M(\bnu)=\Ex_{\bnu^*} [m_{\bnu}(\by_i)]$.  Let $\mathcal{D}_{0}=\{\bnu_0 \in {\mathcal{D}} \ | \  \Ex [m_{\bnu_0}]=\sup_{\bnu} \Ex[m_{\bnu}]\}$. By Jessen's inequality and the identifiablity of the MPLN, we get that $\mathcal{D}_{0}$ only contains one element $\bnu^*$.  Then we need the following two conditions and Lemma  \ref{lem:wald}. The proof of Lemma \ref{lem:wald} can be found in \cite{van2000asymptotic}.
\begin{itemize}
	\item[\bf (SC1)] 
	$\limsup_{\bnu_n\rightarrow\bnu}m_{\bnu_n}(\by)\leq m_{\bnu}(\by)\ \mbox{ for all } \bnu  \mbox{ and a.s. }\by$. 
	\item[\bf (SC2)]  For all sufficiently small ball $U\subset {\mathcal{D}}$, $\by\mapsto \sup_{\bnu\in U}m_{\bnu}(\by)$ is measurable and satisfies $\Ex(\sup_{\bnu\in U}m_{\bnu}(\by))<\infty.$
\end{itemize}

\begin{lemma}[Wald's consistency]\label{lem:wald}
	Assume that Condition (SC1-SC2) hold for $m_{\bnu}(\by)$. Suppose that $ \hat{\bnu}_n$ is any sequence of random vectors such that  $M_n( \hat{\bnu}_n)\geq M_n(\bnu_0)-o_p(1)$  for some $\bnu_0\in \mathcal{D}_{0}$. Then for any $ \epsilon>0$, and every compact set $K\subset {\mathcal{D}}$,  as $n\rightarrow \infty $, we have
	$$\pr(d(\hat{\bnu}_n,\mathcal{D}_0)\geq\epsilon\wedge\hat{\bnu}_n\in K)\rightarrow 0,$$
	where $d(\hat{\bnu}_n,\mathcal{D}_0) = \inf_{\bnu_0\in \mathcal{D}_{0}} ||\hat{\bnu}_n - \bnu_0 ||_2$.
\end{lemma}

\begin{lemma} \label{lem:consis}
	Assume that the estimator $\hat{\bnu}_n$ minimizes  (\ref{prob:opt})  in parameter space ${\mathcal{D}}$ and $\lambda_n$ goes to zero.  Then, for any $\epsilon>0$, as $n\rightarrow \infty$,  we have:
	$$\pr(||\hat{\bnu}_n-\bnu^*||_2 \geq\epsilon)\rightarrow 0, \mbox{ as } n \rightarrow \infty. $$
	
\end{lemma}

\begin{lemma} [Uniform law of large numbers] \label{lem:ULLN}
	For all $g,i,j,g',i',j'$ we have 
	$$\pr\left\lbrace \lim_{n \rightarrow \infty} \sup_{\bnu \in {\mathcal{D}}}\left| \frac{1}{n} \sum_{k=1}^n\F_{(g,i,j)(g',i',j')}(\bnu, \by_k) - \D_{(g,i,j)(g',i',j')}(\bnu)\right|  = 0\right\rbrace  = 1,$$
	where $\D(\bnu) = \Ex_{\bnu^*}(\F(\bnu, \by)).$
	Furthermore,  we have 
	$$\pr\left\lbrace \lim_{n \rightarrow \infty} \sup_{\bnu \in {\mathcal{D}}}\left\|  \frac{1}{n} \sum_{k=1}^n\F(\bnu,  \by_k) - \D(\bnu)\right\| _2 = 0\right\rbrace  = 1.$$
	
\end{lemma}

\begin{remark} \label{cont}
	By the proof of Lemma \ref{lem:ULLN}, we know that there exists a function $F_0(\by)$ such that for any $\bnu$,
	$|\F_{(g,i,j,g',i',j')}(\nu, \by)| \leq F_0(\by)$  and $\Ex(F_0(\by)) < \infty$ .  Then, by the  dominated convergence theorem, $\D(\bnu) = \Ex_{\bnu^*}(\F(\bnu, \by))$ is continuous. 
\end{remark}

Let $\Delta_n = \hat{\bnu}_{n} - \bnu^{*}$ .  We define 
$$\delta \mathcal{L}_n = \mathcal{L}_n(\bnu^{*} + \Delta_n) -\mathcal{L}_n(\bnu^{*}) - <\nabla\mathcal{L}_n(\bnu^{*}), {\Delta_n}>,$$ 
where $<\nabla\mathcal{L}_n(\bnu^{*}), {\Delta_n}> = \nabla\mathcal{L}_n(\bnu^{*})\trans {\Delta_n}$.  According to Lemma \ref{lem:ULLN}, we can prove the following lemma.
\begin{lemma}\label{lem:delta}
	Under Condition (C1-C3), with high probability,  we have
	$$\delta \mathcal{L}_n \geq \frac{\kappa}{3} ||{\Delta_n}||_2^2,$$
	where $\kappa =  \lambda_{min}(\bGamma^*)$.
	
\end{lemma}

~\\
\noindent
{\bf{Proof of Theorem 2.}} 
Define
$$\mathcal{F}(\Delta) = \mathcal{L}_n(\bnu^{*} + \Delta) -\mathcal{L}_n(\bnu^{*}) + \lambda_n( \mathcal{R}(\bnu^* + \Delta) - \mathcal{R}(\bnu^*)).$$ 
Since $\mathcal{F}(0) = 0$ and $\hat{\bnu}_n$ minimizes (\ref{prob:opt}), 
we must have $\mathcal{F}(\Delta_n) \leq 0.$
Further, by Lemma \ref{lem:delta}, with high probability, 	$\delta \mathcal{L}_n \geq \frac{\kappa}{3} ||\Delta_n||_2^2$. Then, combining with Cauchy's inequality and the convexity of the lasso penalty, with high probability, we get
\begin{align*}
	\mathcal{F}(\Delta_n) 
	&= \delta \mathcal{L}_n + <\nabla\mathcal{L}_n(\bnu^{*}), \Delta_n>+ {\lambda_n}( \mathcal{R}(\bnu^* + \Delta_n) - \mathcal{R}(\bnu^*)) \\
	&\geq \frac{\kappa}{3} ||\Delta_n||_2^2 - (||\nabla\mathcal{L}_n(\bnu^{*})||_{\infty}  + {2\lambda_n})||\Delta_n||_1 \\
	&\geq \frac{\kappa}{3} ||\Delta_n||_2^2 -\sqrt{Gp(p+1)/2} (||\nabla\mathcal{L}_n(\bnu^{*})||_{\infty}  + {2\lambda_n})||\Delta_n||_2 .
\end{align*}
When 
$$||\Delta_n||_2  >  \frac{3}{\kappa} \sqrt{Gp(p+1)/2} (||\nabla\mathcal{L}_n(\bnu^{*})||_{\infty} + {2\lambda_n}),$$
we have $\mathcal{F}(\Delta_n) > 0$. By the fact that $\mathcal{F}(\Delta_n) \leq 0$.  we obtain
$$||\Delta_n||_2  \leq   \frac{3}{\kappa} \sqrt{Gp(p+1)/2} (||\nabla\mathcal{L}_n(\bnu^{*})||_{\infty} + {2\lambda_n}).$$

~\\
\noindent
{\bf{Proof of Lemma \ref{lem:consis}}. }We use  Lemma \ref{lem:wald} to prove this lemma.  We first check Condition (SC1-SC2).  Note that  $m_{\bnu}(\by)= \ell(\bnu, \by) = \log(L_M(\bnu, \by)) + C(\by) $ where $C(\by) = \sum_{j=1}^p \log y_j! - 2^{-1}p \log(2\pi)$ and $L_M(\bnu, \by)$ is continuous at all $  \bnu \in {\mathcal{D}}$ for any fixed $\by$. Condition (SC1) thus follows.

For Condition (SC2), we first check the measurability of $\sup_{\bnu\in U}m_{\bnu}(\by)$ for any small ball $U$. 
Let $Q_U=\{\bnu |\bnu\ \mbox{ is a rational point }, \bnu \in U\}$. Then $Q_U$ has a countable  number of elements.  From the measurability of $m_{\bnu}(\by)$, we get that $\sup_{\bnu\in Q_U}m_{\bnu}(\by)$ is measurable. 
On the other hand, from the continuity of $m_{\bnu}(\by)$ in $ \bnu$, we  get
$$\sup_{\bnu\in Q_U}m_{\bnu}(\by)=\sup_{\bnu\in U}m_{\bnu}(\by),$$
and hence $\sup_{\bnu\in U}m_{\bnu}(\by)$ is measurable.
Finally, we  prove $\Ex [\sup_{\bnu\in U}m_{\bnu}(\by)]<\infty$. We have
$$\begin{array}{ll}
	&\quad \log(L_M(\bnu, \by)) \\
	&= \log\left(\sum_{g=1}^{G}\pi_g \int {\exp\left( \sum_{j=1}^{p}\left( x_{j}y_{j}-\exp\left( x_{j}\right)\right) -\frac{1}{2}(\bx-\bmu_g)^T\bTheta_g(\bx-\bmu_g) \right)}\det(\bTheta_g)^{1/2}d\bx\right)\\
	&\leq \log\left(\sum_{g=1}^{G}\pi_g \int {\exp\left( \bx^T\by -\frac{1}{2}(\bx-\bmu_g)^T\bTheta_g(\bx-\bmu_g)\right)}\det(\bTheta_g)^{1/2}d\bx\right)\\
	&=\log\left( \sum_{g=1}^{G}\pi_g \exp\left( \frac{1}{2}\by^T\bTheta_g^{-1}\by+\by^T\mu_g\right) \right) + 2^{-1}p \log(2\pi) . 
\end{array}$$
and $\sup_{\bnu\in {\mathcal{D}}}\frac{1}{2}\by^T\bTheta_g^{-1}\by+\by^T\bmu_g\leq\|\by\|_2^2/(2M)+M\|\by\|_1$. Then, we have 
$$\sup_{\bnu\in {\mathcal{D}}}\log(L_M(\bnu, \by))\leq \|\by\|^2_2/(2M)+M\|\by\|_1 + 2^{-1}p \log(2\pi).$$
Also,  we have 
$$C(\by) = \sum_{j=1}^p \log y_j!  - 2^{-1}p \log(2\pi) \leq  \sum_{j=1}^p  y_j \log(y_j + 1)- 2^{-1}p \log(2\pi). $$
Since any polynomial of a MPLN random variable $\by$ is integrable, we prove  $\Ex(\sup_{\bnu\in U}m_{\bnu}(\by))<\infty.$
In addition, we have
$$\begin{array}{ll}
	M_n(\hat{\bnu}_n)&\geq \lambda_n \sum_{g=1}^{G}\|\hat{\bTheta}_g\|_{1,\rm off}+ M_n(\bnu^*)-\lambda_n \sum_{g=1}^{G}\|\bTheta_{g}^*\|_{1,\rm off}\\
	&\geq M_n(\bnu^*)-\lambda_n \sum_{g=1}^{G}\|\bTheta_{g}^*\|_{1,\rm off}\\
	&\geq M_n(\bnu^*)-C\lambda_n\\
	&=M_n(\bnu^*)-o(1).
\end{array}$$
Thus, all conditions in Lemma \ref{lem:wald} are satisfied. 
Finally, note that $\mathcal{D}_{0}$ only contains one element $\bnu^*$. Taking $K = {\mathcal{D}} $,  we get, for any $\epsilon>0$,  
$$\pr(||\hat{\bnu}_n-\bnu^*||_2 \geq\epsilon)\rightarrow 0, \mbox{ as } n \rightarrow \infty. $$

~\\
\noindent
{\bf Proof of Lemma \ref{lem:ULLN}. }By Theorem  16(a) in \cite{ferguson2017course},  for any $g,i,j,g',i',j'$ 
we only need to verify  there exists a function $F_0(\by)$ such that 
$|\F_{(g,i,j)(g',i',j')}(\bnu, \by)| \leq F_0(\by)$ and $\Ex_{\bnu^*}(F_0(\by)) < \infty$.   The existence of such function is guaranteed by Lemma \ref{lem:MPLNFisher} .

~\\
\noindent
{\bf{Proof of Lemma \ref{lem:delta}. }}By the definition of $\mathcal{L}_n$ and Taylor expansion, we have 
\begin{align*}
	\delta \mathcal{L}_n &= \mathcal{L}_n(\bnu^{*} + \Delta_n) -\mathcal{L}_n(\bnu^{*}) - <\nabla\mathcal{L}_n(\bnu^{*}),\Delta_n> \\
	&= -\frac{1}{n} \sum_{i=1}^n\Delta_n \trans \F(\check{\bnu}, \by_i)\Delta_n \\
	&=\Delta_n \trans (\D(\check{\bnu})- \frac{1}{n} \sum_{i=1}^n F(\check{\bnu}, \by_i)){\Delta_n} + 
	{\Delta_n} \trans (\D(\bnu^*)-\D(\check{\bnu})){\Delta_n}
	+ \Delta_n \trans (-\D(\bnu^*)){\Delta_n}
\end{align*} 
where $\check{\bnu} = \bnu^{*} + \theta \Delta_n, 0 \leq\theta \leq 1.$
By Lemma \ref{lem:ULLN}, we get, for any $\epsilon$,  there exists $N$ such that when $n > N$, with probability $1-\epsilon$, 
\begin{equation}\label{equ:delta1}
	\Delta_n \trans (\D(\check{\bnu})- \frac{1}{n} \sum_{i=1}^n \F(\check{\bnu}, \by_i))\Delta_n \geq -\frac{\kappa}{3} ||\Delta_n||_2^2. 
\end{equation}
Also, by the continuity  of $\D(\bnu)$ (Remark \ref{cont}) and Wielandt-Hoffman Theorem \citep{bhatia2013matrix}, there exists a constant $\tau$ such that when $||\Delta_n||_2 < \tau$,  
we have $$  \lambda_{min} (\D(\bnu^*)-\D(\check{\bnu})) \geq -\frac{\kappa}{3}. $$ It follows that 
\begin{equation}\label{equ:delta2}
	\Delta_n \trans (\D(\bnu^*)-\D(\check{\bnu}))\Delta_n \geq -\frac{\kappa}{3} ||\Delta_n||_2^2. 
\end{equation}
Finally, by $\kappa = \lambda_{min}(\bGamma^*)$, we have 
\begin{equation}\label{equ:delta3}
	\Delta_n \trans (-\D(\bnu^*))\Delta_n \geq \kappa ||\Delta_n||_2^2.
\end{equation}
Combining the above inequalities (\ref{equ:delta1}-\ref{equ:delta3}), we have  when  $||\Delta_n||_2 < \tau$, with high probability, 
$$\delta \mathcal{L}_n \geq \frac{\kappa}{3} ||\Delta_n||_2^2.$$ 
Finally, since $\tau$ is a constant, by the consistency of the MPLN, we have 
$\pr(||{\Delta}_n|||_2 < \tau) \rightarrow 1 \mbox{ as } n \rightarrow \infty$. Then, we get 
$\delta \mathcal{L}_n \geq \frac{\kappa}{3} ||{\Delta}_n|||_2^2.$
Thus we complete the proof.

\subsubsection{Proof of Theorem 3}
In this subsection,  we simplify the notation and use $S$ and $S^c$ to denote $S(\bnu^*)$ and $S^c(\bnu^*)$, respectively. Define
$\hat{\bGamma}^*_i = -\F(\bnu^*, \by_i) $
and $\hat{\bGamma}^* = \frac{1}{n} \sum_{i=1}^n \hat{\bGamma}^*_i$, which is an estimator of $\bGamma^*$. 
We write $\alpha = 1-||\bGamma^*_{S^cS}{(\bGamma^*_{SS})}^{-1}||_{1,\infty}$. 	By Condition (C4), $\alpha > 0$.
\begin{lemma} \label{samplegamma}
	For any $\epsilon > 0$, with high probability, $\hat{\bGamma}_{SS}^*$ is invertible and
	$||\hat{\bGamma}^*_{S^cS}{(\hat{\bGamma}^*_{SS})}^{-1}||_{1,\infty} \leq 1-\alpha/2.$
\end{lemma}

\begin{lemma}\label{lem:KKT}
	For any $\lambda_n > 0$, with high probability, the solution $\hat{\nu}_n$ to the optimization problem (\ref{prob:opt}) is characterized by
	$$-n^{-1}\sum_{i=1}^n\mathcal{S}^M(\hat{\bnu}_n, \by_i)  + \lambda_n {\hat{\bK}} = 0,$$
	where ${\hat{\bK}}$ is the subdifferential of $\mathcal{R}(\bnu)$ at $\hat{\bnu}_n$.
\end{lemma}

Next, we construct the primal-dual witness solution $(\widetilde{\bnu}_n, {\widetilde{\bK}} )$. Let $\widetilde{\bnu}_n$ be  the solution to the restricted  optimization problem
$$\widetilde{\bnu}_n \in \arg \min_{\bnu_{S^c} = 0,  \bnu \in {\mathcal{D}}} \mathcal{L}_n(\bnu) + {\lambda_n} \mathcal{R}(\bnu).$$
Denote 
\begin{equation} \label{eq:tildeK}
	\widetilde{\bK} = \lambda_n^{-1} \left\lbrace n^{-1}\sum_{i=1}^n\mathcal{S}^M(\widetilde{\bnu}_n, \by_i)  \right\rbrace. 
\end{equation} 

Then, we have
\begin{equation} \label{equ:res}
	-n^{-1}\sum_{i=1}^n\mathcal{S}^M(\widetilde{\bnu}_n, \by_i)  + \lambda_n {\widetilde{\bK}} = 0.
\end{equation}
Define
$\widetilde{\Delta} _n= \bnu^* - \widetilde{\bnu}_n$. 
We rewrite (\ref{equ:res}) as  
$$-n^{-1}\sum_{i=1}^n\mathcal{S}^M(\widetilde{\bnu}_n, \by_i)  + n^{-1}\sum_{i=1}^n\mathcal{S}^M(\bnu^*, \by_i)-n^{-1}\sum_{i=1}^n\mathcal{S}^M(\bnu^*, \by_i) + \hat{\bGamma}^*\widetilde{\Delta} _n- \hat{\bGamma}^*\widetilde{\Delta} _n+ \lambda_n {\widetilde{\bK}}  = 0.$$
Let 
$${\R} = -n^{-1}\sum_{i=1}^n\mathcal{S}^M(\widetilde{\bnu}_n, \by_i)  + n^{-1}\sum_{i=1}^n\mathcal{S}^M(\bnu^*, \by_i) + \hat{\bGamma}^*\widetilde{\Delta}_n,$$
and ${\bW} = -n^{-1}\sum_{i=1}^n\mathcal{S}^M(\bnu^*, \by_i)$.
Then, we have 
\begin{equation} \label{KKT}
	{\R} - \hat{\bGamma}^*\widetilde{\Delta} _n+ {\bW} + \lambda_n {\widetilde{\bK}} = 0.
\end{equation}
Since we  restrict  the solution to the set of true support, similarly to Theorem 1, we can proceed analogously to the proof of 
\begin{equation} \label{eq:dual}
	\pr\left[||\widetilde{\nu}_n-\nu^*||_2 \leq  \frac{3}{\kappa} \sqrt\frac{G{p(p+1)}}{{2}}\left(n^{-1} ||\nabla{\ell}_n(\bnu^*)||_{\infty} + 2{\lambda_n}{}\right)\right] \rightarrow 1, \mbox{ as } n \rightarrow \infty.
\end{equation}
Considering the local convex property of loss function at $\bnu^*$ in Lemma \ref{lem:delta}, if we verify that with high probability, the strict dual feasibility condition
$||\widetilde{\bK}||_{\infty} \leq 2$ holds,  then we prove that with high probability $\widetilde{\bnu}_n $ is equal to $\hat{\bnu}_n$. Then, the model can recover all zeros. Similarly to Lemma \ref{lem:KKT}, by the fact that $\widetilde{\bnu}_n$ is the restricted construction and the definition of $\widetilde{\bK}$ in (\ref{eq:tildeK}), we have 
\begin{equation}\label{eq:tildeKS}
	||\widetilde{\bK}_{S}||_{\infty} \leq 2 .
\end{equation} 
Therefore, we only need to show that $||\widetilde{\bK}_{S^c}||_{\infty} \leq 2 $.  
Note that the infinity norm of $\widetilde{\bK}_{S^c}$ is less than 2 instead of 1. The reason is  $\mathcal{R}(\bnu) =  \sum_{g=1}^G ||\bTheta_g||_{1,\rm off} = 2\sum_{g=1}^G \sum_{i<j} |\bTheta_{g,ij}|$. 
Hence, we aim to  verify  the strict dual feasibility. 

\begin{lemma}[Strict dual feasibility] \label{dualf}
	Under Condition (C1-C4),  suppose 
	that $\hat{\bGamma}_{SS}^*$ is invertible and
	$$||{\bW}||_{\infty} + ||{\R}||_{\infty} < \frac{\alpha \lambda_n}{4},\, \left\| \hat{\bGamma}^*_{S^cS}{\left(\hat{\bGamma}^*_{SS}\right)}^{-1}\right\|_{1,\infty} \leq 1-\alpha/2.$$
	Then, the matrix $\widetilde{\bK}$ satisfies 
	$$||\widetilde{\bK}_{S^c}||_{\infty} \leq 2.$$
\end{lemma}

\begin{lemma} \label{lem:remainder}
	$\frac{||{\R}||_{\infty}}{||\widetilde{\Delta}_n||_2}\rightarrow 0$, in probability as $n \rightarrow  \infty$.
\end{lemma}

Applying  Chebyshev's inequality, the following lemma is clear. 
\begin{lemma} \label{noiseterm}
	Let $a_n$ be any sequence such that ${a_n}{\sqrt{n}} \rightarrow \infty, a_n > 0$ and $a_n \rightarrow 0$. Then, we have 
	$$\pr(||{\bW}||_{\infty} \leq a_n) \rightarrow 1, \mbox{ as } n \rightarrow \infty.$$
\end{lemma}

\noindent
{\bf Proof of Theorem 3. }Note that ${\bW} =-n^{-1} \nabla \ell_n(\bnu^{*})$ and $\widetilde{\Delta} _n= \bnu^* - \widetilde{\bnu}_n$. 
Applying Lemma \ref{lem:remainder}, we have 
$  ||{\R}||_{\infty} = o_p(1) ||\widetilde{\Delta}_n||_2. $ 
By (\ref{eq:dual}), we have 
$$||{\R}||_{\infty} = o_p(1)  \left( \frac{3}{\kappa} \sqrt\frac{G{p(p+1)}}{{2}} \left(||{\bW}||_{\infty} +2 {\lambda_n}\right)\right) .$$
It follows that 
$$||{\bW}||_{\infty} + ||{\R}||_{\infty} =||{\bW}||_{\infty} +  o_p(1)\left( \frac{3}{\kappa} \sqrt\frac{G{p(p+1)}}{{2}} \left(||{\bW}||_{\infty} +2 {\lambda_n}\right)\right) = (1+o_p(1))||{\bW}||_{\infty} + o_p(1)\lambda_n.$$
In order to show   $||{\bW}||_{\infty} + ||{\R}||_{\infty} < \frac{\alpha \lambda_n}{4},$ we only need to show that with high probability, 
$$||{\bW}||_{\infty} \leq (1+o_p(1))^{-1}\left(\frac{\alpha}{4}-o_p(1)\right)\lambda_n.$$
By the  choice of $\lambda_n$ such that $\lambda_n \rightarrow 0$ and ${\sqrt{n}\lambda_n} \rightarrow \infty$,
applying Lemma \ref{noiseterm}, we have 
$\pr (||{\bW}||_{\infty} \leq \left(1+o_p(1))^{-1}\left(\frac{\alpha}{4}-o_p(1)\right)\lambda_n\right) \rightarrow 1.$  
By Lemma  \ref{samplegamma}, we have with high probability 	$\hat{\bGamma}_{SS}^*$ is invertible and 
$ \left\|\hat{\bGamma}^*_{S^cS}{\left(\hat{\bGamma}^*_{SS}\right)}^{-1} \right\|_{1,\infty} \leq 1-\alpha/2.$
Then, by Lemma \ref{dualf}, with high probability, we have  the strict dual feasibility condition holds. Hence, the witness solution $\widetilde{\bnu}_n$ is equal to the original solution $\hat{\bnu}_n$. 
Since $\widetilde{\bnu}_n$ is the restricted solution,  with high probability, $\widetilde{\bnu}_n$ can recover all zeros. It follows that $\hat{\bnu}_n$ can recover all zeros.  Finally, since $ ||\hat{\bnu}_n-\bnu^*||_2  \rightarrow 0 $ in probability, with high probability, $\hat{\bnu}_n$ can recover all non zeros.

~\\
\noindent
{\bf Proof of Lemma \ref{lem:KKT}.  }Since $\bnu^*$  is an interior point of $\mathcal{D}$ ,  we only need to prove with high probability, the maximum value will not be taken at $\partial{\mathcal{D}}$.  By Jessen's inequality and the identifiablity of the MPLN, we have  
$\sup_{\bnu \in \partial{\mathcal{D}}} 
\Ex_{\bnu^*}\left[\log\frac{p\left(\by_i;\bnu, \{\bmu_g\}_{g=1}^G\right)}{p\left(\by_i;\bnu^*, \{\bmu_g\}_{g=1}^G\right)}\right] < 0$. By uniform law of large numbers, we have  with high probability,
$$\sup_{\bnu \in \partial{\mathcal{D}}}\left\lbrace n^{-1}\sum_{i=1}^n \log \, p\left(\by_i;\bnu, \{\bmu_g\}_{g=1}^G\right) - n^{-1}\sum_{i=1}^n \log \, p\left(\by_i;\bnu^*, \{\bmu_g\}_{g=1}^G\right)\right\rbrace  < 0.$$
Thus we complete the proof.  

~\\
\noindent
{\bf Proof of Lemma \ref{dualf}. }We split $\widetilde{\Delta}_n$ into $\widetilde{\Delta}_{nS} $ and $\widetilde{\Delta}_{nS^c} $. 
By $\widetilde{\Delta}_{nS^c} = 0$,  (\ref{KKT})  can be rewritten as two blocks of linear equations 
\begin{equation}\label{part2}
	{\R}_S - \hat{\bGamma}_{SS}^*\widetilde{\Delta}_{nS} + {\bW}_S + \lambda_n {\widetilde{\bK}}_{S} = 0,
\end{equation}
\begin{equation}\label{part}
	{\R}_{S^c} - \hat{\bGamma}_{S^cS}^*\widetilde{\Delta}_{nS} + {\bW}_{S^c} + \lambda_n {\widetilde{\bK}}_{S^c} = 0.
\end{equation}
From (\ref{part2}), we have
$$\widetilde{\Delta}_{nS} ={\left(\hat{\bGamma}^*_{SS}\right)}^{-1}\left({\R}_S + {\bW}_S + \lambda_n {\widetilde{\bK}}_S\right).$$
Substituting this expression into (\ref{part}), with high probability, we have
$${\widetilde{\bK}}_{S^c} = \lambda_n^{-1}	\left\lbrace -{\R}_{S^c} + \hat{\bGamma}_{S^cS}^*{\left(\hat{\bGamma}^*_{SS}\right)}^{-1}\left({\R}_S + {\bW}_S + \lambda_n {\widetilde{\bK}}_S\right) - {\bW}_{S^c} \right\rbrace .$$
Let $\bA$ be a matrix and $\mathbf{a}$ be a vector.
By the fact that $||\bA \mathbf{a}||_{\infty} \leq ||\bA||_{1,\infty} ||\mathbf{a}||_{\infty}$, we have
\begin{align*}
	||{\widetilde{\bK}}_{S^c}||_{\infty} &\leq {\lambda_n}^{-1}\left(1 + \bigg|\bigg|\hat{\bGamma}_{S^cS}^* {\left(\hat{\bGamma}^*_{SS}\right)}^{-1}\bigg|\bigg|_{1,\infty}\right)\left(||{\bW}||_{\infty} + ||{\R}||_{\infty}\right)  + \bigg|\bigg|\hat{\bGamma}_{S^cS}^* {\left(\hat{\bGamma}^*_{SS}\right)}^{-1}\bigg|\bigg|_{1,\infty} ||{\widetilde{\bK}}_{S}||_{\infty} \\
	&\leq  2  (1-\alpha/2) + \frac{(2-\alpha/2)}{\lambda_n}(||{\bW}||_{\infty} + ||{\R}||_{\infty}) \\
	&\leq 2.
\end{align*}
Here we use the inequality that $||{\widetilde{\bK}}_{S}||_{\infty} \leq 2$ in (\ref{eq:tildeKS}). Thus we complete the proof.

~\\
\noindent 
{\bf Proof of Lemma \ref{lem:remainder}. }By the mean value theorem, we have
\begin{align*}
	{\R} &= -n^{-1}\sum_{i=1}^n\mathcal{S}^M(\widetilde{\bnu}_n, \by_i)  + n^{-1}\sum_{i=1}^n\mathcal{S}^M(\bnu^*, \by_i) + \hat{\bGamma}^*\widetilde{\Delta} _n\\
	&= \left(\hat{\bGamma}^* + \frac{1}{n} \sum_{i=1}^n \F(\check{\bnu}, \by_i) \right)\widetilde{\Delta} _n,
\end{align*}
where $\check{\bnu} = \bnu^{*} + \theta (-\widetilde{\Delta} _n), 0 \leq\theta \leq 1.$
Since $\frac{{||\R||_{\infty}}}{||\widetilde{\Delta} _n||_2} \leq \frac{||{\R}||_2}{||\widetilde{\Delta} _n||_2}$,
we have 
$$\frac{{||\R||_{\infty}}}{||\widetilde{\Delta} _n||_2} \leq \left\|\hat{\bGamma}^*+ \frac{1}{n} \sum_{i=1}^n \F(\check{\bnu}, \by_i)  \right\|_2.$$
By the triangle inequality, we have	
\begin{align*}
	&\quad \bigg|\bigg|\hat{\bGamma}^*+ \frac{1}{n} \sum_{i=1}^n  \F(\check{\bnu}, \by_i) \bigg|\bigg|_2 \\
	&= \bigg|\bigg|\hat{\bGamma}^*+\D(\bnu^*)-\D(\bnu^*) +\D(\check{\bnu})-\D(\check{\bnu}) + \frac{1}{n} \sum_{i=1}^n \F(\check{\bnu}, \by_i) \bigg|\bigg|_2 \\
	&\leq ||\hat{\bGamma}^*+\D(\bnu^*)||_2+||-\D(\bnu^*) +\D(\check{\bnu})||_2+\bigg|\bigg|-\D(\check{\bnu}) + \frac{1}{n} \sum_{i=1}^n  \F(\check{\bnu}, \by_i) \bigg|\bigg|_2.
\end{align*}
By $||\bnu^*-\check{\bnu} ||_2^2 \rightarrow 0$ in probability and Lemma \ref{lem:ULLN}, we have $\frac{||{\R}||_{\infty}}{||\widetilde{\Delta} _n||_2}\rightarrow 0$ in probability.

\subsection{Simulation}
\subsubsection{Details of the data generation process}
For each simulation dataset, we first independently generate the precision matrix for each of the 3 latent normal distributions according to one of the four graph structures. When generating the precision matrices, the diagonal elements are set as 1 plus a small positive number to guarantee positive definiteness. Then, we generate the mean vectors $\bmu_1,\bmu_2,\bmu_3$ for the latent normal distributions. The first $p_d$ elements of $\bmu_g$ ($g=1,2,3$) are independently sampled from $\{{{v}_{1}},\left({{v}_{1}}+{{v}_{2}}\right)/2,{{v}_{2}}\}$. The remaining $p-p_d$ elements are shared among $\bmu_1,\bmu_2, \bmu_3$ and are independently sampled from $\{{v}_{3},{v}_{4}\}$. We set $\left(v_1,v_2,v_3,v_4\right)$ as  $(2.4,-0.1,0.9,-0.1)$  in the low dropout case (about $10\%$ zeros) and $(1.4,-1.1,-0.1,-1.1)$ in the high dropout case  (about $40\%$ zeros). We vary $p_d$ to control the mixing degree of the three cell types. The library sizes $\bl = (l_1,\cdots,l_n)$ are independently generated from a log-normal distribution $\log \mbox{N}(\log 10,0.05)$. With these model parameters, we finally generate the observed expression $\bY_1,\cdots,\bY_n$ from the MPLN model. We calculate the Adjusted Rand Index (ARI) between the true cell-type label and  cell-type label given by the K-means clustering \citep{hartigan1979algorithm} of the normalized data $\tilde{\bY} \triangleq \log (\bY+1) - (\log \hat{\bl}) \mathbf{1}_{p}^{T}$ with $\hat{l}_i = \sum_{j=1}^{p} Y_{i j} / 10^4$ $(i = 1,2,\cdots,n)$. We vary $p_d$ such that the low-level mixing data have an ARI value in  $\left(0.9,1\right]$, the middle-level mixing data have an ARI value in $\left(0.75,0.85\right]$ and the high-level mixing data have an ARI value in $\left(0.65,0.75\right]$.

\subsubsection{Parameter selection for different algorithms}

The network inference methods can be classified as dense-network methods and sparse-network methods depending on their reported networks are dense networks or sparse networks. The dense-network methods include PPCOR, GENIE3 and PIDC. They report a non-negative connected weight for every gene pair and the network can be inferred by choosing edges with the largest connected weights. The sparse-network methods, including VMPLN, VPLN, Glasso and LPGM, usually have tuning parameters to control the sparsity of the estimated networks, and thus report a list of precision matrices related to tuning parameters. LPGM reports a stability score for every gene pair and the network can be inferred by choosing edges with the largest stability score.

We first use default parameters for the dense-network methods, including PPCOR, GENIE3 and PIDC. For the sparse-network methods including VMPLN, VPLN, Glasso and LPGM, we select their tuning parameters using their default criteria, i.e. ICL for VMPLN, Bayesian information criterion (BIC) for VPLN and Glasso, and stability for LPGM \citep{meinshausen2010stability}.

Then, we select the parameters such that the  estimated network density is around 20\%. For the sparse-network methods including VMPLN, VPLN, Glasso and LPGM, we first tune their tuning parameters such that the densities of the estimated networks are 20\%. For dense-network methods including PPCOR, GENIE3 and PIDC, we select the confidence score cutoffs such that the estimated network densities are 20\%.

\subsection{Real data Analysis}
\subsubsection{Silver standard construction for benchmarking on scRNA-seq data}
Silver standards are constructed using the IFNB1-stimulated group and the $3^\prime$ batch for the Kang data and Zheng data, respectively. The gene pairs that occur in the public GRN databases are taken as potential regulatory relationships.  Note that each of these  potential regulatory relationships involves at least one TF. Then, for each cell type in the construction batch, we calculate the Spearman's $\rho$ correlation between the gene pairs having potential regulatory relationships. If a gene pair has a significant Spearman's $\rho$ correlation, we consider the gene pair having a true regulatory relationship and add the edge to the silver standard edge set of the cell type.  

\subsubsection{The COVID-19 dataset}
We select top 2000 HVG genes for each patient and use the union of the HVG genes from all patients for VMPLN analysis. The gene set of interest for GRN inference is selected as the overall top 1000 HVG genes as defined by Seurat \citep{stuart2019comprehensive}. The edges that do not appear in the public GRN databases listed in Table \ref{GRNdatabase} are set to 0. We only focus on the GRN among genes in the gene set of interest for GRN inference. We perform VMPLN analysis for each patient separately and select the parameters such that the density of the estimated networks (i.e. the number of inferred edges divided by the number of edges in the prior GRN set) was ~5\%. Then, for each macrophage group, we weighted average the estimated partial correlations from each moderate patients with the number of cells as weight to get the GRN under the moderate condition. Similarly, we obtain the GRN for every macrophage group under the severe condition.

\makeatletter
\section{Supplementary Table}

\begin{table}[H]
	\renewcommand{\arraystretch}{1}
	\centering
	\caption{The cell numbers of each cell type in the two benchmarking datasets.}
	\renewcommand\arraystretch{0.9}
	\begin{tabular}{|c|c|c|}
		\hline
		\textbf{Dataset} & \textbf{Cell type} & \textbf{The number of cells}\\
		\hline
		\multirow{13}{*}{\textbf{Kang data} \citep{kang2018multiplexed}} & \text{CD14+ monocytes} & 2147\\
		\cline{2-3}
		\multirow{13}{*}{} & \text{CD16+ monocytes} & 537\\
		\cline{2-3}
		\multirow{13}{*}{} & \text{Dendritic cells} & 214\\
		\cline{2-3}
		\multirow{13}{*}{} & \text{CD4+ naive T cells} & 1526\\			
		\cline{2-3}
		\multirow{13}{*}{} & \text{CD4+ memory T cells} & 903\\						
		\cline{2-3}
		\multirow{13}{*}{} & \text{CD8+ T cells} & 462\\		
		\cline{2-3}
		\multirow{13}{*}{} & \text{Natural killer cells} & 321\\								
		\cline{2-3}
		\multirow{13}{*}{} & \text{Activated T cells} & 333\\		
		\cline{2-3}
		\multirow{13}{*}{} & \text{B cells} & 571\\		
		\cline{2-3}
		\multirow{13}{*}{} & \text{Activated B cells} & 203\\		
		\cline{2-3}
		\multirow{13}{*}{} & \text{Megakaryocytes} & 121\\		
		\cline{2-3}
		\multirow{13}{*}{} & \text{Plasmacytoid dendritic cells} & 81\\								
		\cline{2-3}
		\multirow{13}{*}{} & \text{Erythrocytes} & 32\\		
		\hline
		\multirow{9}{*}{\textbf{Zheng data} \citep{zheng2017massively}} & \text{FCGR3A+ monocytes} & 355\\
		\cline{2-3}
		\multirow{9}{*}{} & \text{CD14+ monocytes} & 2176\\
		\cline{2-3}
		\multirow{9}{*}{} & \text{Natural killer cells} & 290\\
		\cline{2-3}
		\multirow{9}{*}{} & \text{CD8+ T cells} & 1066\\			
		\cline{2-3}
		\multirow{9}{*}{} & \text{CD4+ T cells} & 903\\						
		\cline{2-3}
		\multirow{9}{*}{} & \text{B cells} & 1172\\		
		\cline{2-3}
		\multirow{9}{*}{} & \text{Hematopoietic stem cell} & 7\\								
		\cline{2-3}
		\multirow{9}{*}{} & \text{Megakaryocyte} & 57\\		
		\cline{2-3}
		\multirow{9}{*}{} & \text{Plasmacytoid dendritic cell} & 72\\			
		\hline
	\end{tabular}
\end{table}

\begin{table}[H]
	\renewcommand{\arraystretch}{1}
	\centering
	\caption{The public GRN databases.}
	\renewcommand\arraystretch{0.7}
	\begin{tabular}{|c|c|c|}
		\hline
		\textbf{Type} & \textbf{Source} & \textbf{Download link}\\
		\hline
		\multirow{2}{*}{\textbf{PPI databases}} & \textbf{STRING} \citep{szklarczyk2019string} & \makecell[l]{https://string-db.org/}\\
		\cline{2-3}
		\multirow{2}{*}{} & \textbf{HumanTFDB} \citep{hu2019animaltfdb}  & \makecell[l]{http://bioinfo.life.hust.edu.cn/\\HumanTFDB\#!/}\\
		\hline
		\multirow{5}{*}{\textbf{ChIP-seq databases}} & \textbf{hTFtarget} \citep{zhang2020htftarget} & \makecell[l]{http://bioinfo.life.hust.edu.cn/\\hTFtarget\#!/}\\
		\cline{2-3}
		\multirow{5}{*}{} & \textbf{ChEA} \citep{lachmann2010chea} & \makecell[l]{https://maayanlab.cloud/\\Harmonizome/dataset/\\CHEA+Transcription+Factor+Targets}\\
		\cline{2-3}
		\multirow{5}{*}{} & \textbf{ChIP-Atlas} \citep{oki2018ch} & \makecell[l]{https://chip-atlas.org/peak\_browser}\\
		\cline{2-3}
		\multirow{5}{*}{} & \textbf{ChIPBase} \citep{zhou2016chipbase} & \makecell[l]{https://rna.sysu.edu.cn/chipbase/}\\
		\cline{2-3}
		\multirow{5}{*}{} & \textbf{ESCAPE} \citep{xu2013escape} & \makecell[l]{http://www.maayanlab.net/\\ESCAPE/download.php}\\
		\hline
		\multirow{2}{*}{\textbf{Integrated databases}} & \textbf{TRRUST} \citep{han2018trrust} & \makecell[l]{https://www.grnpedia.org/trrust/}\\
		\cline{2-3}
		\multirow{2}{*}{} & \textbf{RegNetwork} \citep{liu2015regnetwork}  & \makecell[l]{http://www.regnetworkweb.org/}\\
		\hline
	\end{tabular}\label{GRNdatabase}
\end{table}
\begin{table}[H]
	\renewcommand{\arraystretch}{1}
	\centering
	\caption{The cell number of each cell type in each patient in the COVID-19 dataset.}
	\renewcommand\arraystretch{0.7}
	\begin{tabular}{|c|c|c|c|c|c|}
		\hline
		\textbf{COVID-19 severity} & \textbf{Patient's ID} & \textbf{Group1} & \textbf{Group2} & \textbf{Group3} & \textbf{Group4}\\
		\hline
		\multirow{2}{*}{\textbf{Moderate}} & \text{M1} & 102 & 1183 & 365 & 470\\
		\cline{2-6}
		\multirow{2}{*}{} & \text{M2} & 62 & 963 & 297 & 702\\
		\hline
		\multirow{6}{*}{\textbf{Severe}} & \text{S1} & 1884 & 4104 & 2378 & 1223\\
		\cline{2-6}
		\multirow{6}{*}{} & \text{S2} & 3443 & 6854 & 1452 & 720\\
		\cline{2-6}
		\multirow{6}{*}{} & \text{S3} & 175 & 231 & 112 & 117\\
		\cline{2-6}
		\multirow{6}{*}{} & \text{S4} & 396 & 374 & 212 & 103\\
		\cline{2-6}
		\multirow{6}{*}{} & \text{S5} & 243 & 529 & 152 & 124\\
		\cline{2-6}
		\multirow{6}{*}{} & \text{S6} & 194 & 540 & 101 & 175\\
		\cline{2-6}
		\hline
	\end{tabular}
\end{table}
\newpage

\section{Supplementary Figure}
\textbf{The definition of Early sensitvity}. The early sensitvity is defined as the sensitvity of top $\min\{K,s\}$ edges in the inferred network, where $K$ is the number of edges in the true network, and $s$ is the number of edges in inferred network.
\begin{figure}[htb]
	\begin{center}
		\subfigure[AUPRC ratios]{
			\includegraphics[height = 4.7 cm,width = 14 cm]{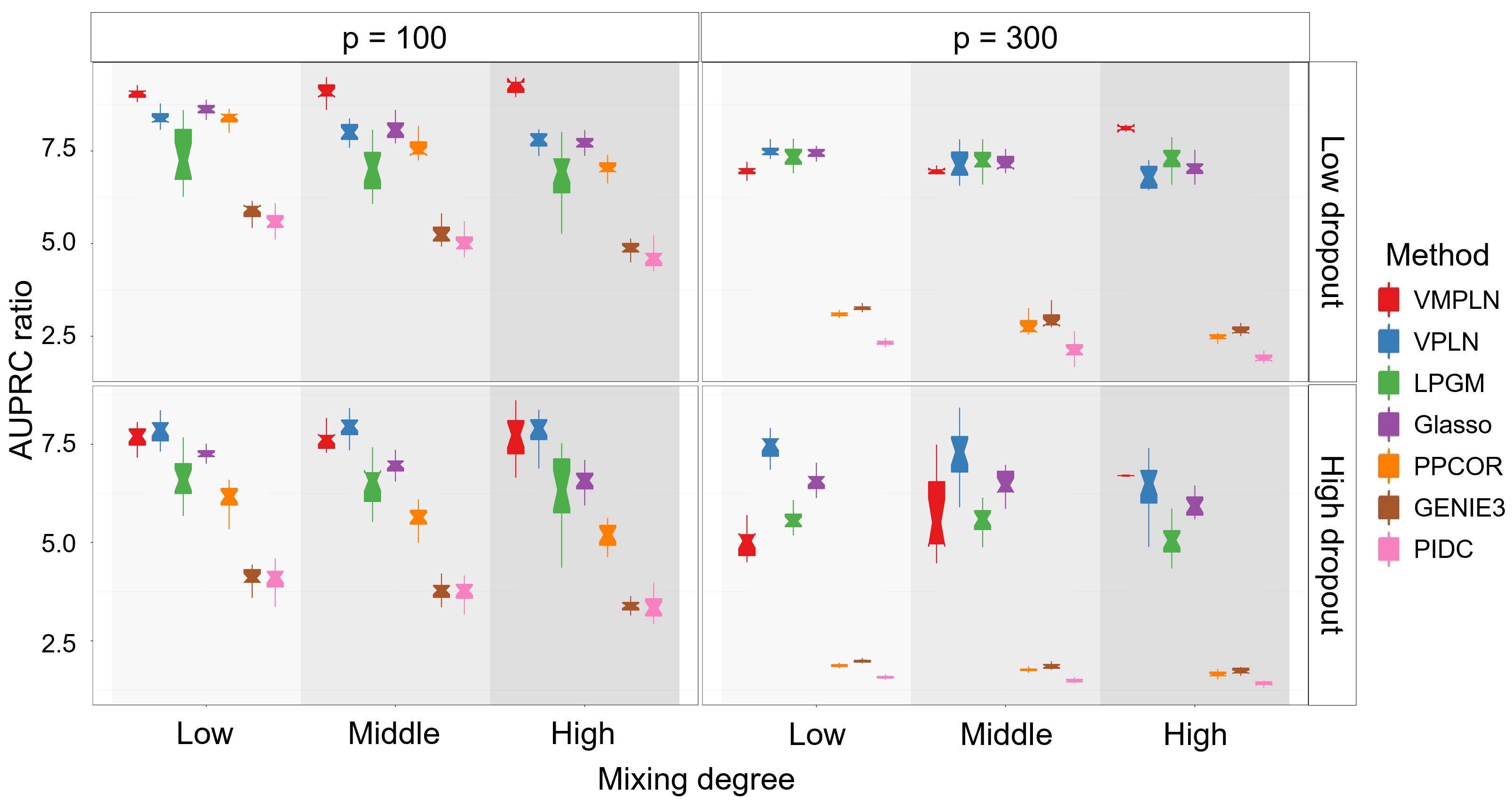}}
		\quad
		\subfigure[EP ratios]{
			\includegraphics[height = 4.7 cm,width = 14 cm]{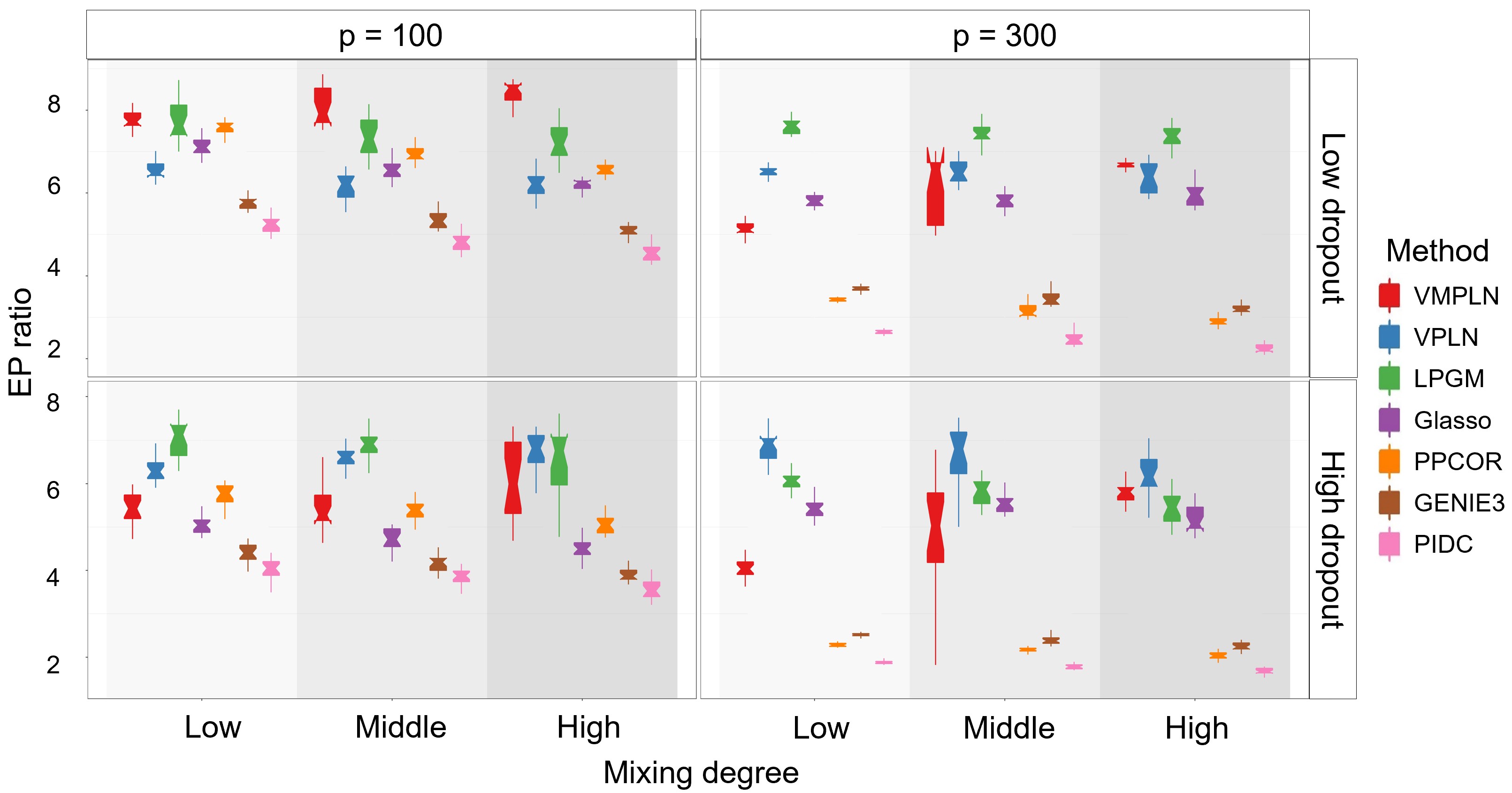}}
		\quad
		\subfigure[Early sensitvity]{
			\includegraphics[height = 4.7 cm,width = 14 cm]{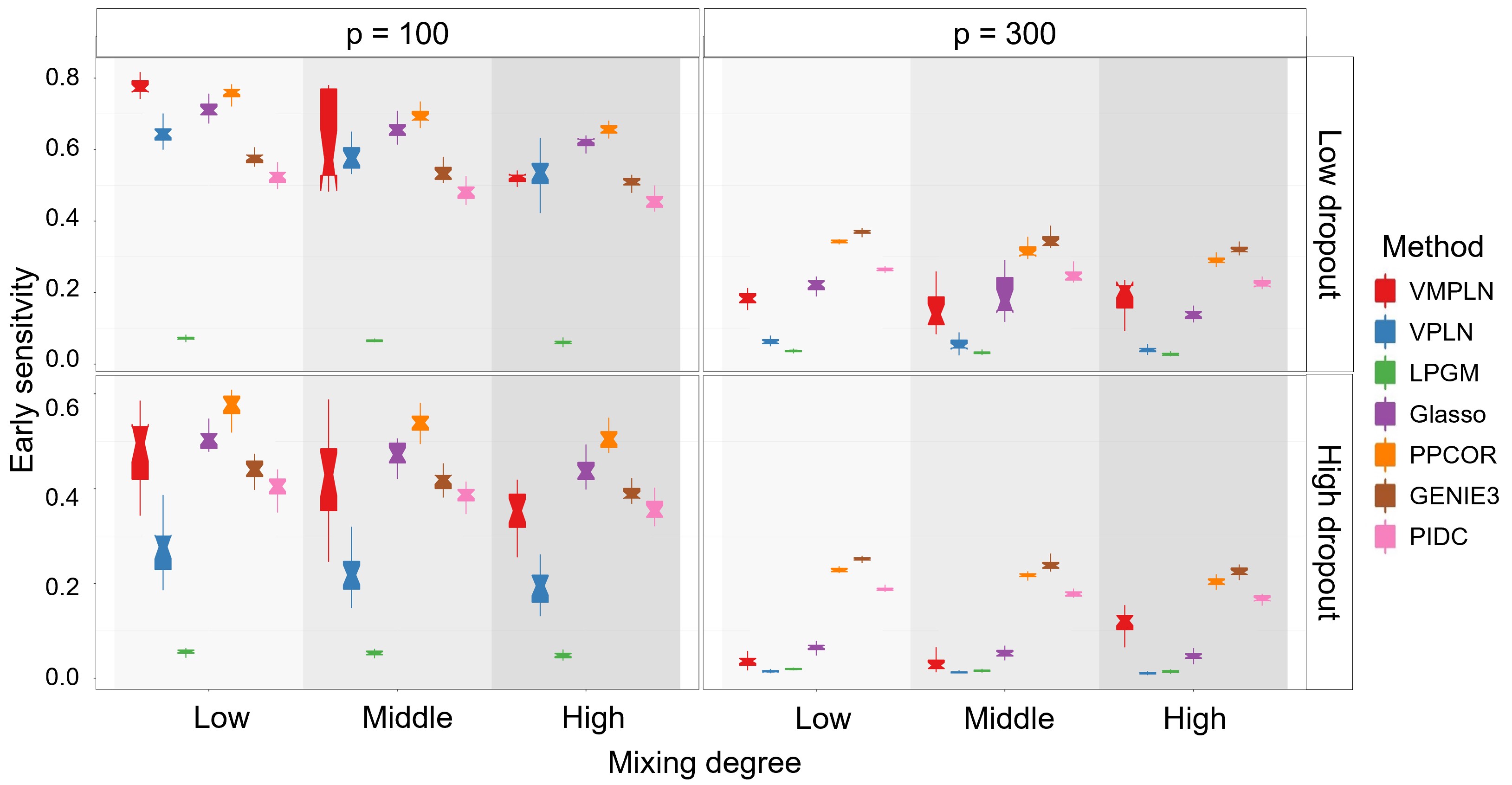}}
	\end{center}
	\caption{The AUPRC ratios, EP ratios and Early sensitvity for the random graph. The parameters are set as their default values or are tuned by their default methods.}
\end{figure}

\begin{figure}[H]
	\begin{center}
		\subfigure[AUPRC ratios]{
			\includegraphics[height = 6.0 cm,width = 14.5 cm]{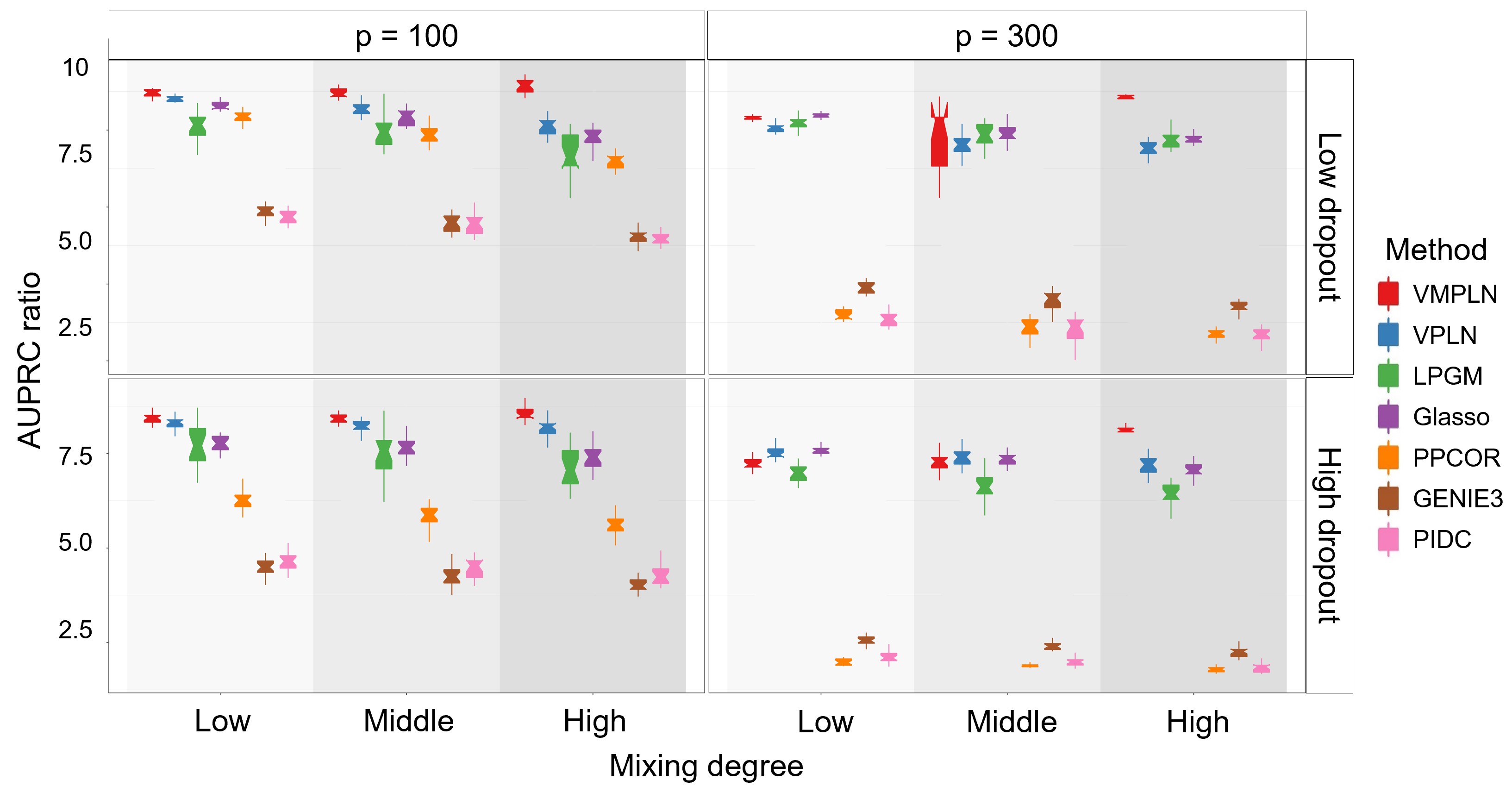}}
		\quad
		\subfigure[EP ratios]{
			\includegraphics[height = 6.0 cm,width = 14.5 cm]{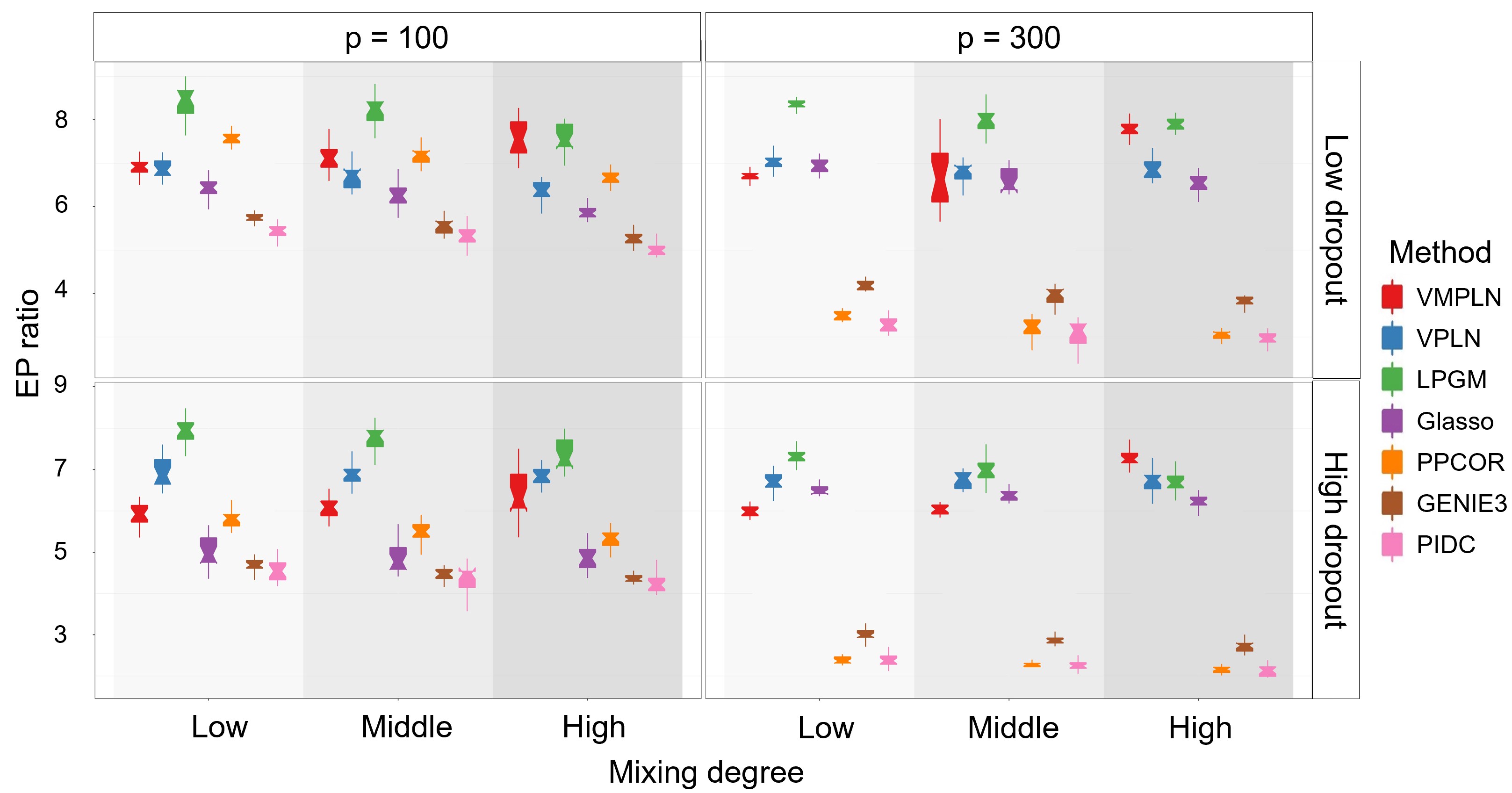}}
		\quad
		\subfigure[Early sensitvity]{
			\includegraphics[height = 6.0 cm,width = 14.5 cm]{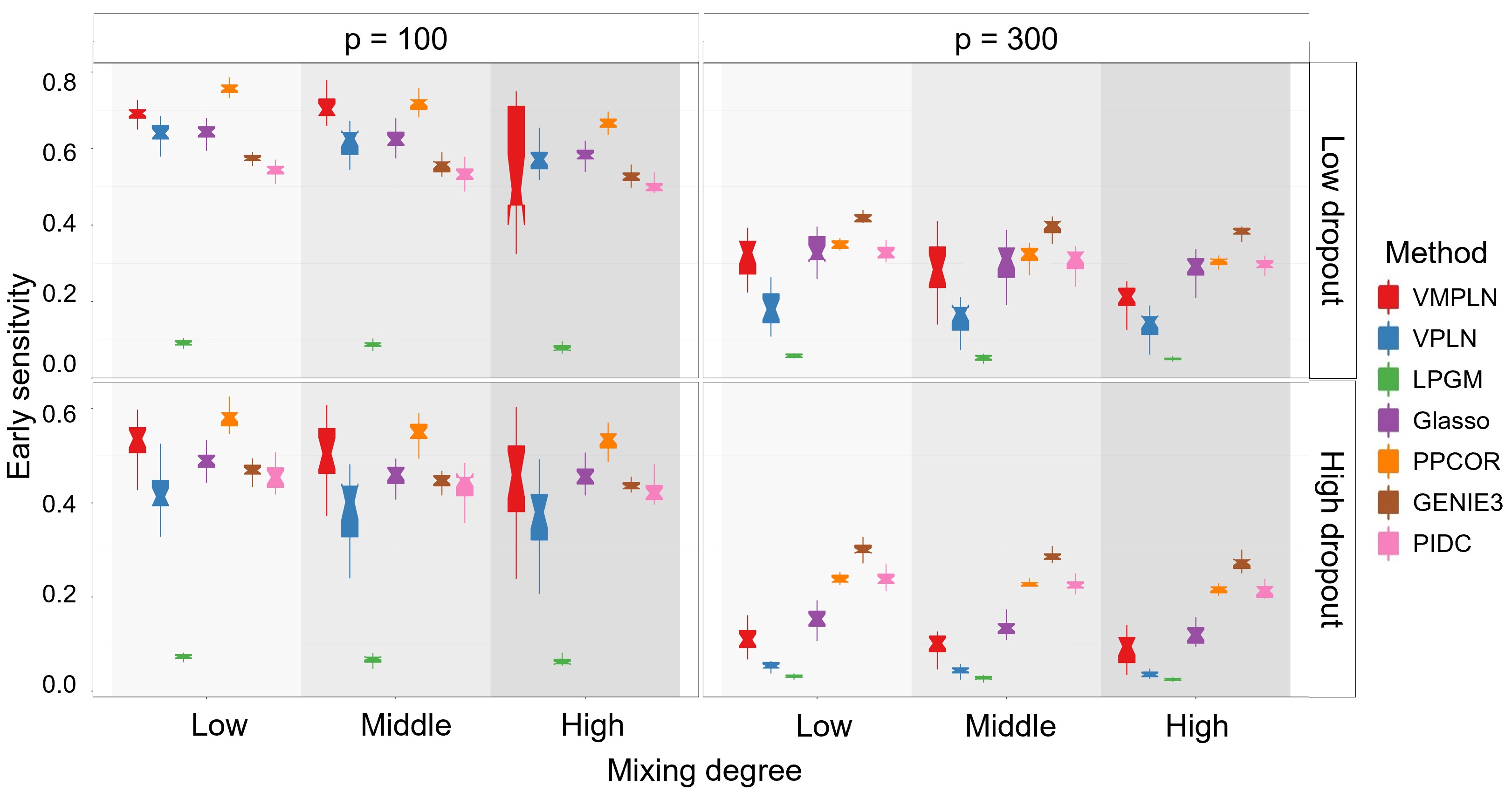}}
	\end{center}
	\caption{The AUPRC ratios, EP ratios and Early sensitvity for the blocked random graph. The parameters are set as their default values or are tuned by their default methods.}
\end{figure}	

\begin{figure}[H]
	\begin{center}
		\subfigure[AUPRC ratios]{
			\includegraphics[height = 6.0 cm,width = 14.5 cm]{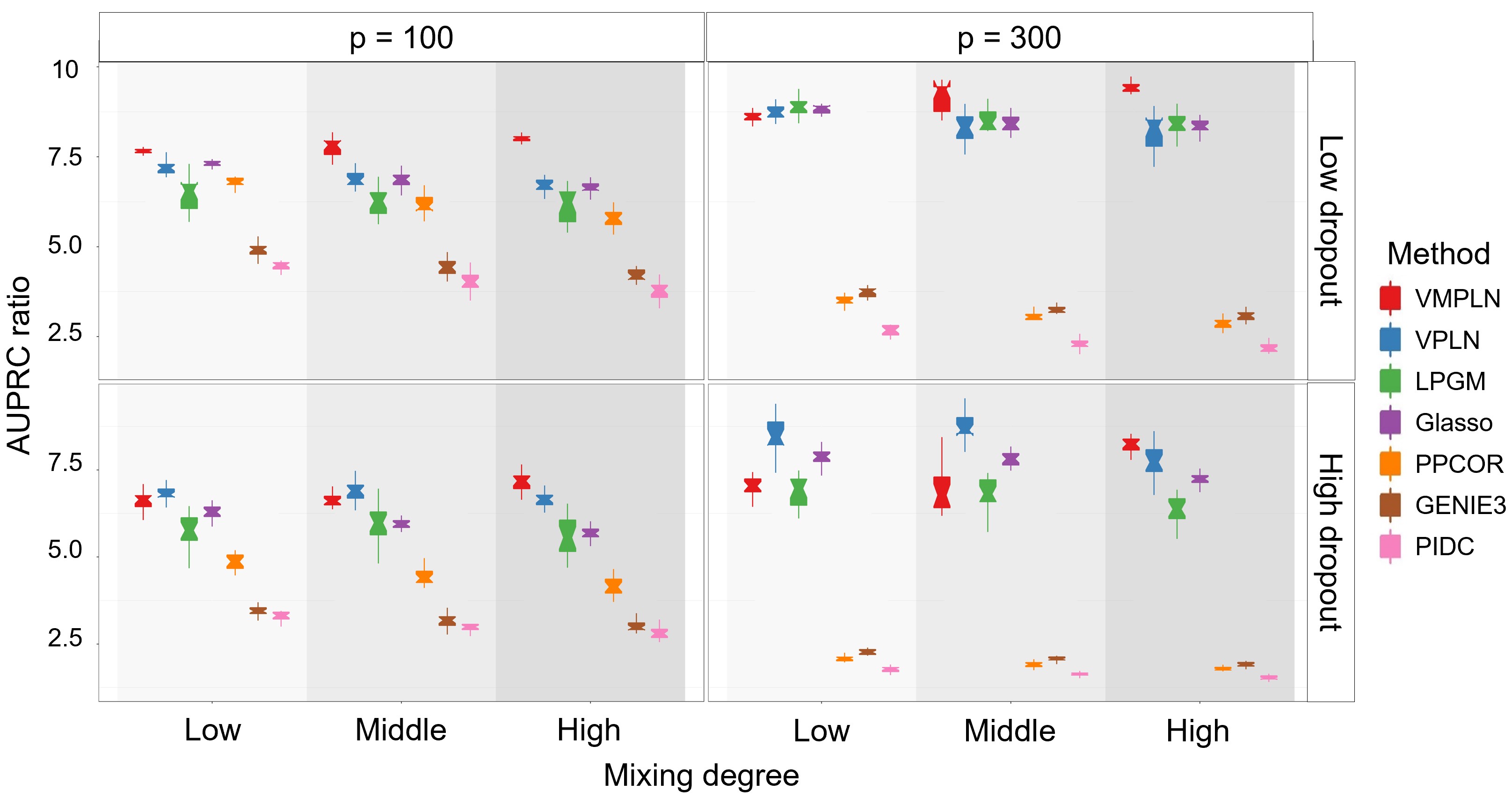}}
		\quad
		\subfigure[EP ratios]{
			\includegraphics[height = 6.0 cm,width = 14.5 cm]{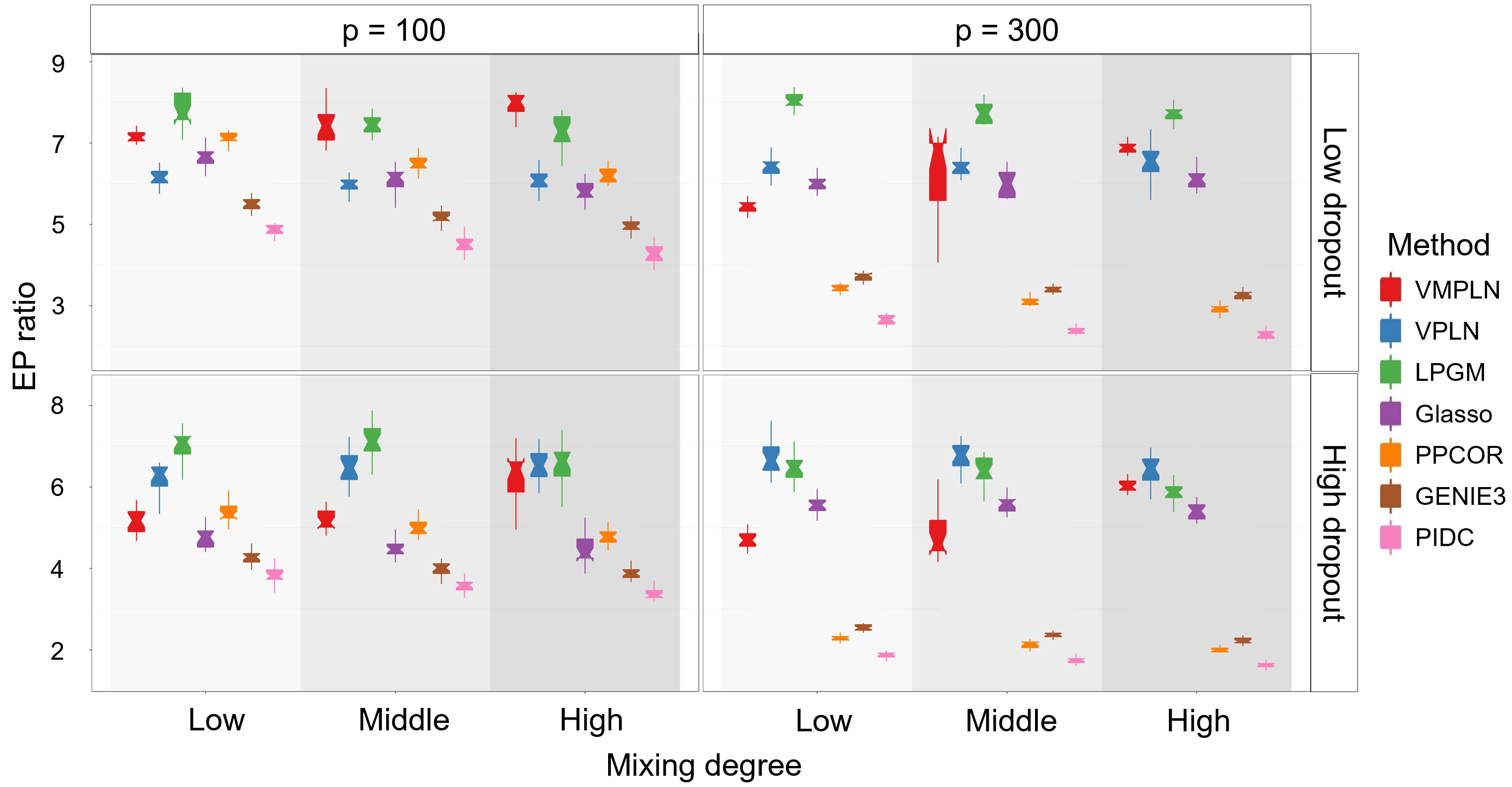}}
		\quad
		\subfigure[Early sensitvity]{
			\includegraphics[height = 6.0 cm,width = 14.5 cm]{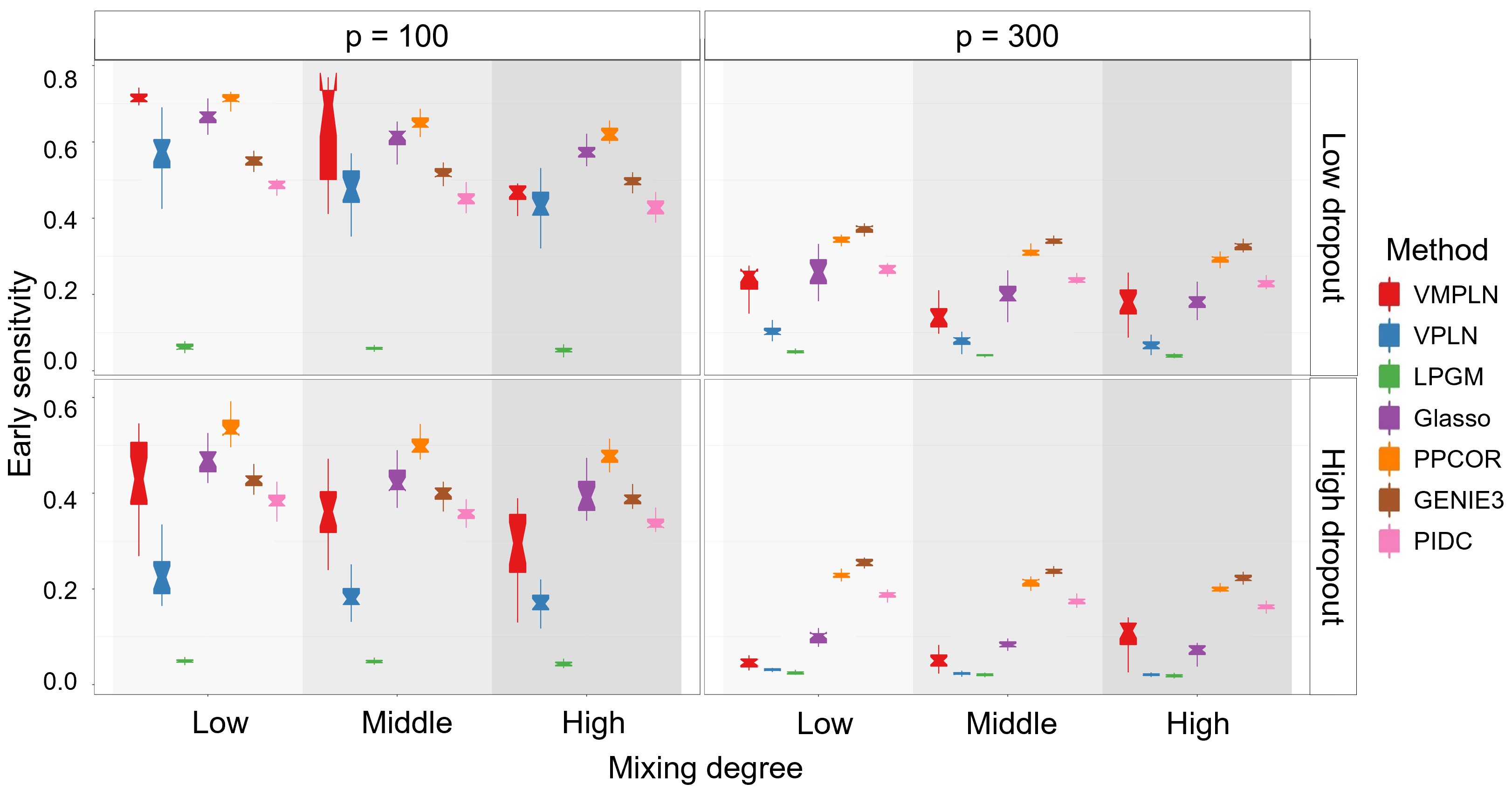}}
	\end{center}
	\caption{The AUPRC ratios, EP ratios and Early sensitvity for the scale-free graph. The parameters are set as their default values or are tuned by their default methods.}
\end{figure}	

\begin{figure}[htb]
	\begin{center}
		\includegraphics[height = 10 cm,width = 16 cm]{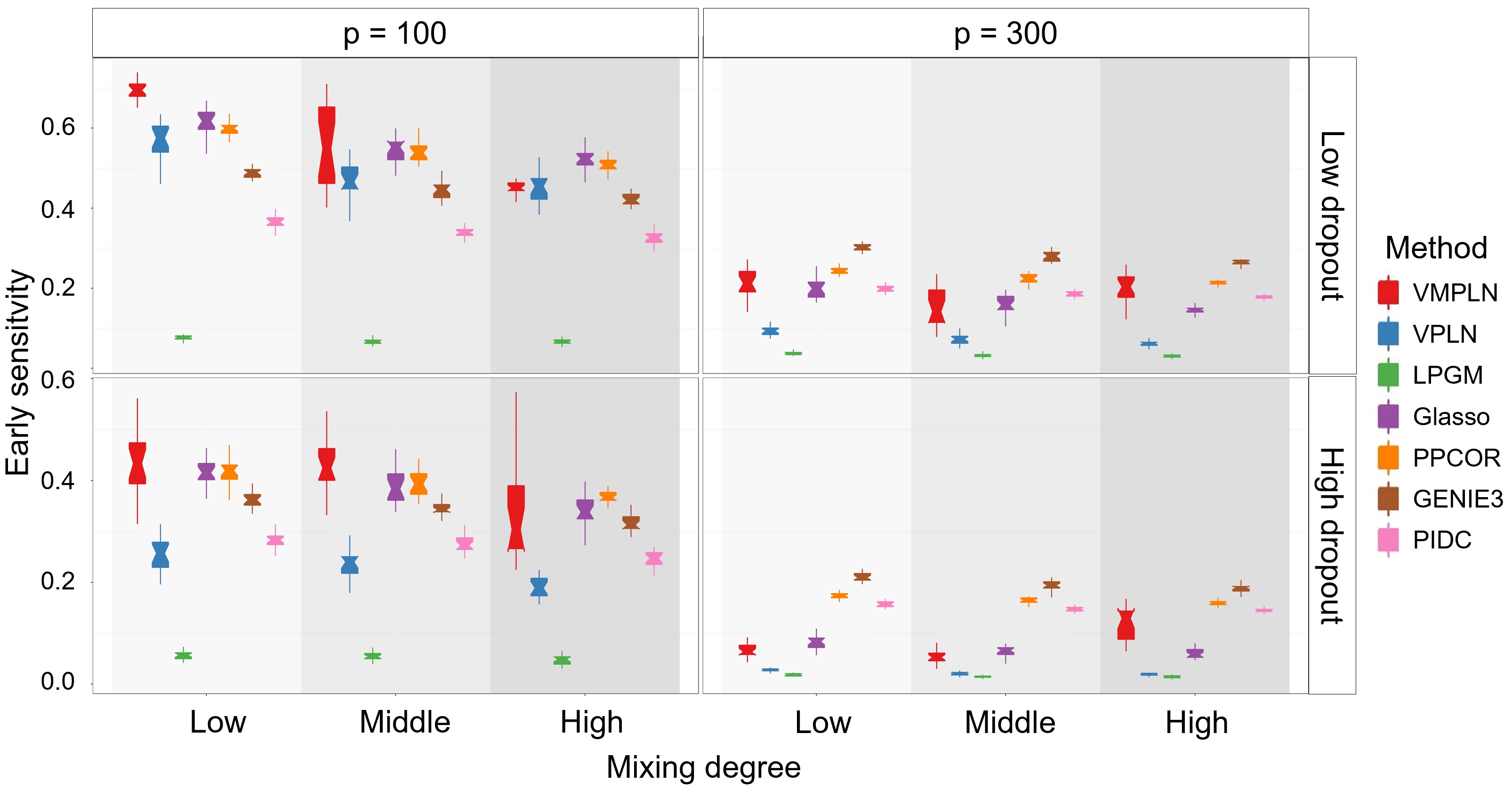}
	\end{center}
	\caption{The Early sensitvity for the hub graph. The parameters are set as their default values or are tuned by their default methods.}
\end{figure}

\begin{figure}[H]
	\begin{center}
		\subfigure[AUPRC ratios]{
			\includegraphics[height = 9 cm,width = 16 cm]{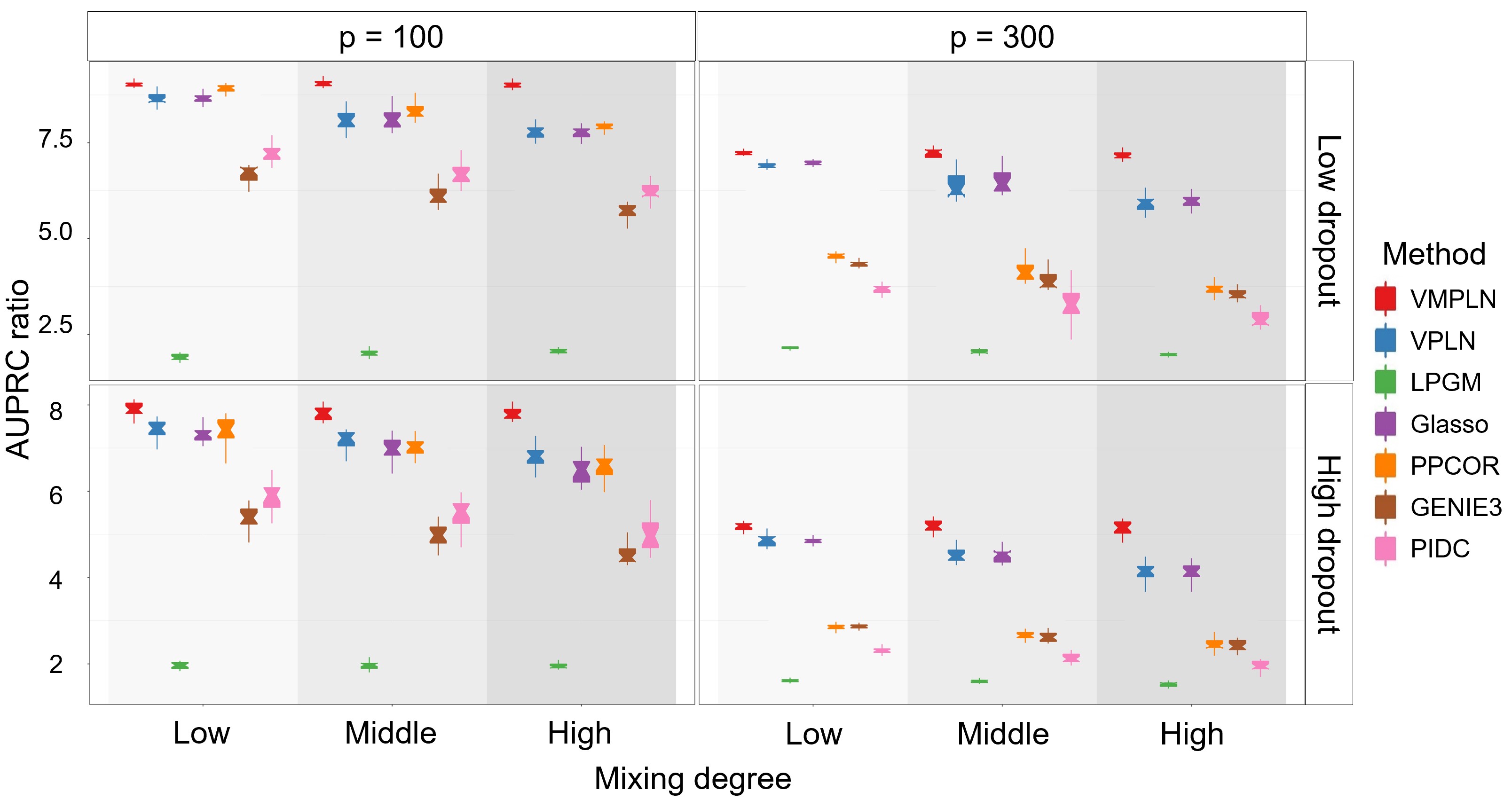}}
		\quad
		\subfigure[EP ratios]{
			\includegraphics[height = 9 cm,width = 16 cm]{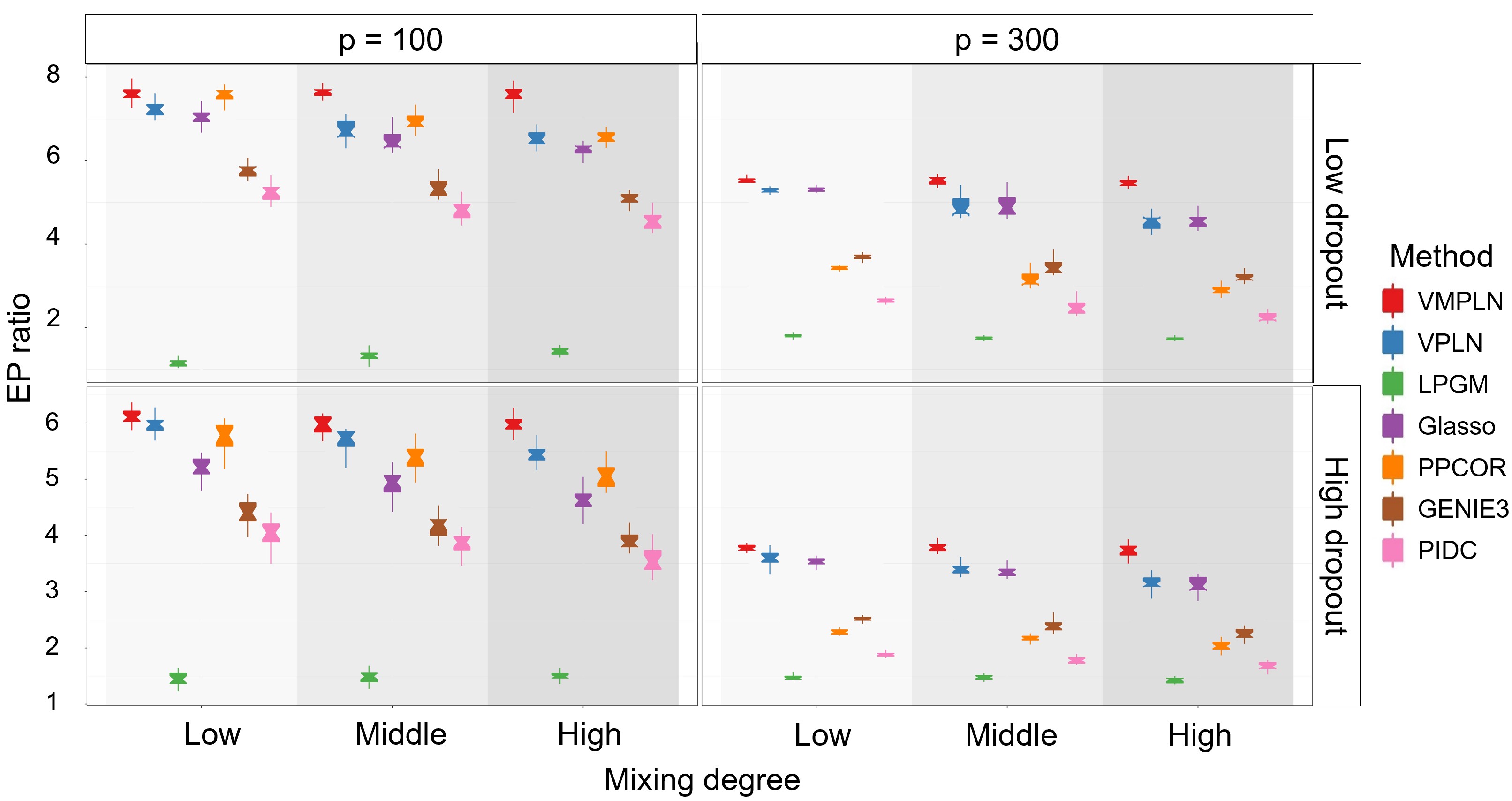}}
		\quad
	\end{center}
	\caption{Similar to Figure 2 for the random graph.}
	
\end{figure}
\begin{figure}[H]
	\begin{center}
		\subfigure[AUPRC ratios]{
			\includegraphics[height = 9 cm,width = 16 cm]{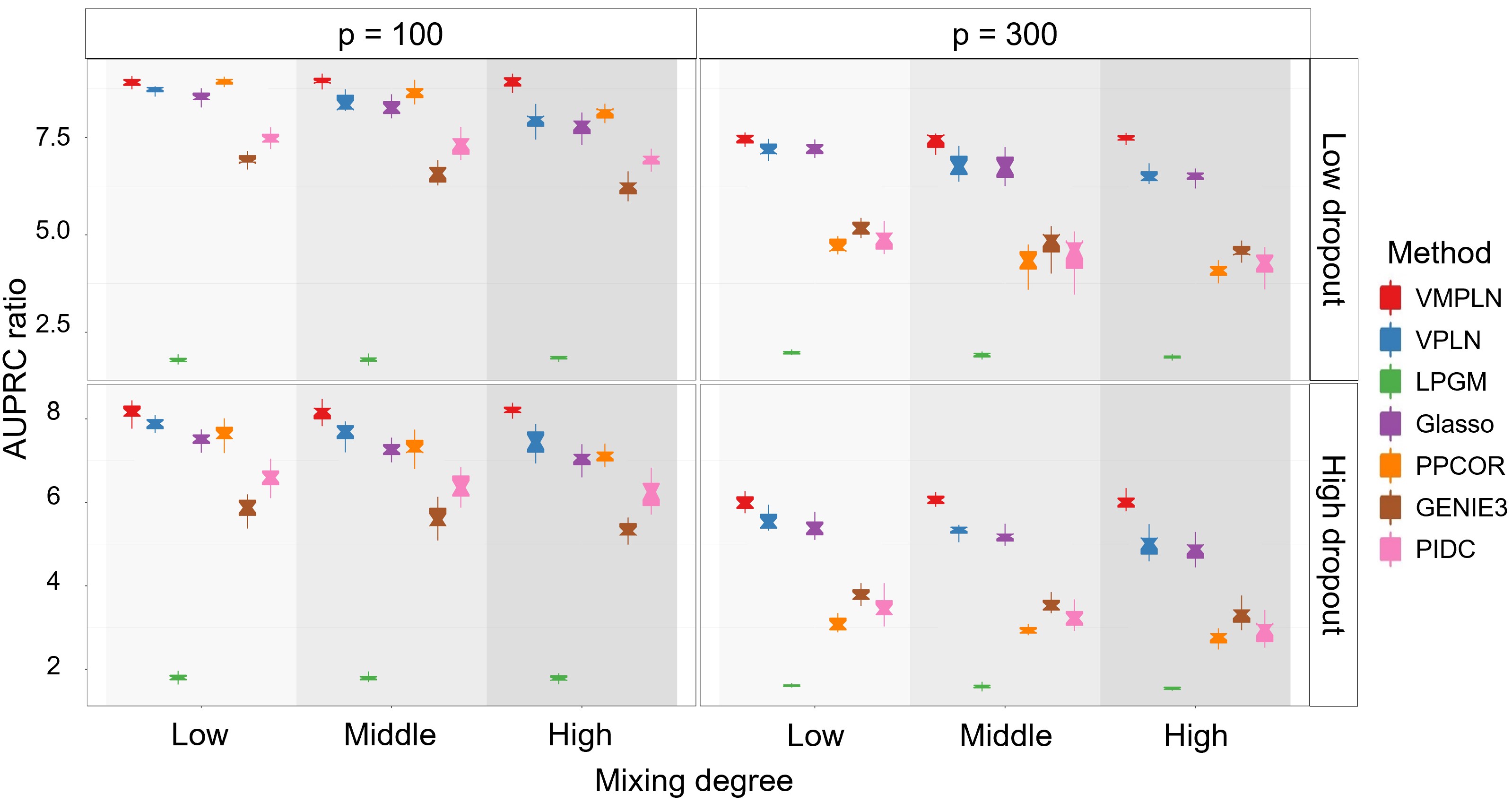}}
		\quad
		\subfigure[EP ratios]{
			\includegraphics[height = 9 cm,width = 16 cm]{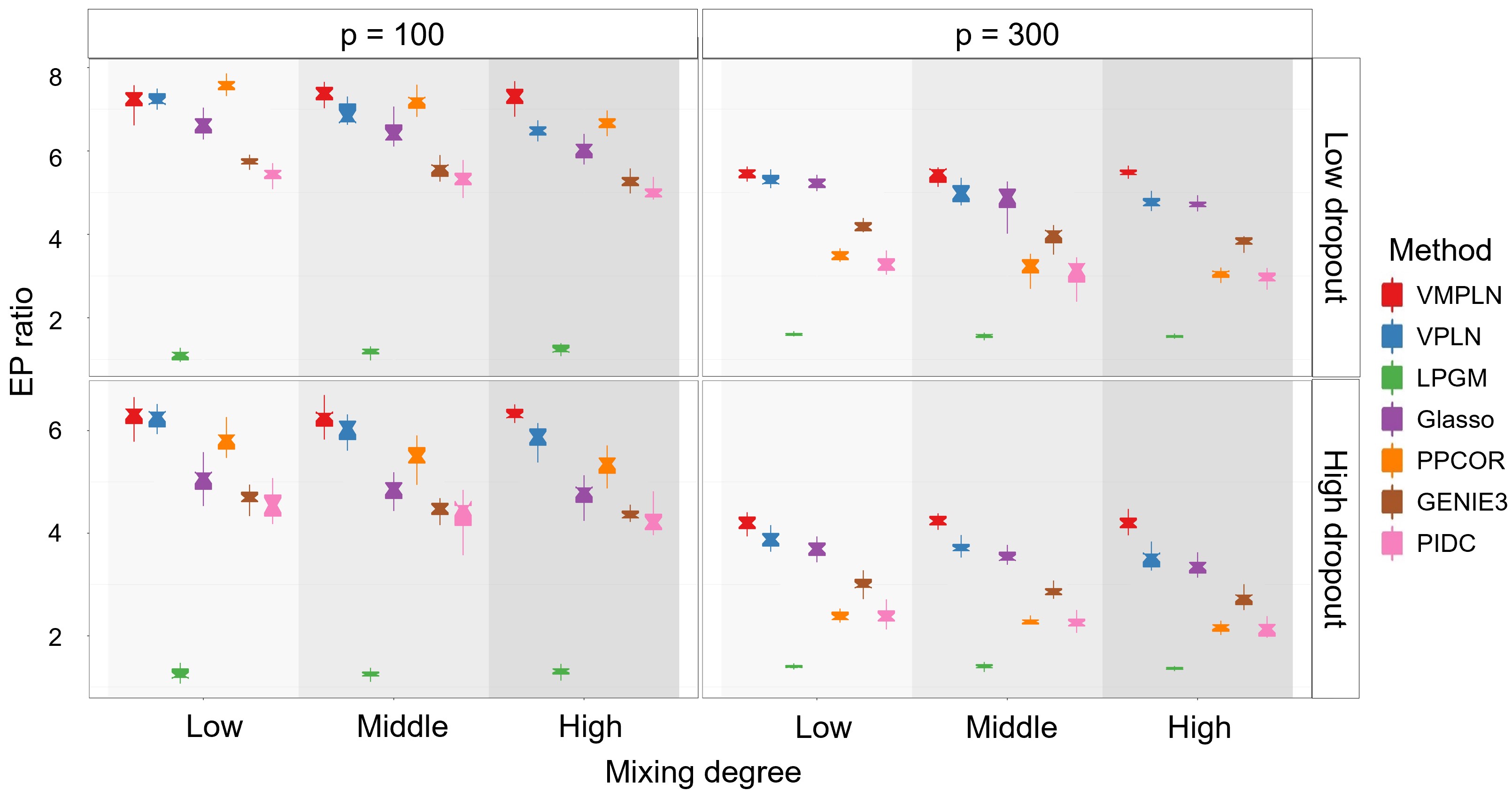}}
		\quad
	\end{center}
	\caption{Similar to Figure 2 for the blocked random graph.}
	
\end{figure}


\begin{figure}[H]
	\begin{center}
		\subfigure[AUPRC ratios]{
			\includegraphics[height = 9 cm,width = 16 cm]{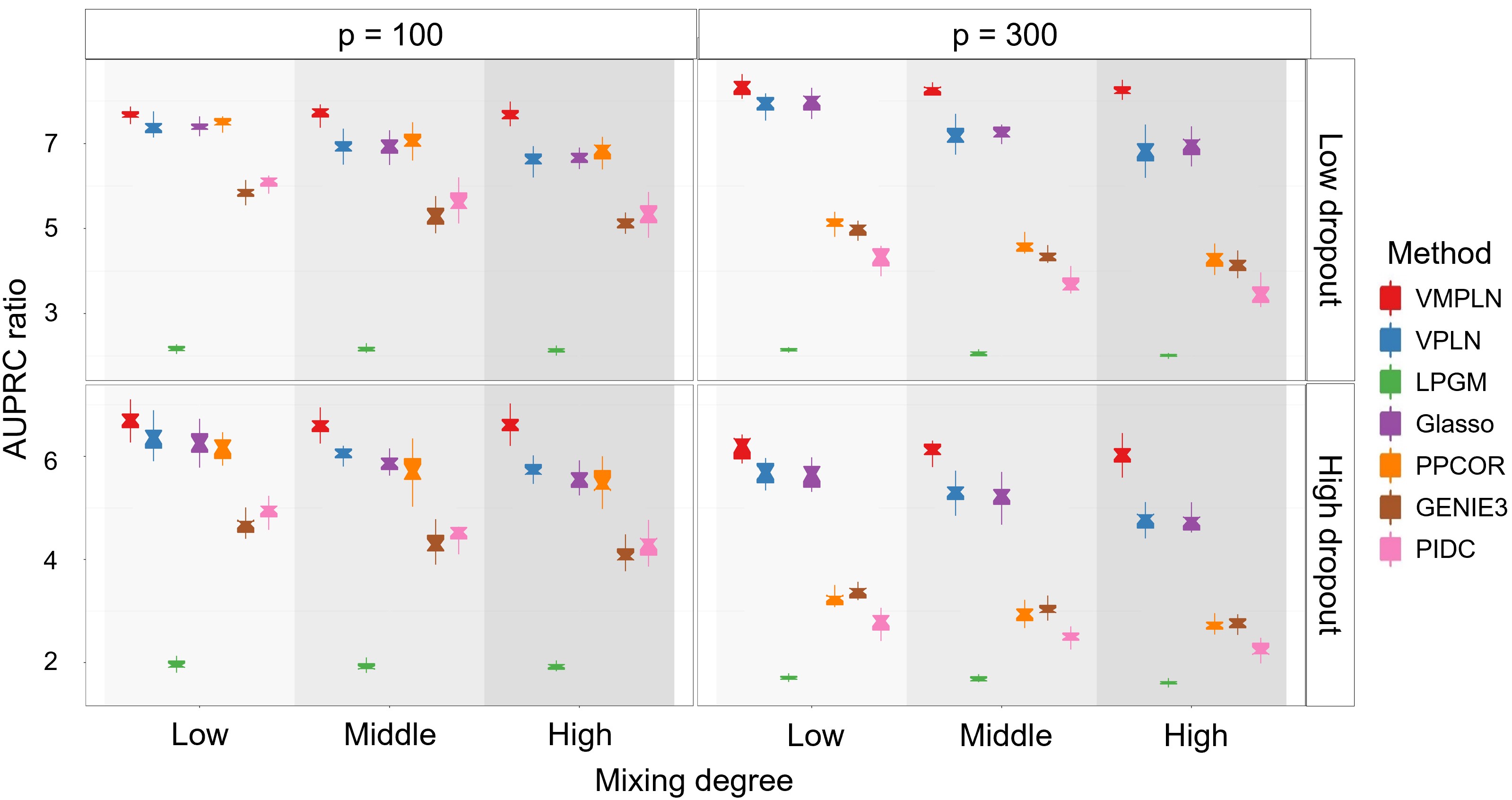}}
		\quad
		\subfigure[EP ratios]{
			\includegraphics[height = 9 cm,width = 16 cm]{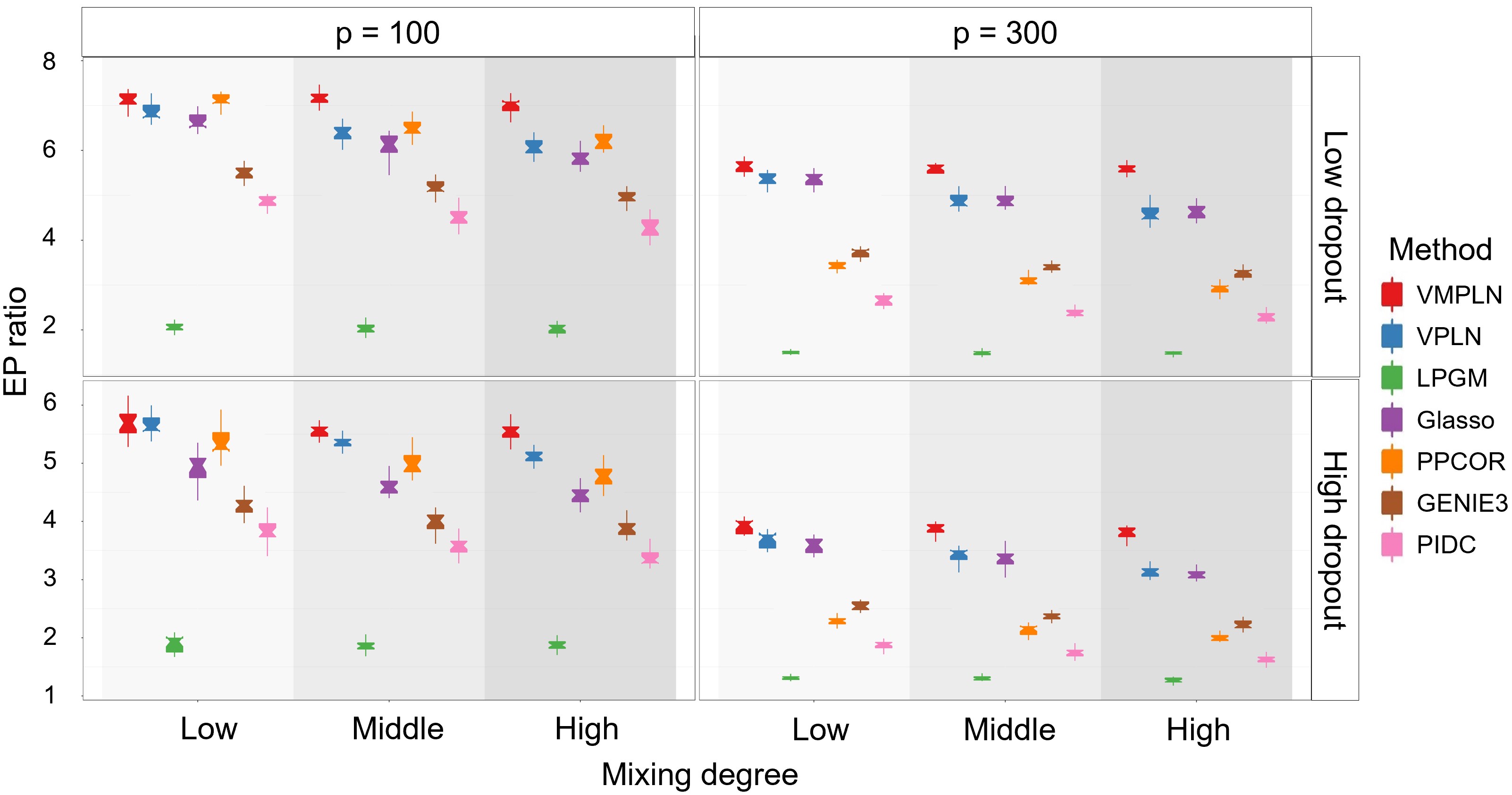}}
		\quad
	\end{center}
	\caption{Similar to Figure 2 for the scale-free graph.}
	
\end{figure}

\begin{figure}[H]
	\begin{center}
		\includegraphics[height = 10 cm,width = 16 cm]{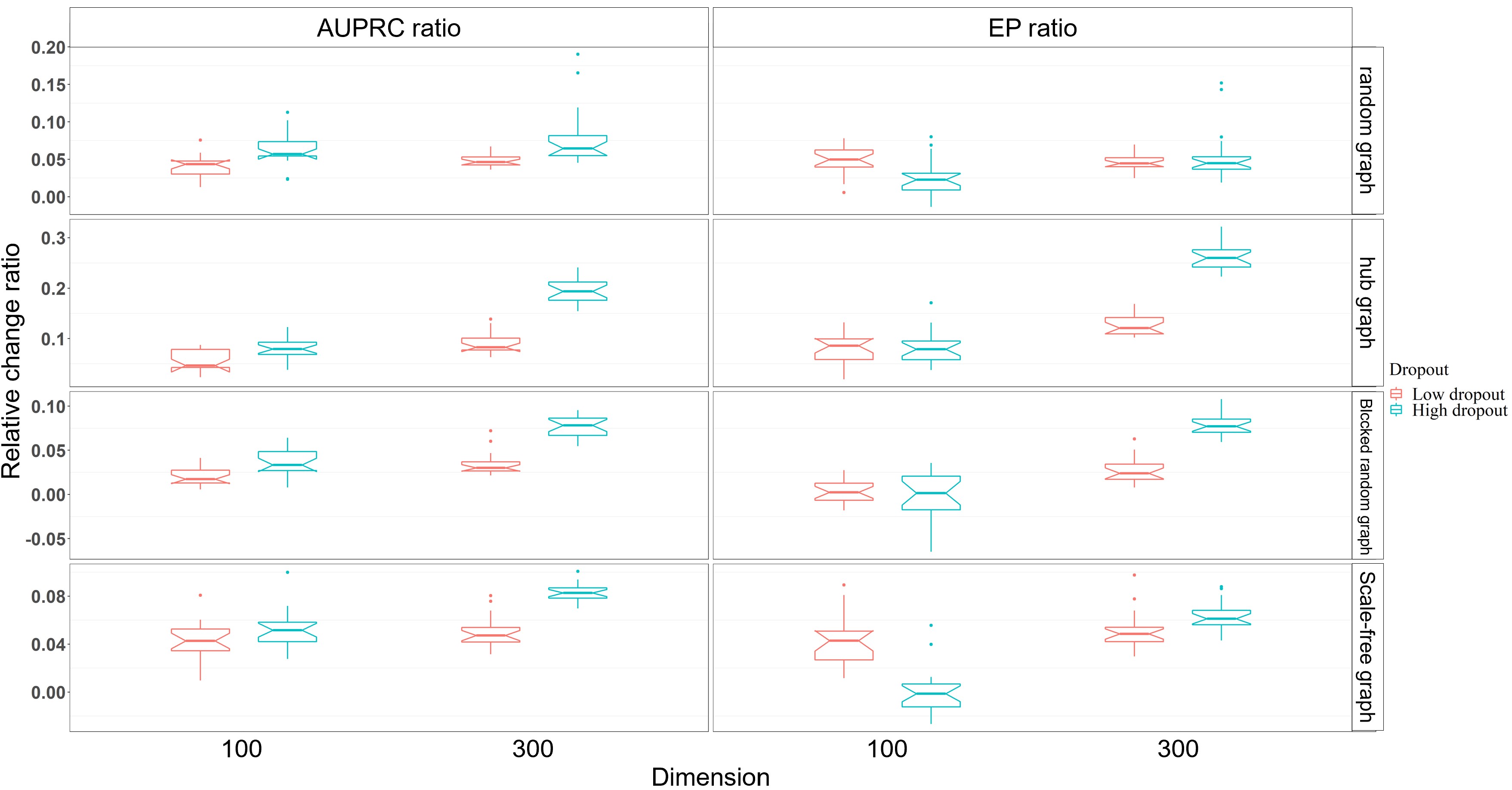}
	\end{center}
	\caption{The relative change ratio of AUPRC ratios and EP ratios between VMPLN and VPLN in the case of low-level mixing data. The relative change ratio of AUPRC ratios between VMPLN and VPLN is defined as the difference of the AUPRC ratios between VMPLN and VPLN divided by the AUPRC ratio of VPLN. The relative change ratio of the EP ratios is similarly defined.}
	
\end{figure}

\begin{figure}[H]
	\begin{center}
		\subfigure[AUPRC ratios]{
			\includegraphics[height = 9 cm,width = 16 cm]{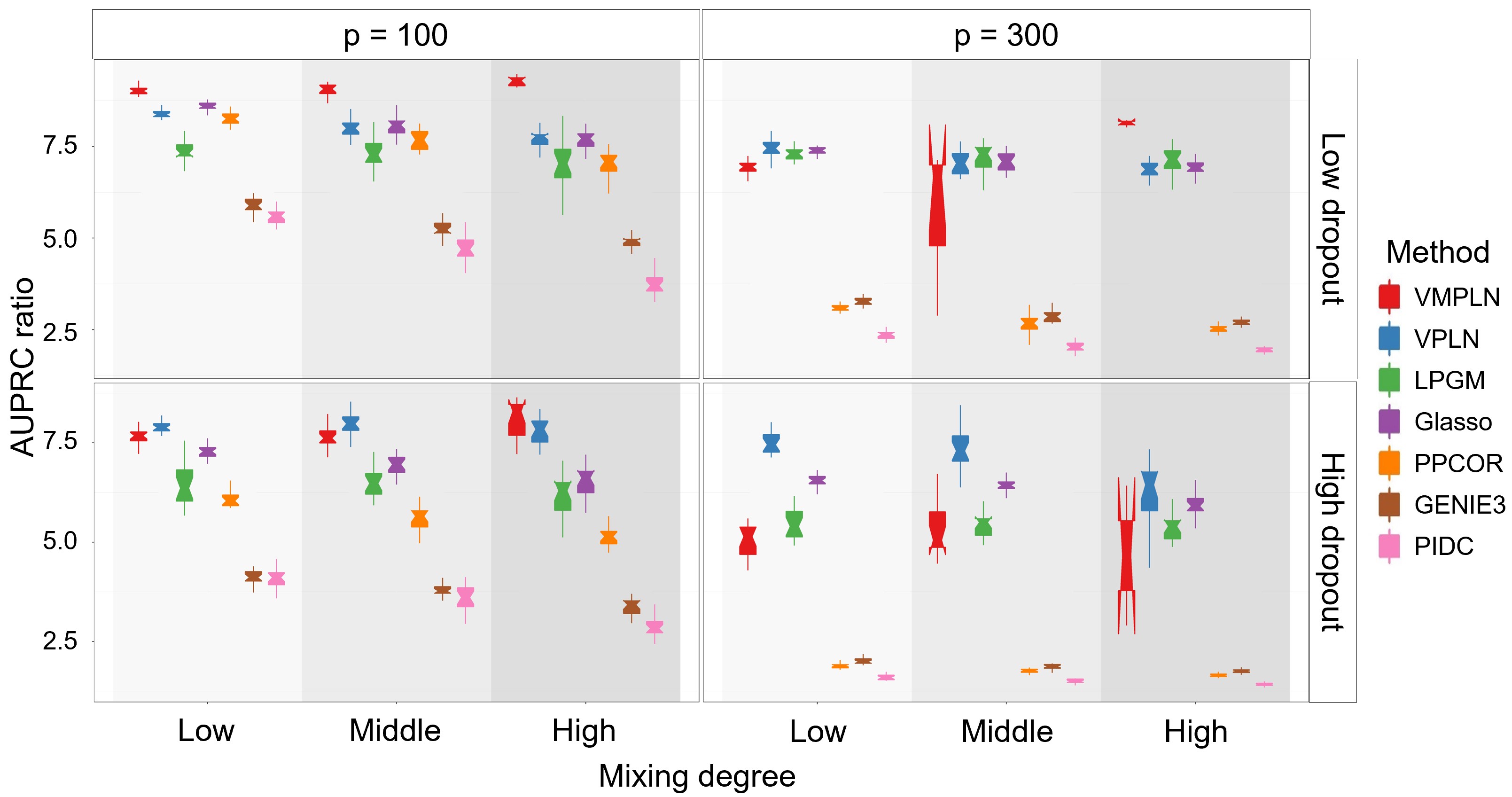}}
		\quad
		\subfigure[EP ratios]{
			\includegraphics[height = 9 cm,width = 16 cm]{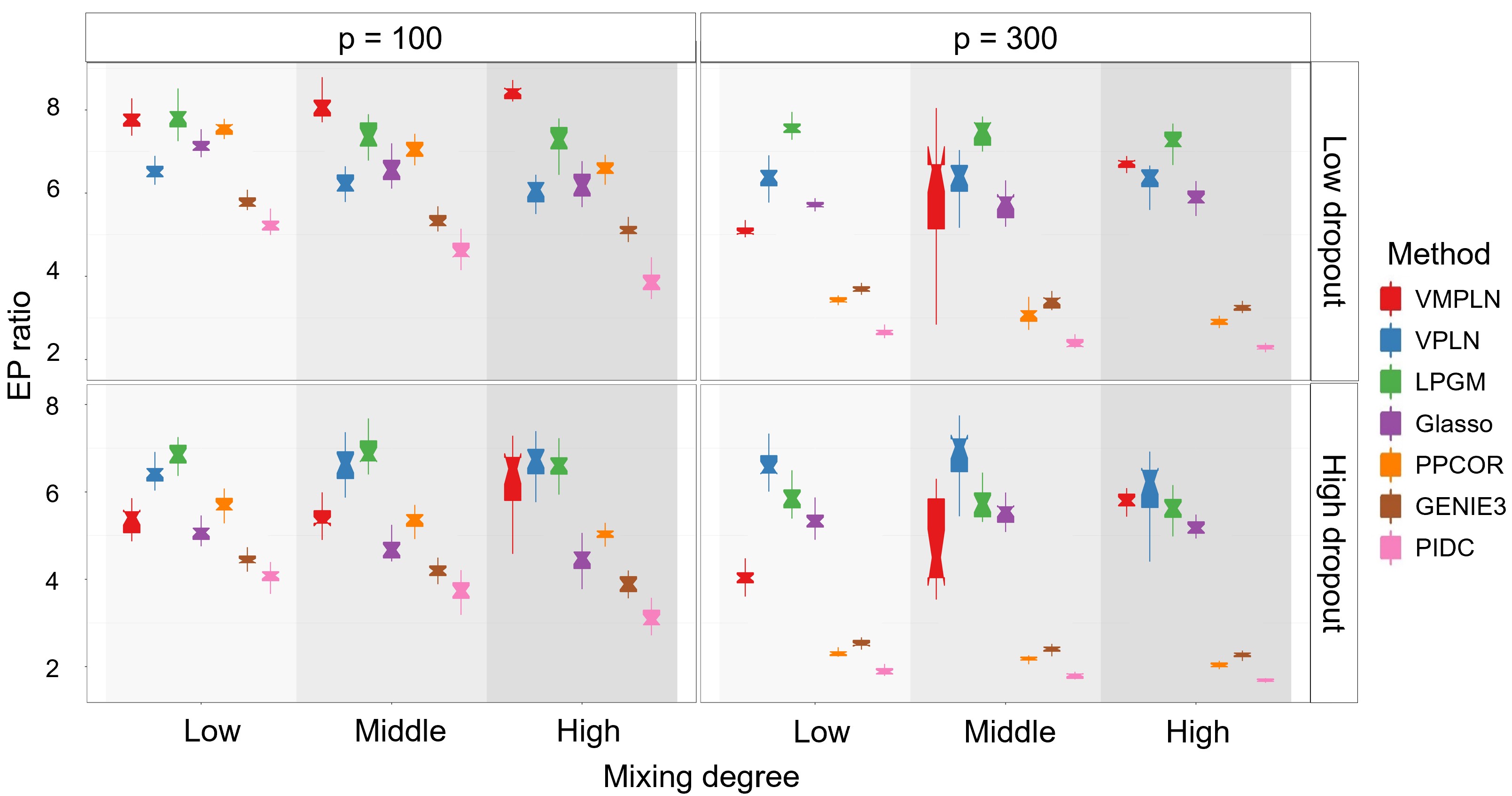}}
		\quad
	\end{center}
	\caption{Similar to Figure 1 for the random graph and data generated from the compositional model.}
\end{figure}

\begin{figure}[H]
	\begin{center}
		\subfigure[AUPRC ratios]{
			\includegraphics[height = 9 cm,width = 16 cm]{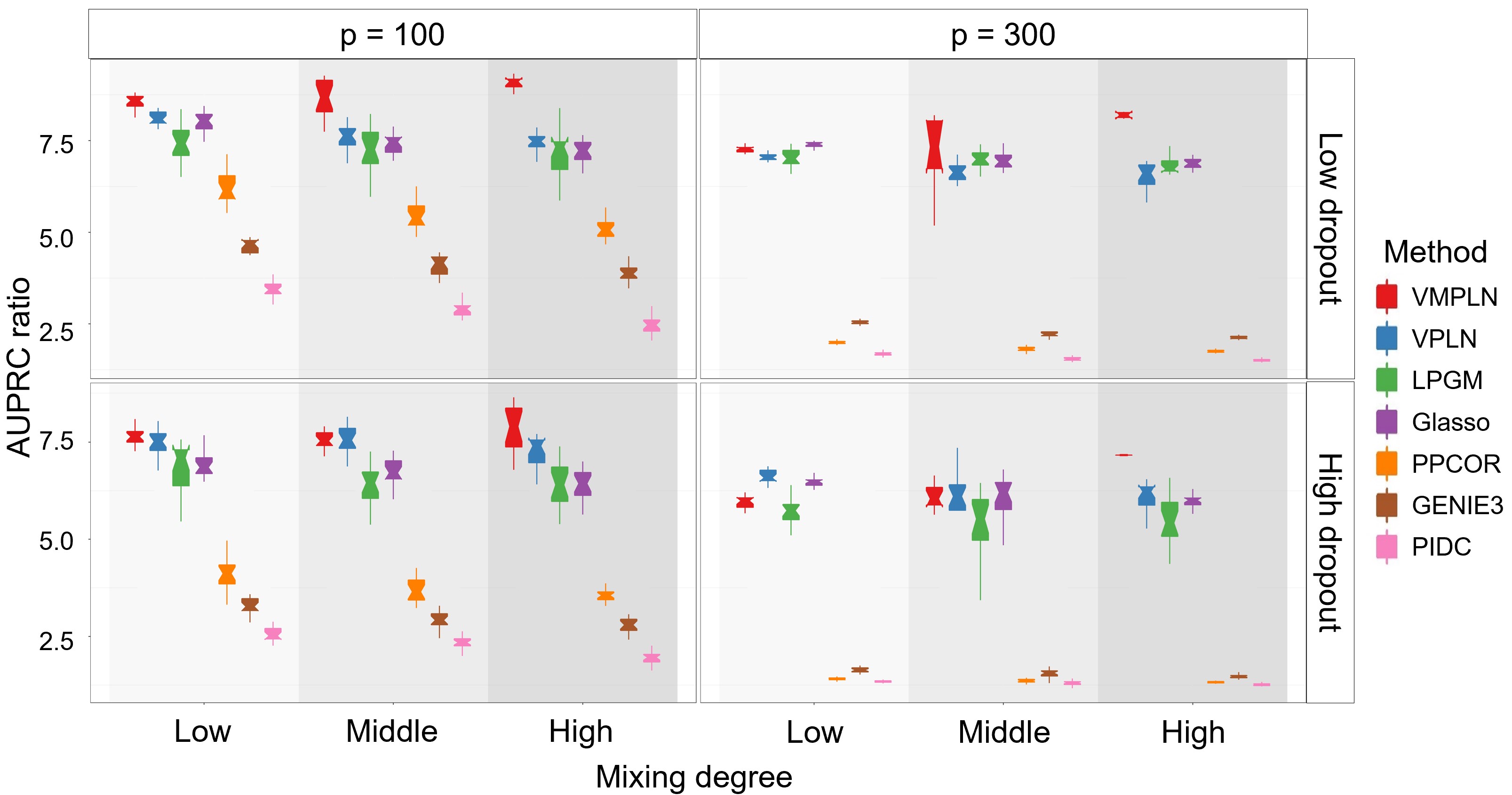}}
		\quad
		\subfigure[EP ratios]{
			\includegraphics[height = 9 cm,width = 16 cm]{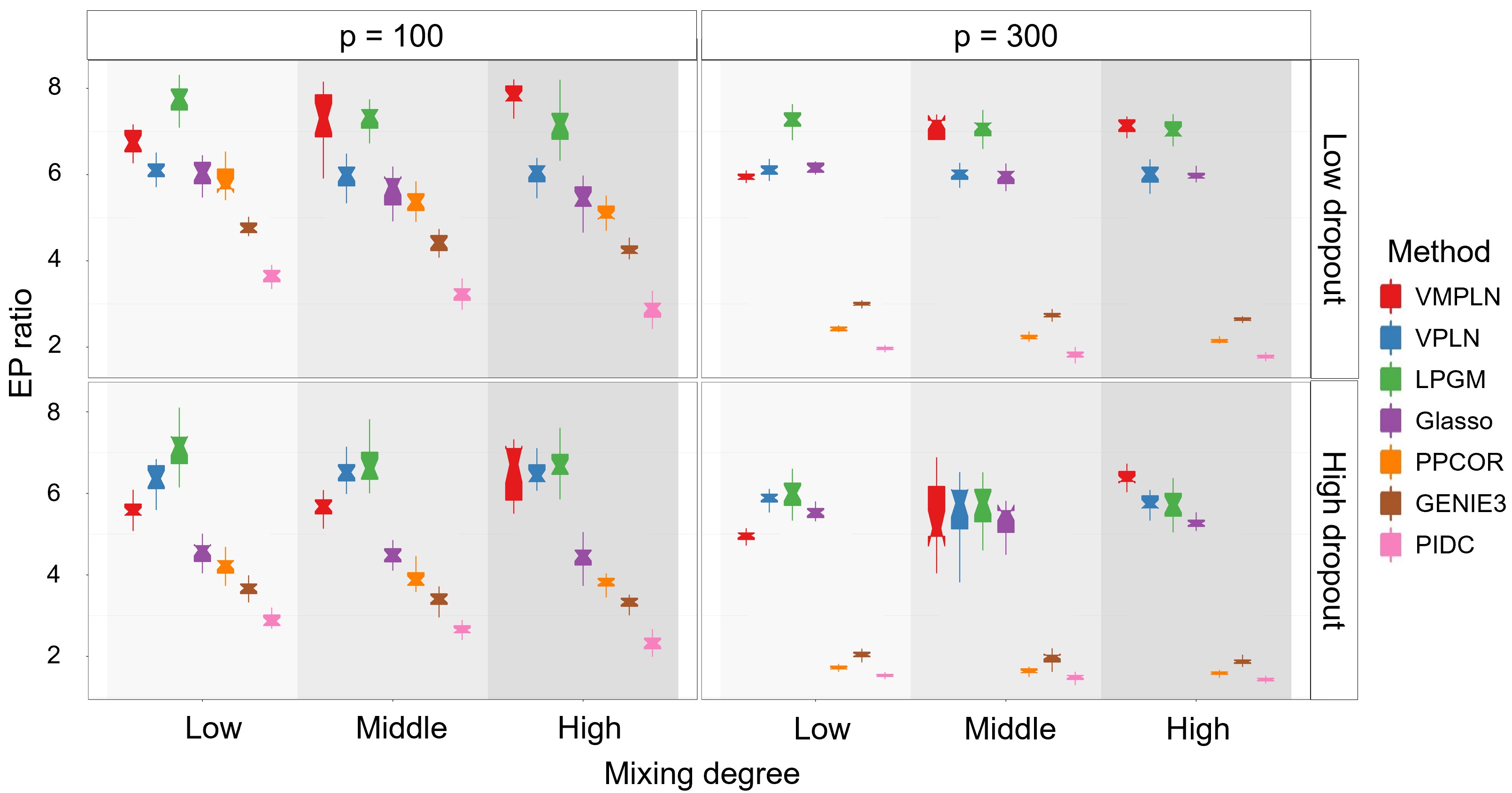}}
		\quad
	\end{center}
	\caption{Similar to Figure 1 for the hub graph and data generated from the compositional model.}	
\end{figure}

\begin{figure}[H]
	\begin{center}
		\subfigure[AUPRC ratios]{
			\includegraphics[height = 9 cm,width = 16 cm]{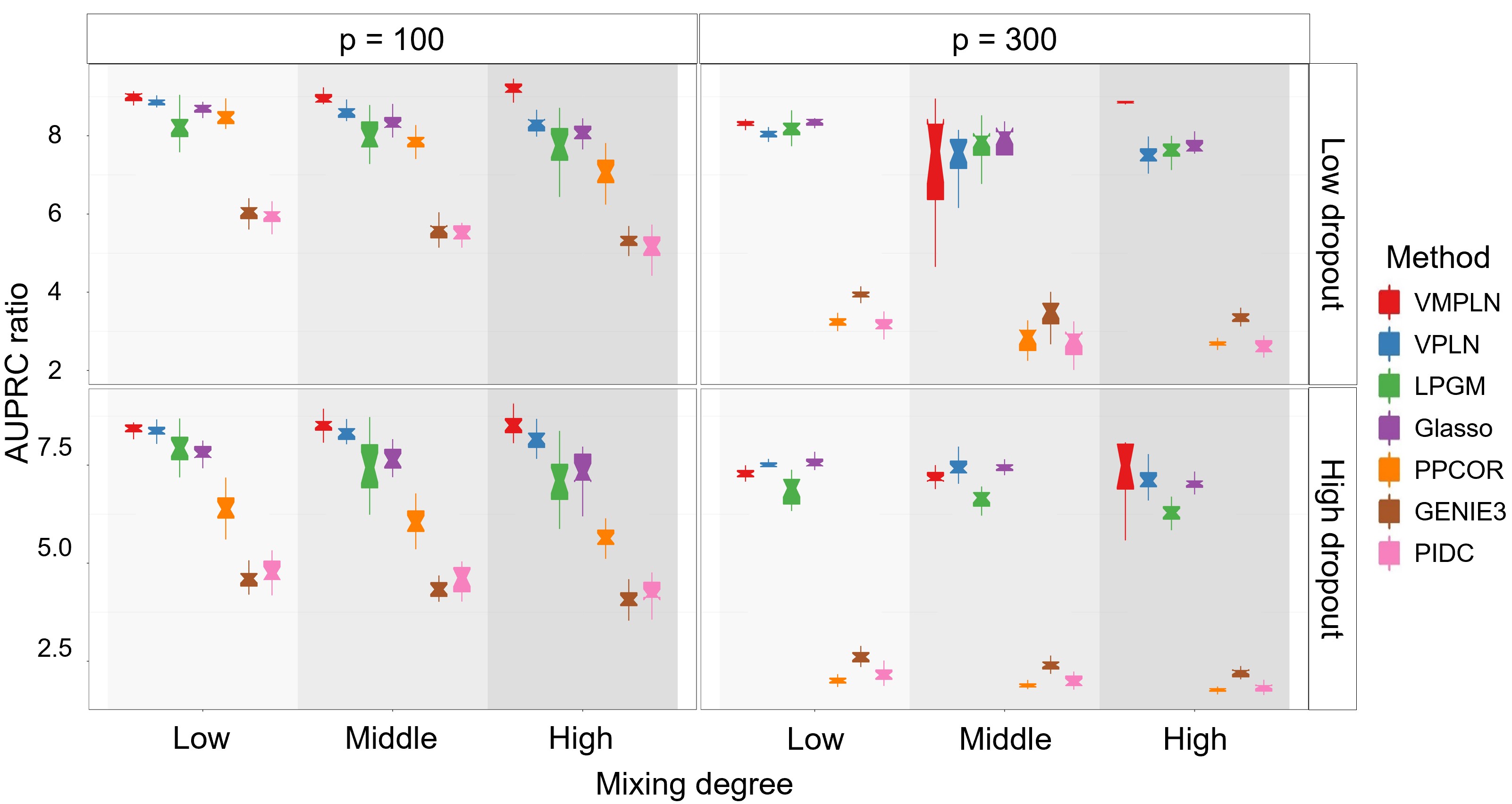}}
		\quad
		\subfigure[EP ratios]{
			\includegraphics[height = 9 cm,width = 16 cm]{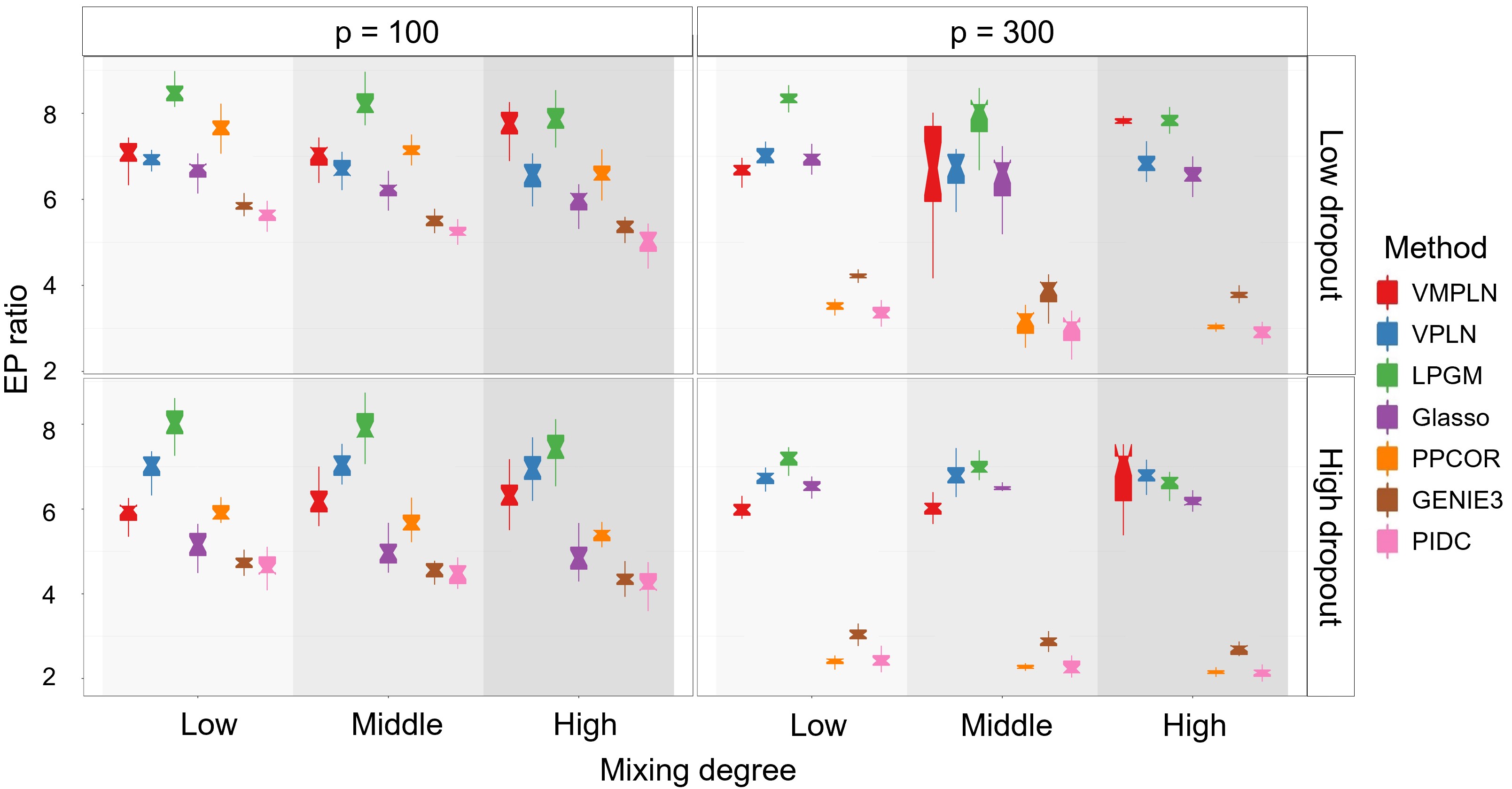}}
		\quad
	\end{center}
	\caption{Similar to Figure 1 for the blocked random graph and data generated from the compositional model.}	
\end{figure}

\begin{figure}[H]
	\begin{center}
		\subfigure[AUPRC ratios]{
			\includegraphics[height = 9 cm,width = 16 cm]{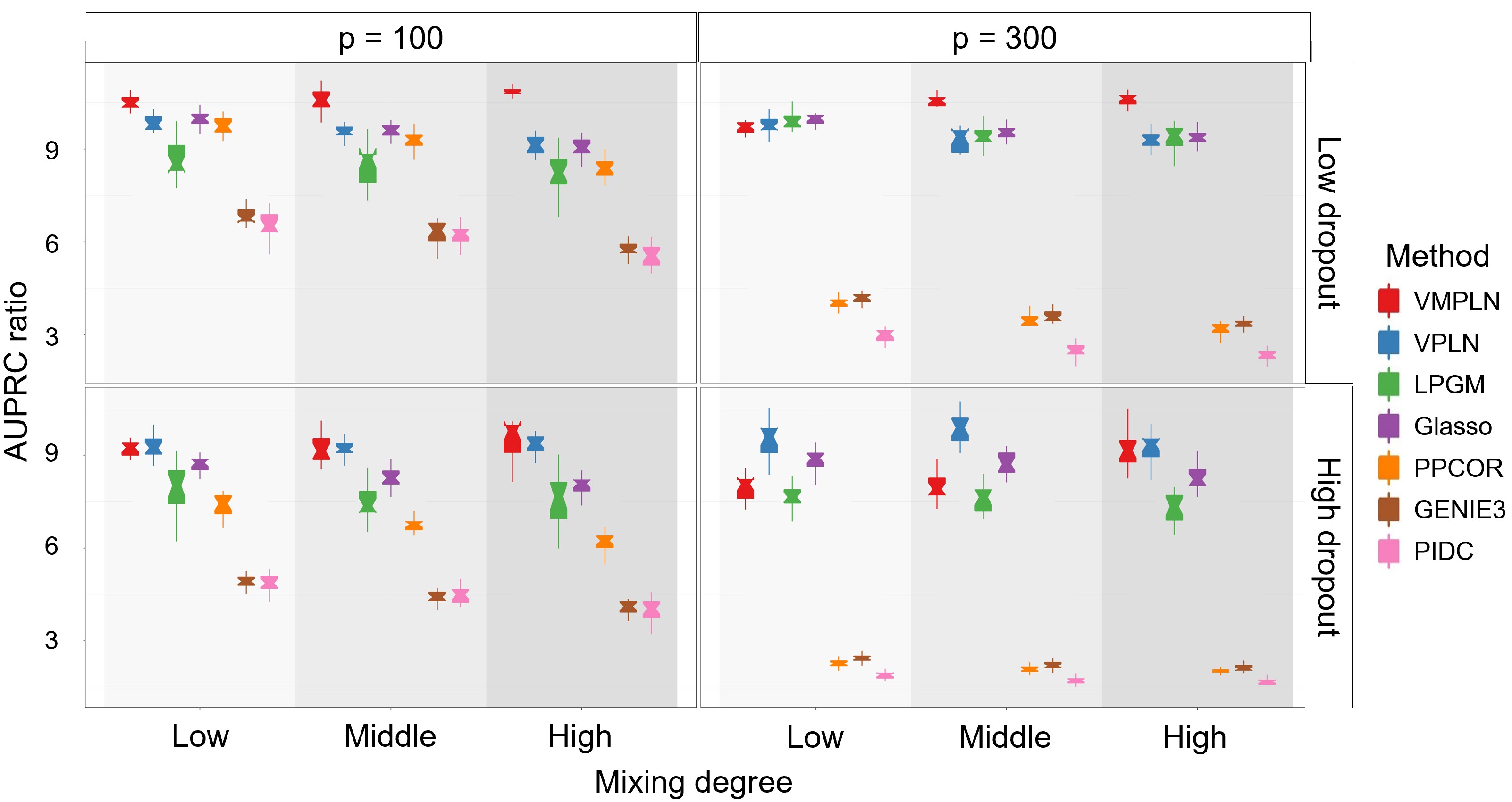}}
		\quad
		\subfigure[EP ratios]{
			\includegraphics[height = 9 cm,width = 16 cm]{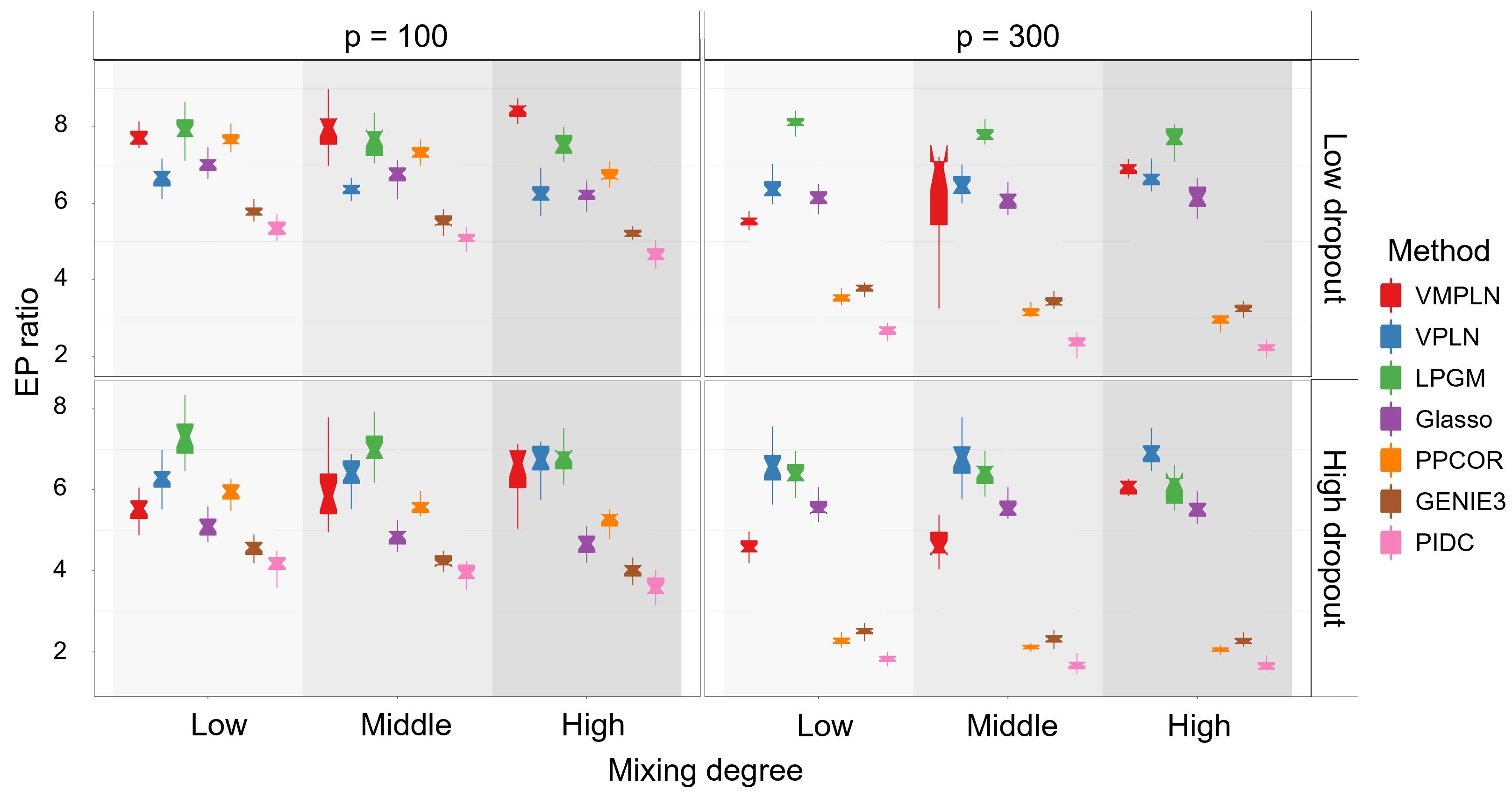}}
		\quad
	\end{center}
	\caption{Similar to Figure 1 for the scale-free graph and data generated from the compositional model.}	
\end{figure}

\begin{figure}[H]
	\begin{center}
		\subfigure[AUPRC ratios]{
			\includegraphics[height = 9 cm,width = 16 cm]{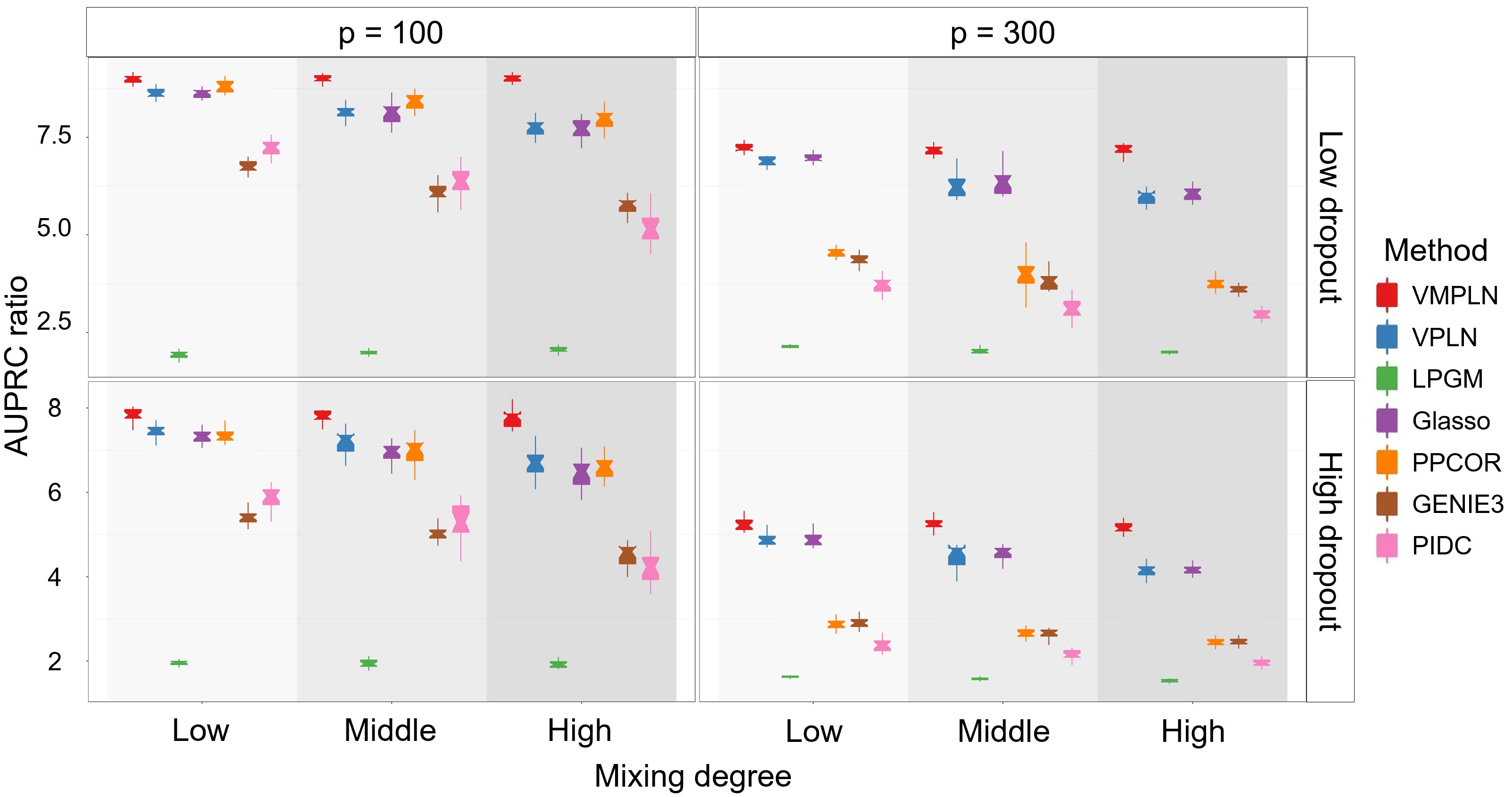}}
		\quad
		\subfigure[EP ratios]{
			\includegraphics[height = 9 cm,width = 16 cm]{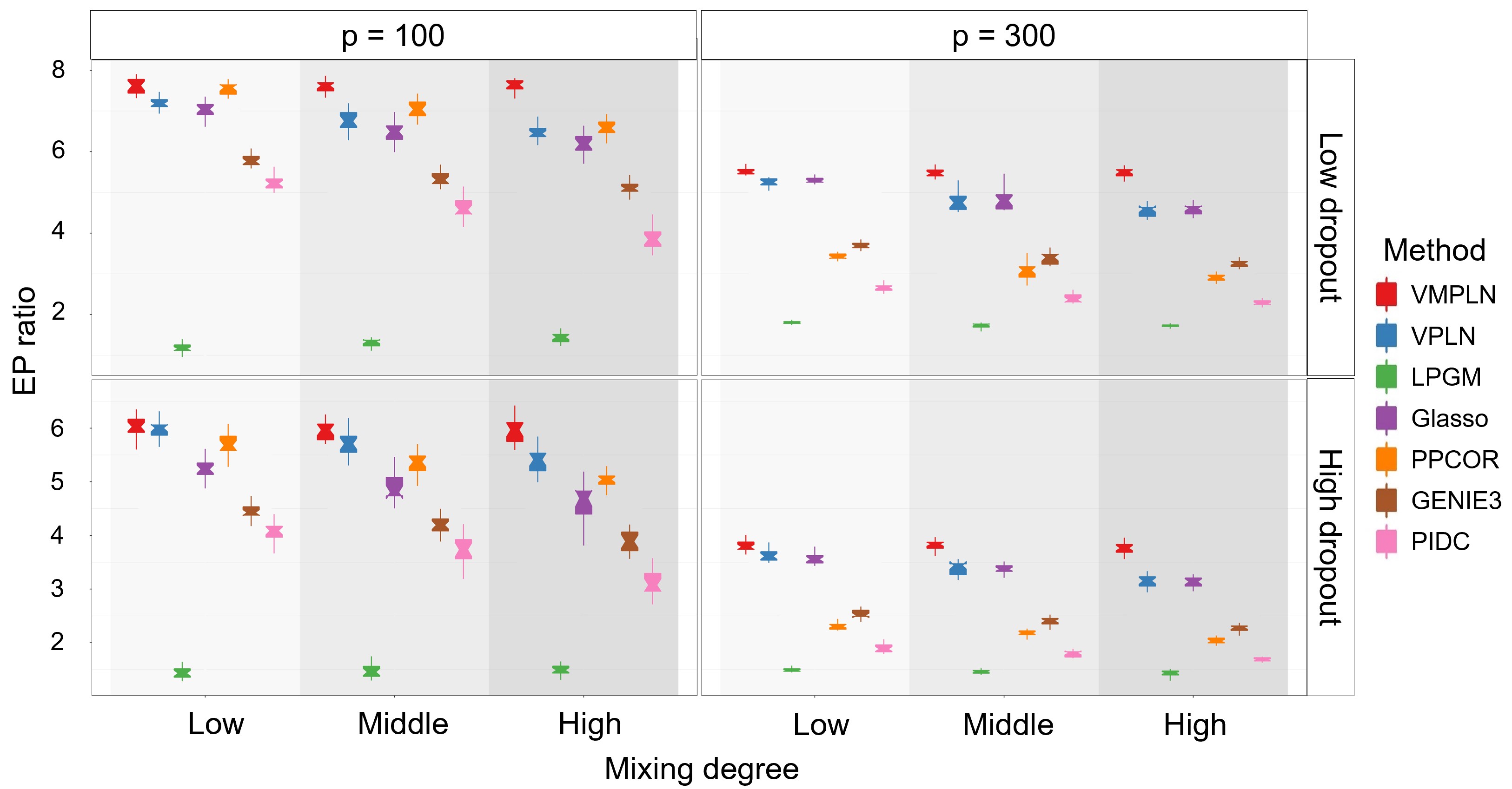}}
		\quad
	\end{center}
	\caption{Similar to Figure 2 for the random graph and data generated from the compositional model.}
\end{figure}

\begin{figure}[H]
	\begin{center}
		\subfigure[AUPRC ratios]{
			\includegraphics[height = 9 cm,width = 16 cm]{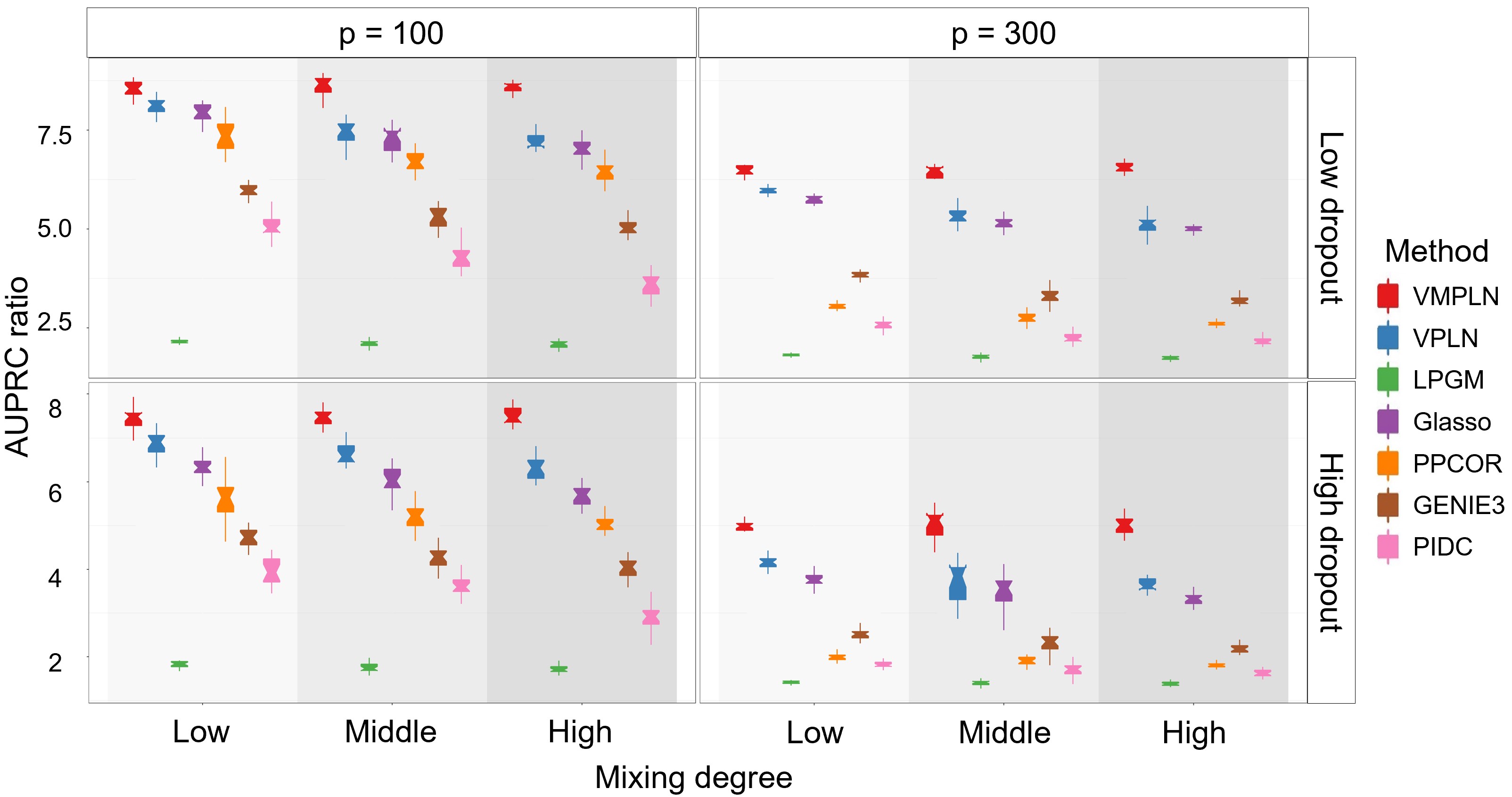}}
		\quad
		\subfigure[EP ratios]{
			\includegraphics[height = 9 cm,width = 16 cm]{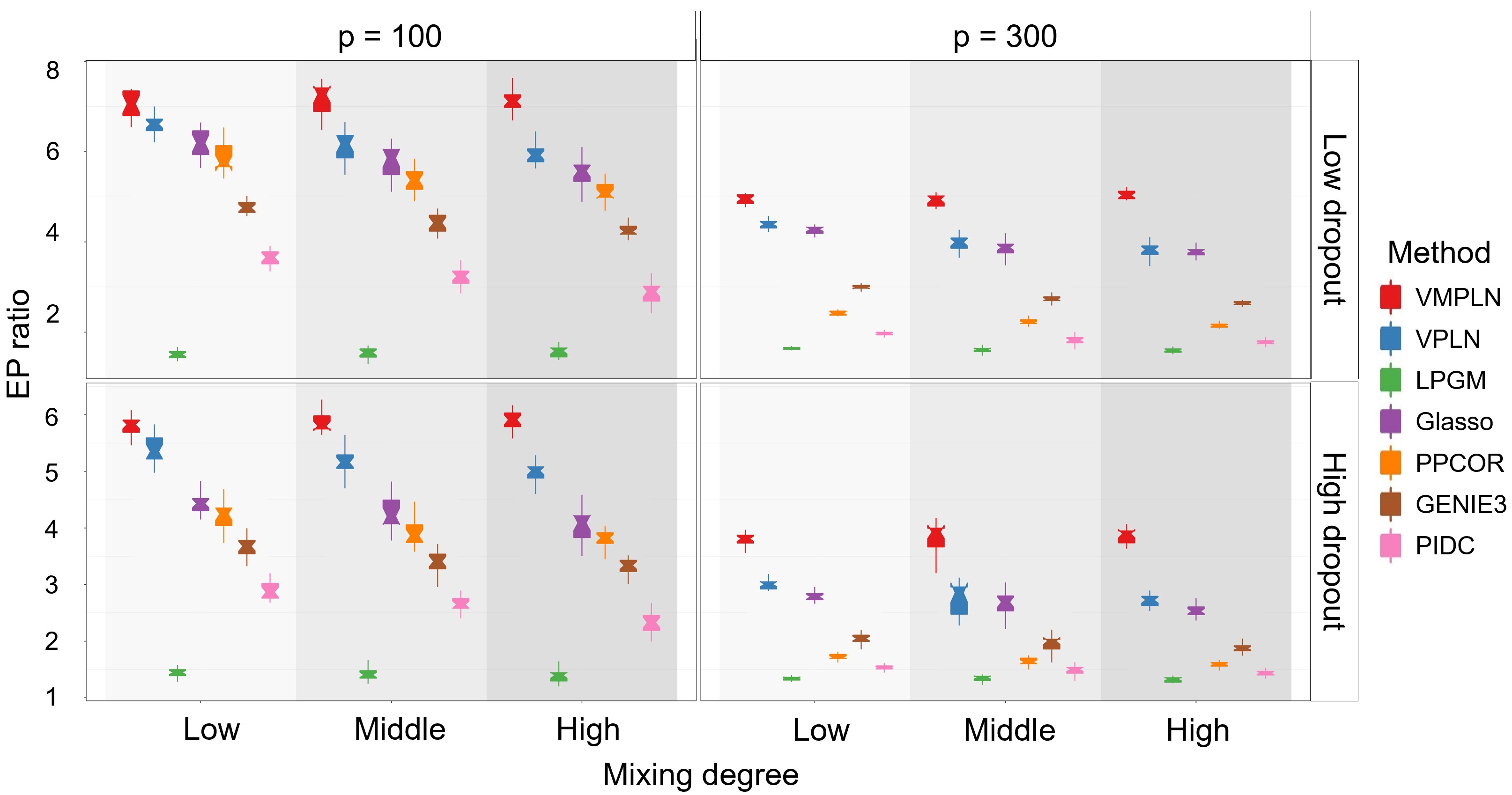}}
		\quad
	\end{center}
	\caption{Similar to Figure 2 for the hub graph and data generated from the compositional model.}	
\end{figure}

\begin{figure}[H]
	\begin{center}
		\subfigure[AUPRC ratios]{
			\includegraphics[height = 9 cm,width = 16 cm]{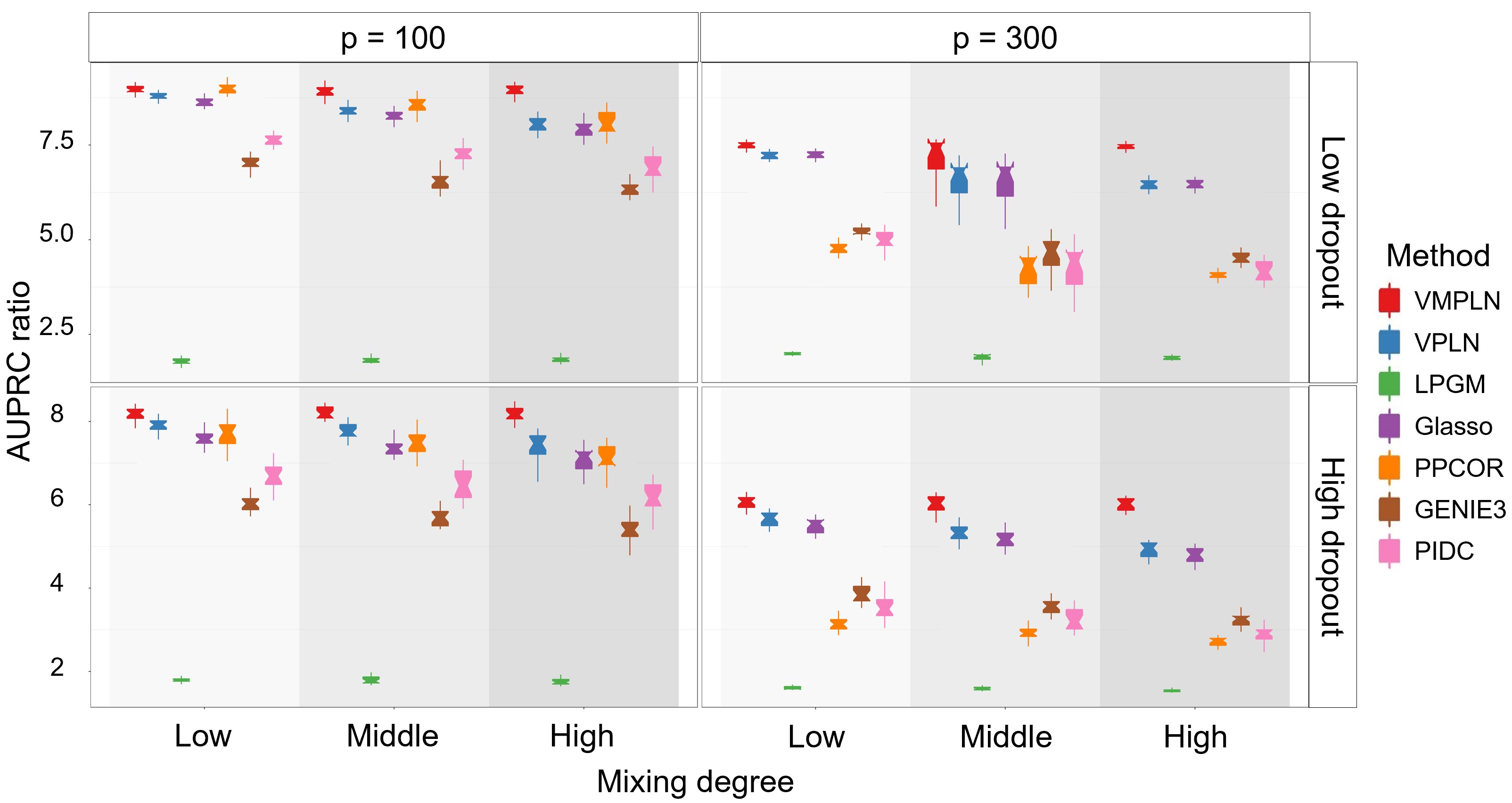}}
		\quad
		\subfigure[EP ratios]{
			\includegraphics[height = 9 cm,width = 16 cm]{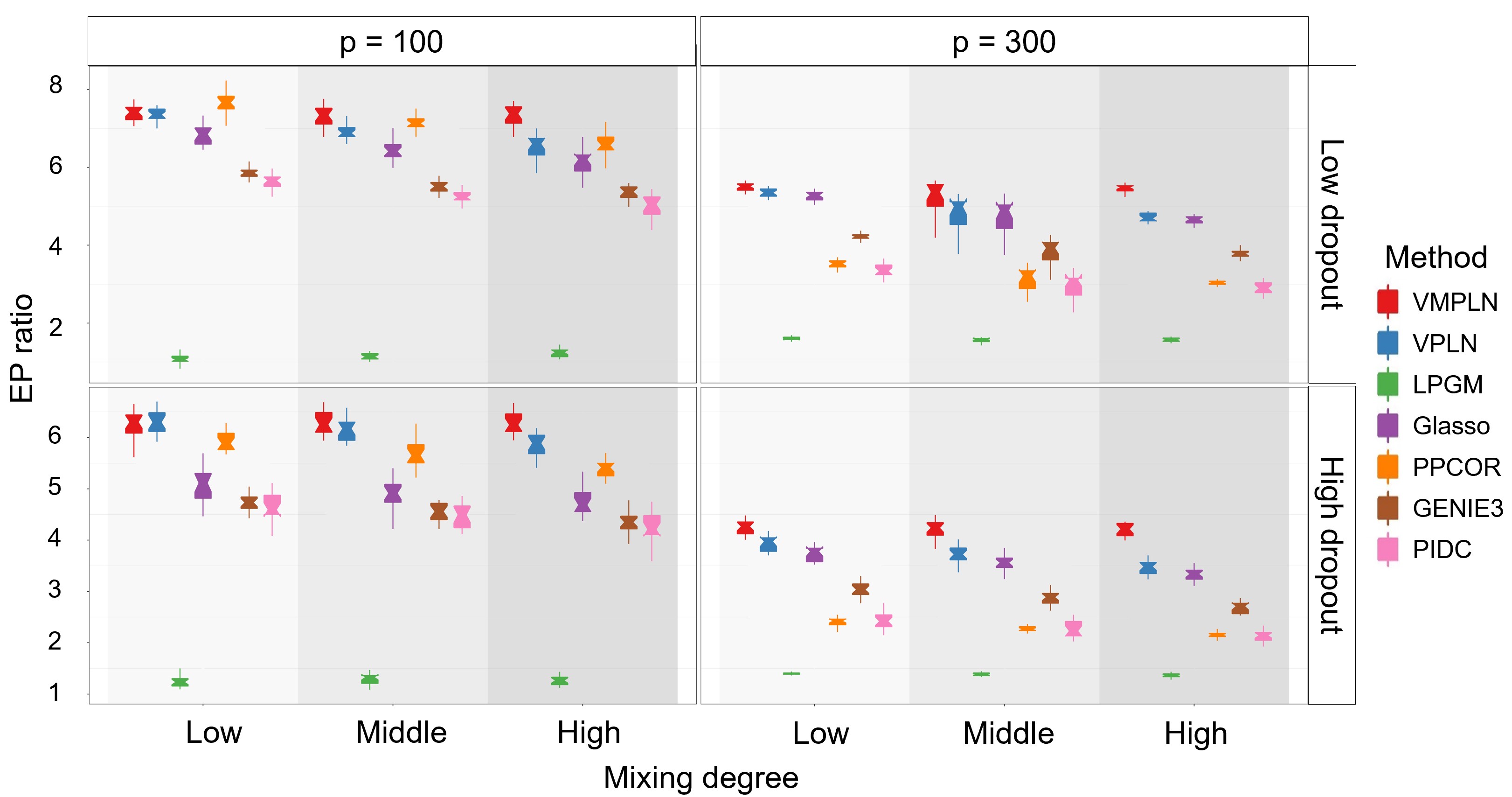}}
		\quad
	\end{center}
	\caption{Similar to Figure 2 for the blocked random graph and data generated from the compositional model.}	
\end{figure}

\begin{figure}[H]
	\begin{center}
		\subfigure[AUPRC ratios]{
			\includegraphics[height = 9 cm,width = 16 cm]{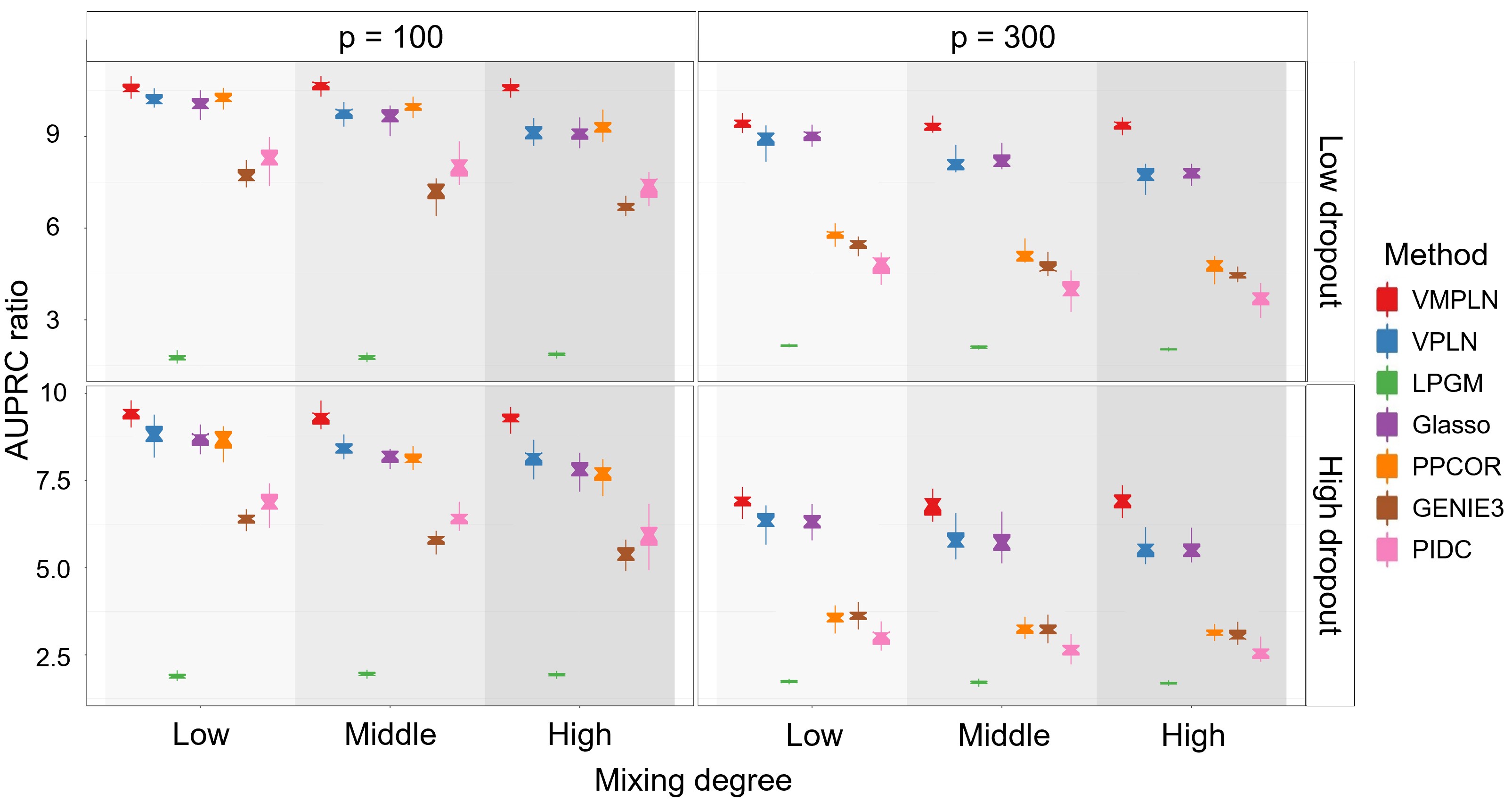}}
		\quad
		\subfigure[EP ratios]{
			\includegraphics[height = 9 cm,width = 16 cm]{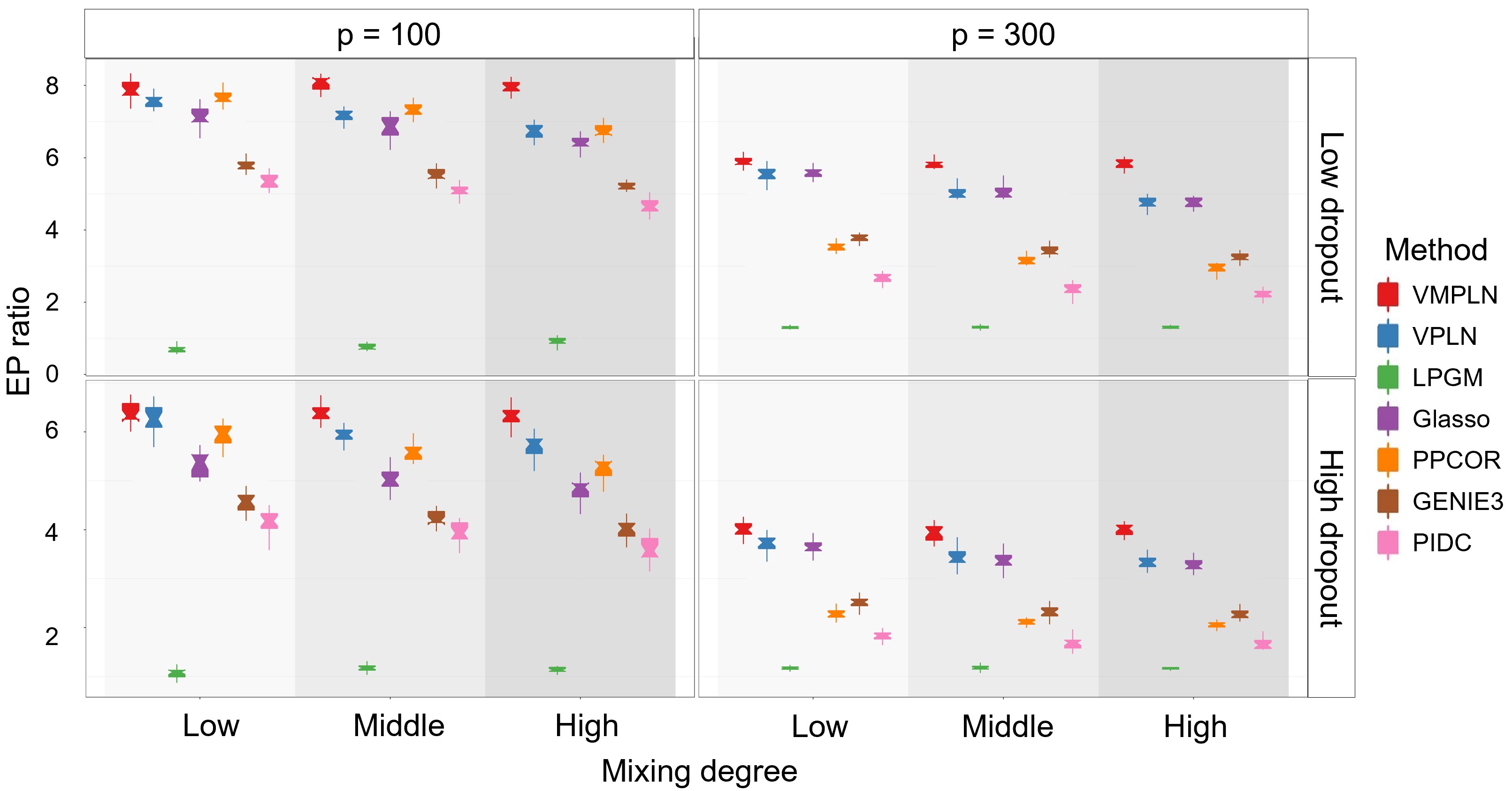}}
		\quad
	\end{center}
	\caption{Similar to Figure 2 for the scale-free random graph and data generated from the compositional model.}	
\end{figure}

\begin{figure}[htb]
	\begin{center}
		\includegraphics[height = 9 cm,width = 16 cm]{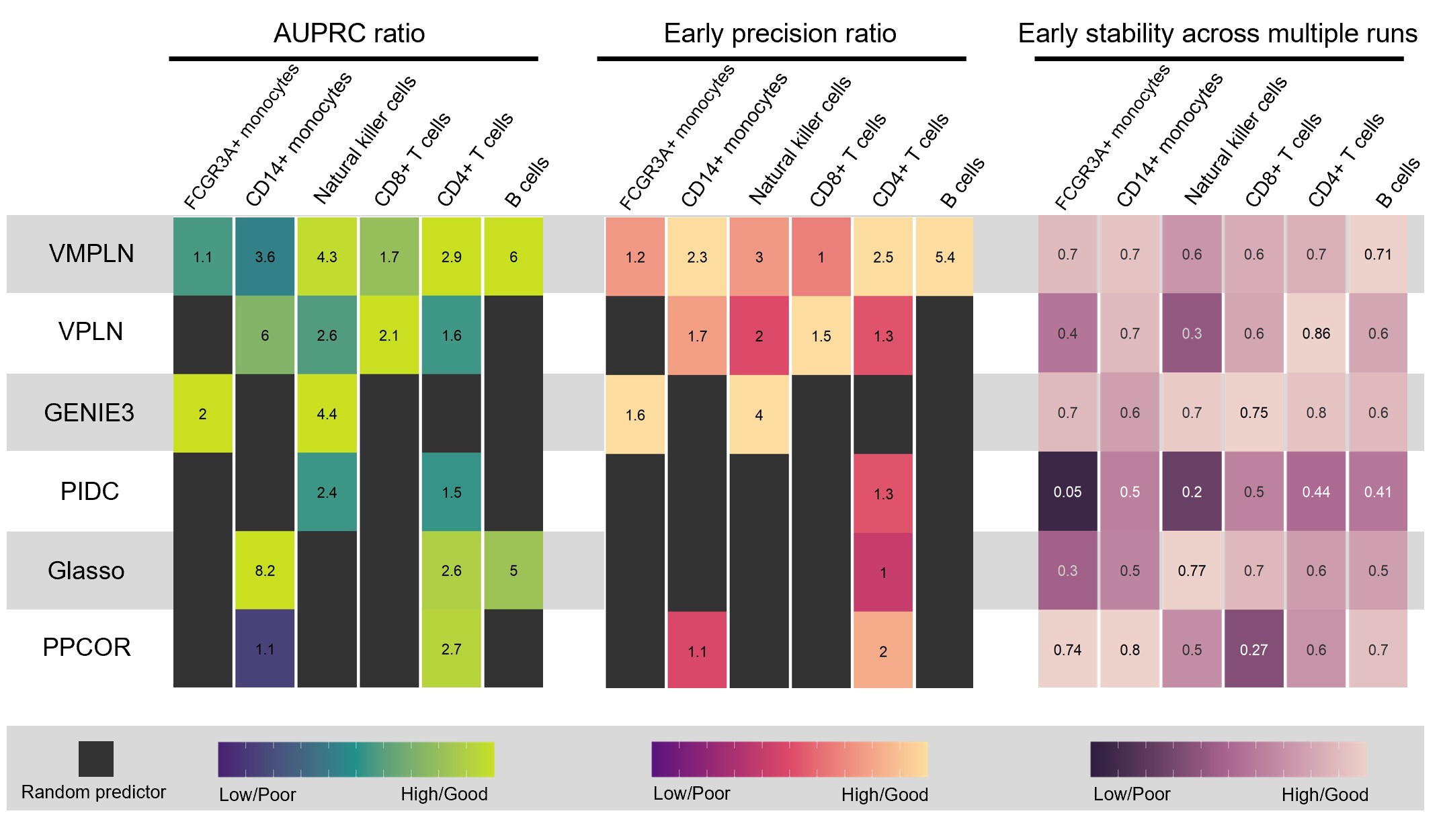}
	\end{center}
	\caption{The performance of the netwok inference algorithms in the Zheng data. The gene set of interest consists of 19 TFs and 481 non-TFs. The colors represent the scaled values of these metrics (scaled to between 0 and 1 within each cell type) and the actual values are marked in the boxes. Black color in the boxes: random predictor performs better.}
\end{figure}

\begin{figure}[H]
	\begin{center}
		\includegraphics[height = 9.0 cm,width = 15 cm ]{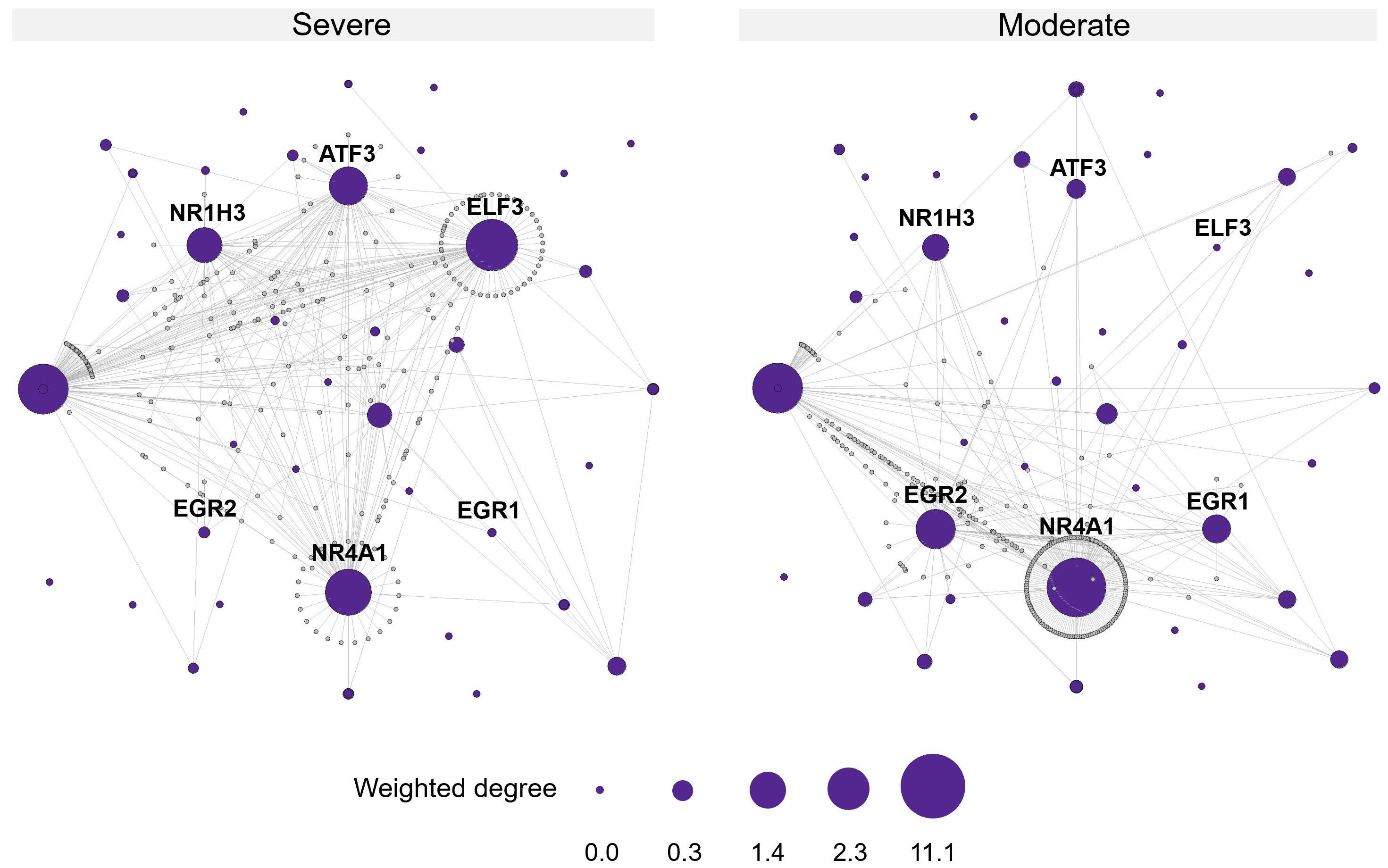}
	\end{center}
	\caption{Similar to Figure 4 a for Group1 macrophages.}
	
\end{figure}

\begin{figure}[H]
	\begin{center}
		\includegraphics[height = 9.0 cm,width = 15 cm ]{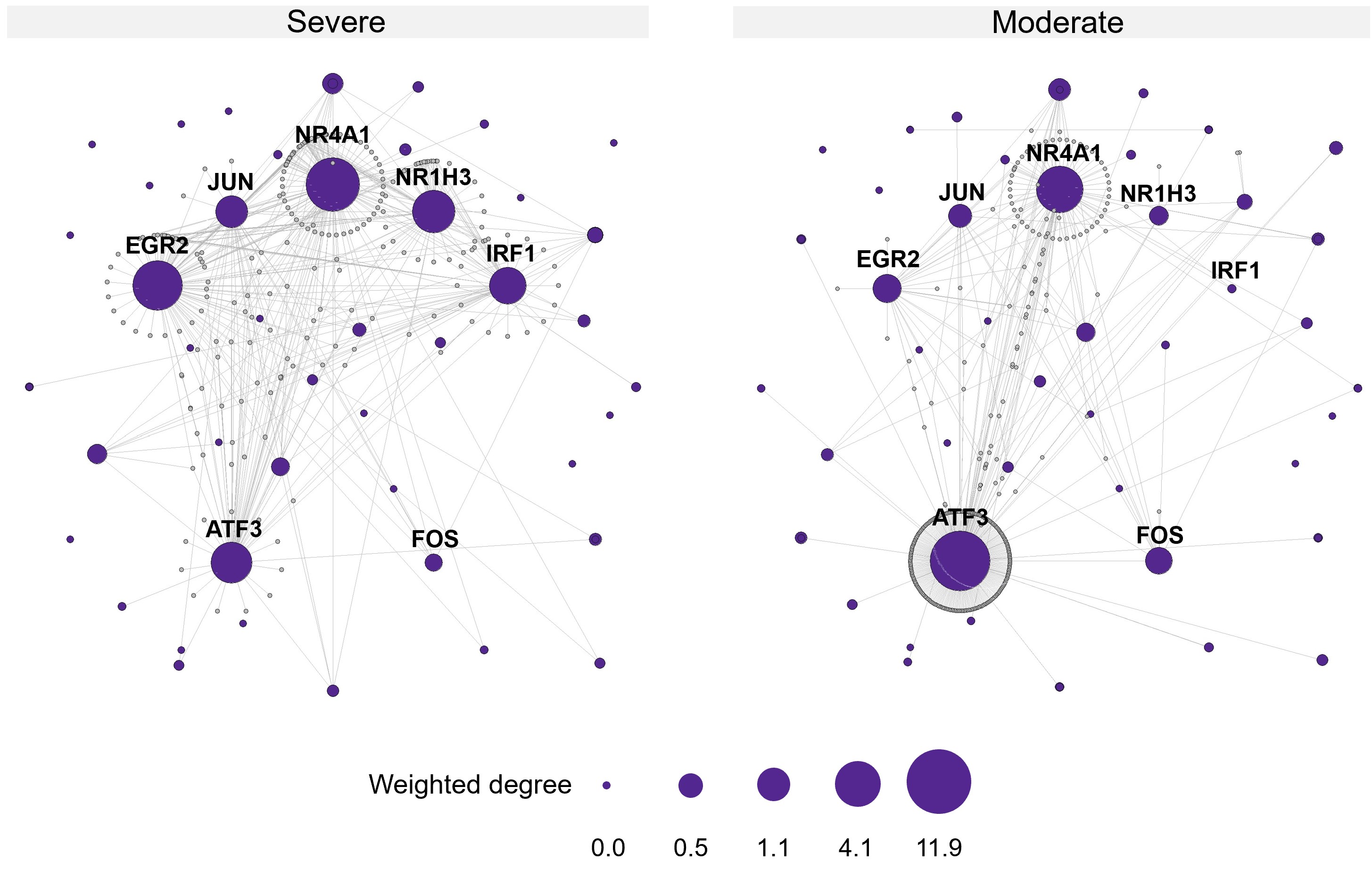}
	\end{center}
	\caption{Similar to Figure 4 a for Group2 macrophages.}
	
\end{figure}

\begin{figure}[H]
	\begin{center}
		\includegraphics[height = 9.0 cm,width = 15 cm ]{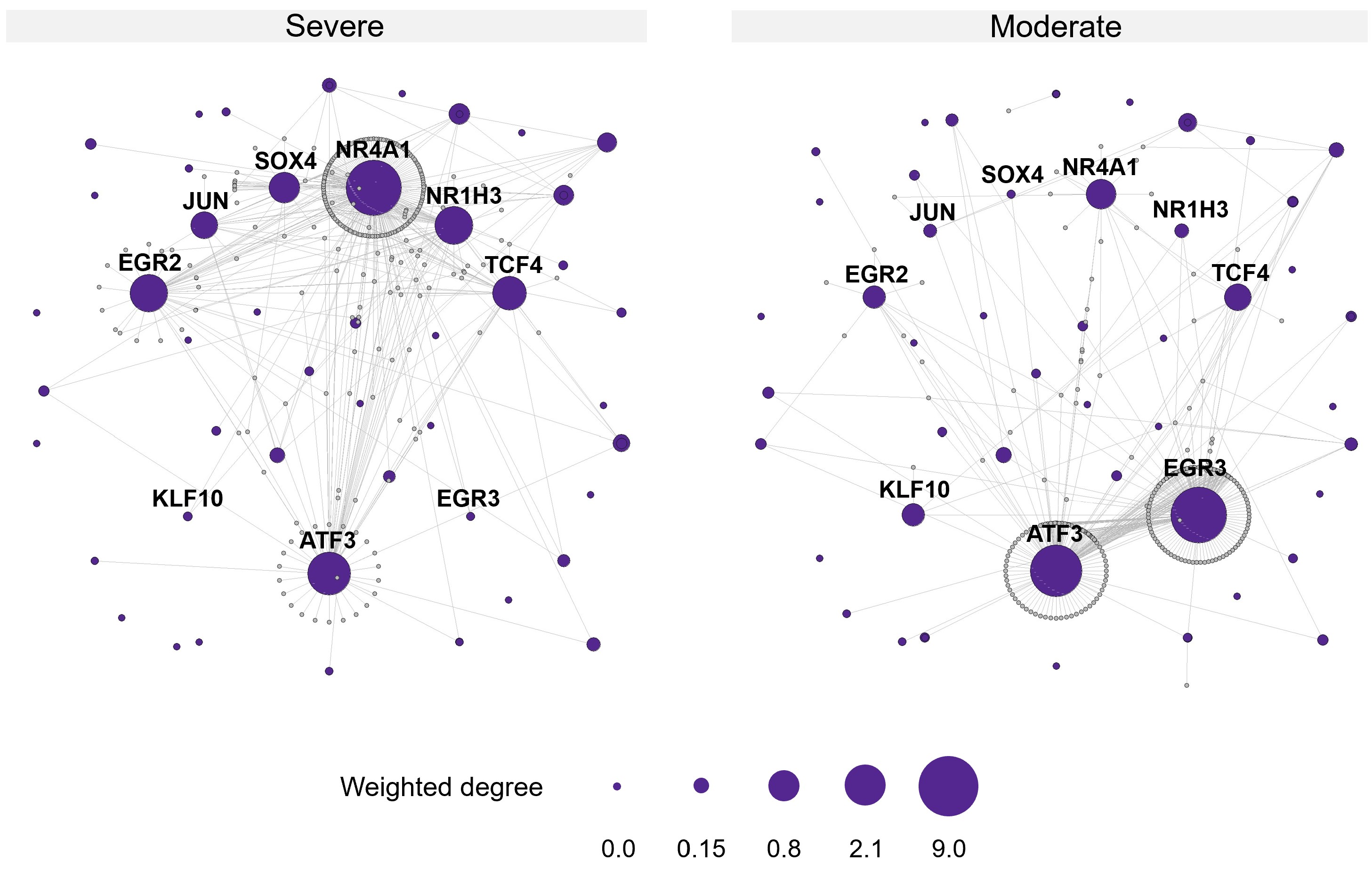}
	\end{center}
	\caption{Similar to Figure 4 a for Group3 macrophages.}
	
\end{figure}

\begin{figure}[H]
	\begin{center}
		\includegraphics[height = 10.0 cm,width = 14 cm ]{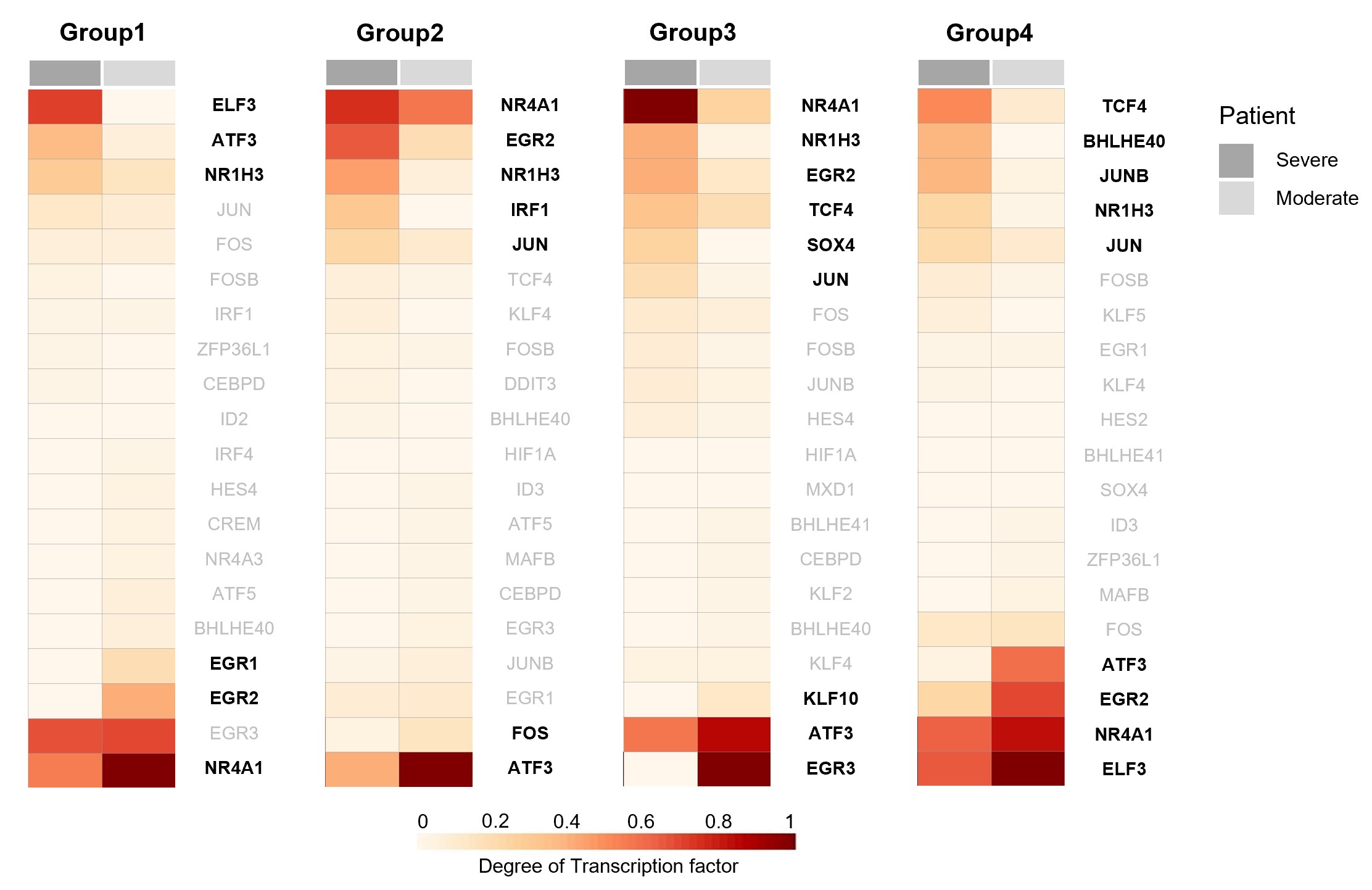}
	\end{center}
	\caption{The scaled node degree of TFs of 4 macrophage groups between severe patients and moderate patients. The node degree is calculated based on the absolute partial correlation matrix induce by the estimated precision matrix. Black bold reprents TFs that exhibit a large degree difference between the GRNs in the moderate and severe patients (greater than 0.2).}
	
\end{figure}

\begin{figure}[H]
	\begin{center}
		\includegraphics[height = 9.5 cm,width = 16 cm ]{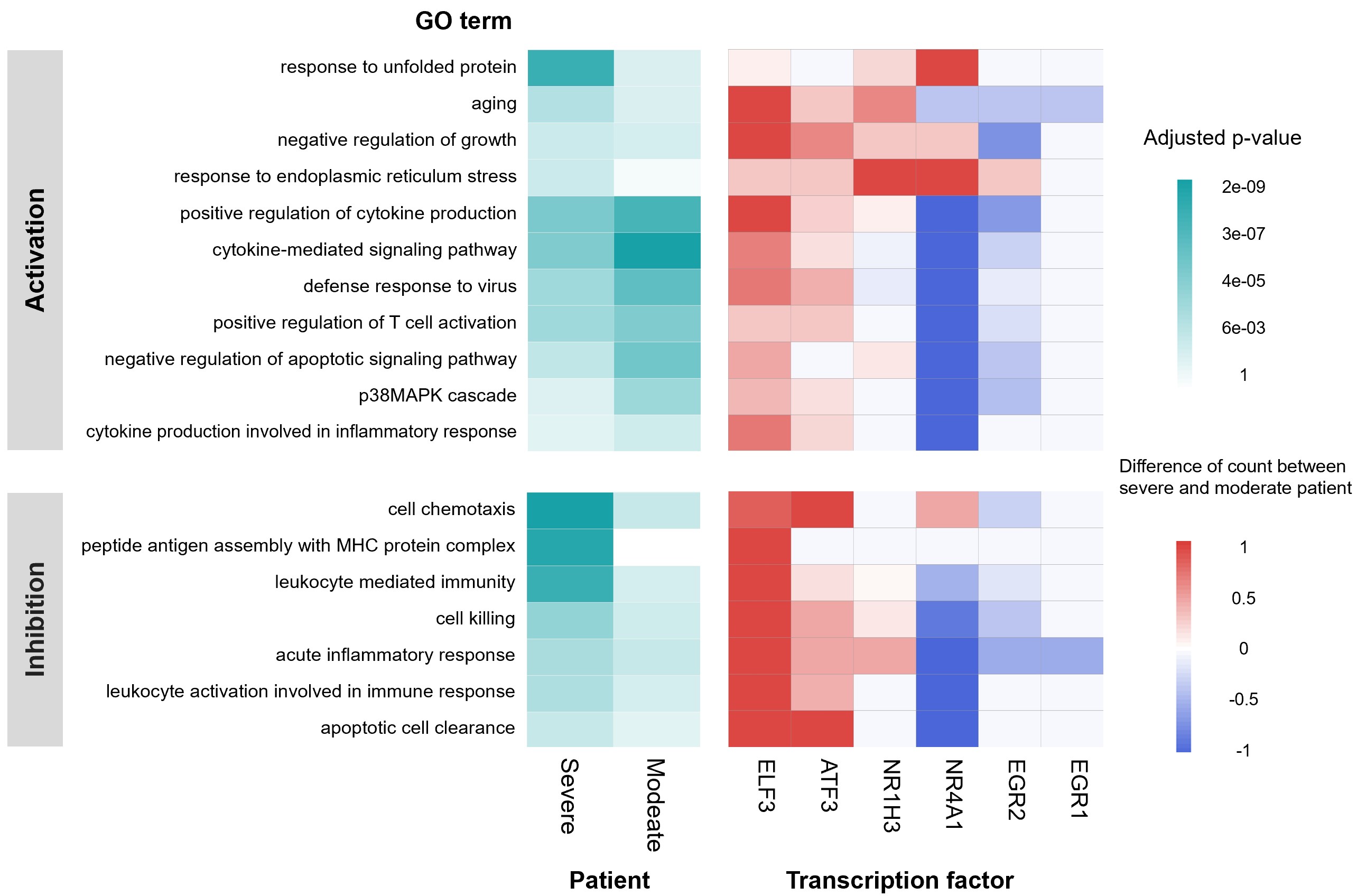}
	\end{center}
	\caption{Similar to Figure 4 b for Group1 macrophages.}
	
\end{figure}

\begin{figure}[H]
	\begin{center}
		\includegraphics[height = 9.5 cm,width = 16 cm ]{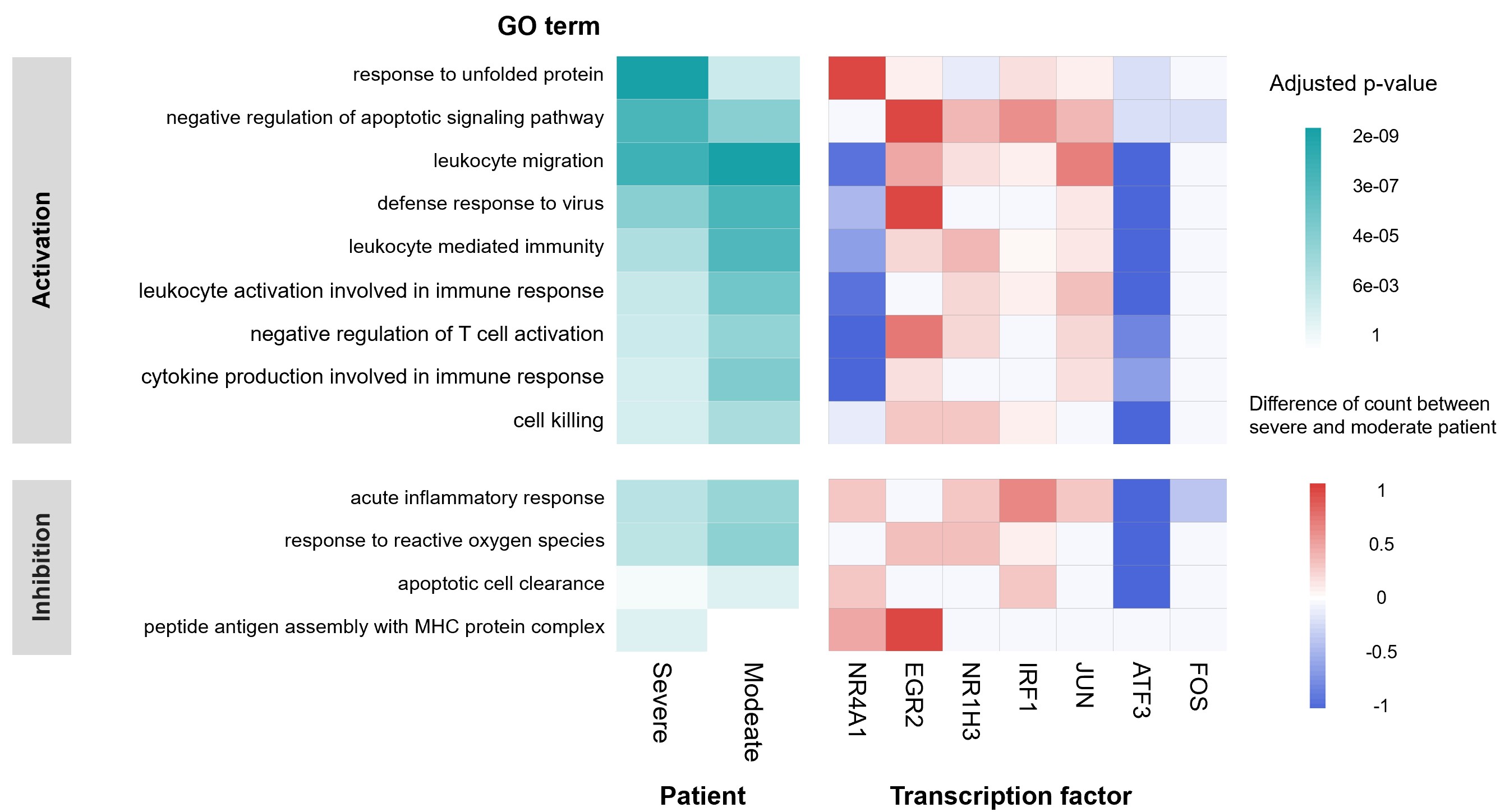}
	\end{center}
	\caption{Similar to Figure 4 b for Group2 macrophages.}
	
\end{figure}

\begin{figure}[H]
	\begin{center}
		\includegraphics[height = 9.5 cm,width = 16 cm ]{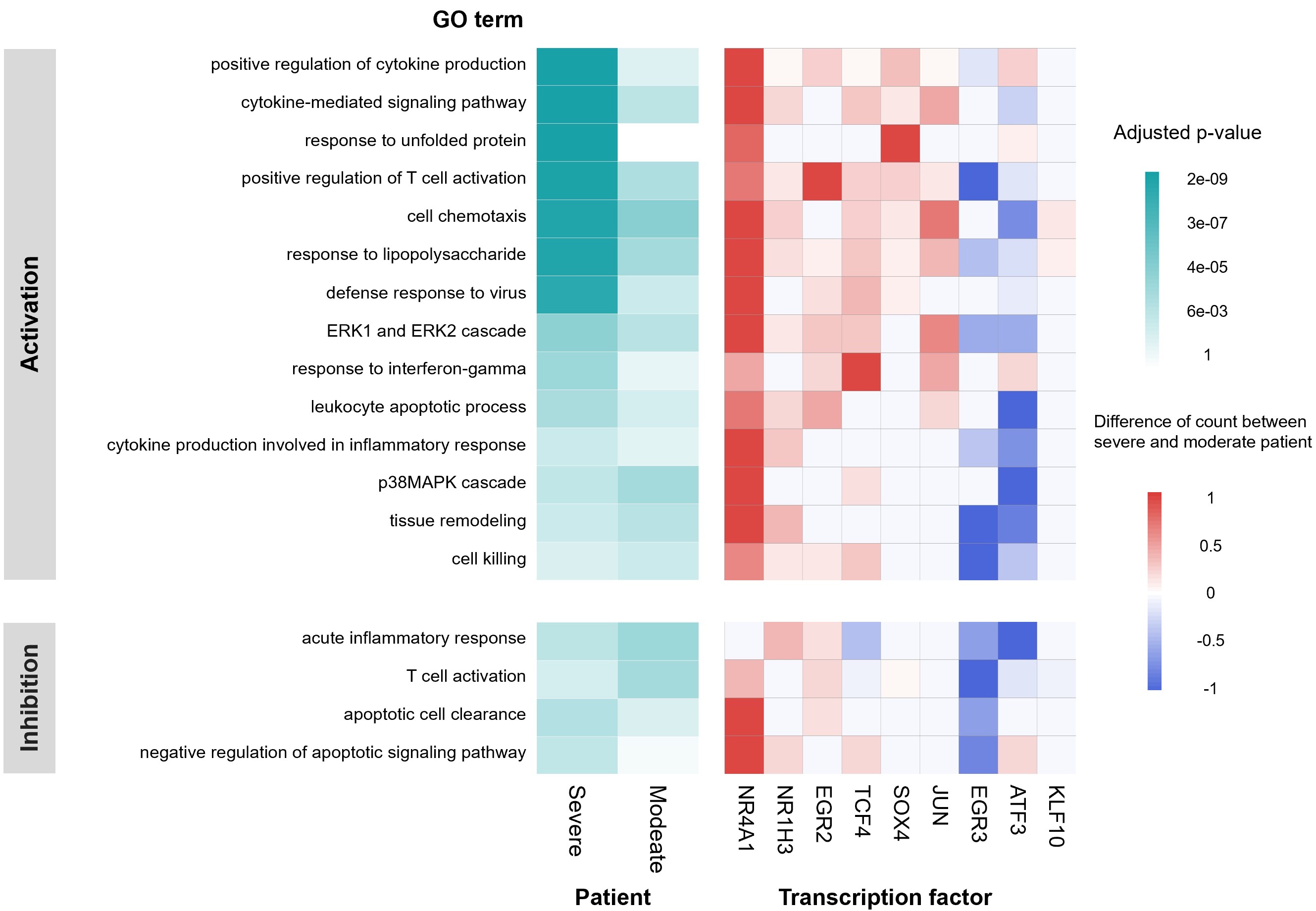}
	\end{center}
	\caption{Similar to Figure 4 b for Group3 macrophages.}
	
\end{figure}

\bibliographystyle{Chicago}

\bibliography{Bibliography-MM-MC}
\end{document}